\documentclass{aa}
\usepackage{natbib}
\bibpunct{(}{)}{;}{a}{}{,}
\usepackage{xstring}
\usepackage{multirow}
\usepackage{tikz}
\usepackage{threeparttable}
\usepackage{makecell}
\usepackage{color}
\usepackage{graphicx} 
\usepackage{txfonts}
\usepackage[breaklinks]{hyperref}

           \defcitealias{pierre16}{XXL~Paper~I}    
\defcitealias{2016A&A...592A...2P}{XXL~Paper~II}   
\defcitealias{2016A&A...592A...3G}{XXL~Paper~III}  
\defcitealias{2016A&A...592A...4L}{XXL~Paper~IV}   
\defcitealias{2016A&A...592A...5F}{XXL~Paper~VI}   
\defcitealias{2016A&A...592A...6P}{XXL~Paper~VII}  
            \defcitealias{baran16}{XXL~Paper~IX}   
          \defcitealias{smolcic16}{XXL~Paper~XI}   
\defcitealias{2016A&A...592A..11K}{XXL~Paper~XII}  
\defcitealias{2016A&A...592A..12E}{XXL~Paper~XIII} 
\defcitealias{2016MNRAS.462.4141L}{XXL~Paper~XV}   
        \defcitealias{butler2018a}{XXL~Paper~XVIII}
          \defcitealias{adami2018}{XXL~Paper~XX}   
     \defcitealias{guglielmo2018a}{XXL~Paper~XXII} 
        \defcitealias{smolcic2018}{XXL~Paper~XXIX} 
          \defcitealias{lucio2018}{XXL~Paper~XXVII}
     \defcitealias{guglielmo2018b}{XXL~Paper~XXX}  
        \defcitealias{butler2018b}{XXL~Paper~XXXI} 

\begin{document} 

\title{The XXL Survey: {\sc XXXIV}. Double irony in XXL-North}
\subtitle{A tale of two radio galaxies in a supercluster at z = 0.14}

\author{C.~Horellou\inst{1} 
\and H.T.~Intema\inst{2} 
\and V.~Smol{\v{c}}i{\'{c}}\inst{3}  
\and A.~Nilsson\inst{1} 
\and F.~Karlsson\inst{1} 
\and C.~Krook\inst{1}
\and L.~Tolliner\inst{1} %
\and C.~Adami\inst{4}
\and C.~Benoist\inst{5} 
\and M.~Birkinshaw\inst{6} 
\and C. Caretta\inst{7}
\and L.~Chiappetti\inst{8} 
\and J.~Delhaize\inst{3} 
\and C.~Ferrari\inst{5} 
\and S.~Fotopoulou\inst{9}
\and V.~Guglielmo\inst{10,}\inst{17} 
\and K.~Kolokythas\inst{11} 
\and F.~Pacaud\inst{12} 
\and M.~Pierre\inst{13,}\inst{14}
\and B.M.~Poggianti\inst{10} 
\and M.E.~Ramos-Ceja\inst{12} 
\and S.~Raychaudhury\inst{11} 
\and H.J.A.~R\"{o}ttgering\inst{2} 
\and C.~Vignali\inst{15,}\inst{16}
   }

    \institute{
    Chalmers University of Technology,
    Dept of Space, Earth and Environment, 
    Onsala Space Observatory, 
    SE-439 92 Onsala, Sweden
    \email{cathy.horellou@chalmers.se}
\and 
Leiden Observatory, Leiden University, Niels Bohrweg 2, 2333 CA, Leiden, The Netherlands
\and 
Department of Physics, Faculty of Science, University of Zagreb,  Bijeni\v{c}ka cesta 32, 10000  Zagreb, Croatia 
\and 
Aix Marseille Universit\'e, CNRS, LAM (Laboratoire d'Astrophysique de Marseille) UMR 7326, 13388, Marseille, France
\and 
Laboratoire Lagrange, Universit\'e C\^ote d'Azur, Observatoire de la C\^ote d'Azur, CNRS, 
Blvd de l'Observatoire, CS 34229, 06304 Nice Cedex 4, France  
\and 
H.H. Wills Physics Laboratory, University of Bristol, Tyndall Avenue, Bristol BS8 1TL, U.K.
\and 
    Departamento de Astronom\'ia, DCNE-CGT, Universidad de Guanajuato; Callej\'on de Jalisco, S/N, Col. Valenciana, 36240, Guanajuato, Gto., Mexico
\and 
    INAF, IASF Milano, via Bassini 15, I-20133 Milano, Italy
\and 
Centre for Extragalactic Astronomy, Department of Physics, Durham University, South Road, Durham DH1 3LE, U.K. 
\and 
    INAF -- Astronomical Observatory of Padova, I-35122 Padova, Italy
\and 
Inter-University Centre for Astronomy and Astrophysics, Pune 411007, India
\and 
Argelander-Institut f\"ur Astronomie, Universit\"at Bonn, Auf dem H\"ugel 71, D-53121 Bonn, Germany
\and 
IRFU, CEA, Université Paris-Saclay, F-91191 Gif sur Yvette, France
\and 
Universit\'e Paris Diderot, AIM, Sorbonne Paris Cit\'e, CEA, CNRS, F-91191 Gif sur Yvette, France
\and 
Dipartimento di Fisica e Astronomia, Alma Mater Studiorum, Universit\`a degli Studi di Bologna,
Via Gobetti 93/2, 40129 Bologna, Italy
\and 
INAF -- Osservatorio di Astrofisica e Scienza dello Spazio,
Via Gobetti 93/3, 40129 Bologna, Italy
\and 
Max-Planck-Institut f\"ur Extraterrestrische Physik, Giessenbachstrasse 1, D-85748 Garching bei M\"unchen, Germany
}
  \date{Received 6 March 2018 / Accepted 12 July 2018}

  \abstract
  {}
   {We show how the XXL multiwavelength survey can be used to shed light on radio galaxies and their environment.} 
   {Two prominent radio galaxies were identified in a visual examination of the mosaic of XXL-North obtained with the Giant Metrewave Radio Telescope at 610~MHz. Counterparts were searched for in other bands. Spectroscopic redshifts from the GAMA database were used to identify clusters and/or groups of galaxies, 
   estimate their masses with the caustic method, and quantify anisotropies in the surrounding galaxy distribution via a Fourier analysis.}
   {Both radio galaxies are of FR~{\sc i} type and are hosted by early-type galaxies at a redshift of 0.138. 
   The first radio source, named the Exemplar, has a physical extent of $\sim 400$~kpc;  
   it is located in the cluster XLSSC~112, which has a temperature of $\sim 2$~keV, a total mass of $\sim10^{14} M_\odot$, and resides in an XXL supercluster with eight known members. 
   The second source, named the Double Irony, is a giant radio galaxy with a total length of about 1.1~Mpc.  
   Its core coincides with a cataloged point-like X-ray source, but no extended X-ray emission from a surrounding galaxy cluster was detected. However, from the optical data we determined that the host is the brightest galaxy in a group that is younger, less virialized, and less massive than the Exemplar's cluster. A friends-of-friends analysis showed that the Double Irony's group is a member of the same supercluster as the Exemplar. There are indications that the jets and plumes of the Double Irony have been deflected by gas associated with the surrounding galaxy distribution. Another overdensity of galaxies (the tenth) containing a radio galaxy was found to be associated with the supercluster.}
   {Radio galaxies can be used to find galaxy clusters/groups that are below the current sensitivity of X-ray surveys.}
   \keywords{
   radiation mechanisms: non-thermal
-- radio continuum: galaxies
-- galaxies: active   
-- galaxies: magnetic fields 
-- methods: data analysis 
               }

  \maketitle
\section{Introduction}
\begin{figure*}
\centering
\includegraphics[width=8.8cm]{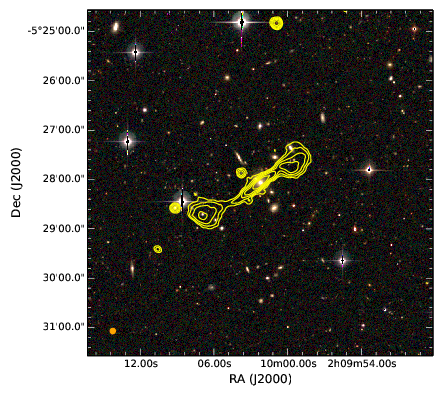}
\includegraphics[width=8.8cm]{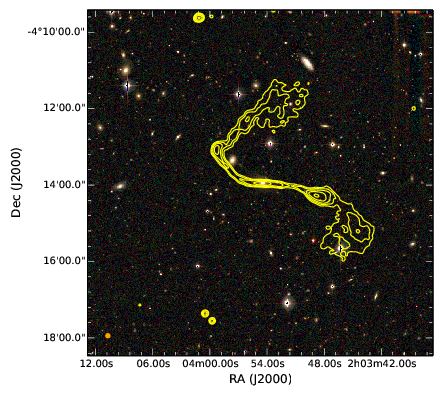} 
\caption{
  {\it Left panel:}  Exemplar. The size of the image is $7'\times7'$. 
  {\it Right panel:}  Double Irony. The size of the image is $9'\times9'$.
  Both images show GMRT 610~MHz isocontours superimposed on three-color ($gri$) CFHT images of the fields. 
  The contour levels increase by a factor of two from 0.3~mJy~beam$^{-1}$ (which corresponds to 5$\sigma$) to 4.8~mJy~beam$^{-1}$. 
  The synthesized beam ($6\farcs5$) is shown in orange in the bottom left corner of each image. 
}
\label{figCFHTcolGMRTcont}  
\end{figure*}

The study of radio galaxies benefits enormously from a multiwavelength approach, from the X-ray to the radio (e.g., \citealt{2006LNP...693...39W}). 
X-ray emission can be produced both by the active galactic nucleus (AGN) and the diffuse surrounding hot gas. 
Optical and infrared imaging and spectroscopy provide valuable information on the host galaxy, 
which most of the time is an early-type galaxy with  very little or no star formation 
(some host galaxies, however, do contain dust and gas in a disk; e.g., \citealt{2016A&A...592A..20D}). 
In the radio, synchrotron radiation directly shows the extent and structure of the emitting plasma that contains relativistic electrons and magnetic fields. 
The jets may cover tens to hundreds of kiloparsecs before terminating in extended lobes. 
These jets and lobes may heat the intergalactic gas, displace material, and create cavities and shocks 
that are seen in the X-ray surface brightness distribution of galaxy clusters (e.g., \citealt{2005Natur.433...45M}). 
This process, called radio-mode AGN feedback,   
is believed to play an important role in preventing gas infall on the host galaxy and regulating its evolution (e.g., \citealt{croton06}).

The XXL survey (\citealt{pierre16}, hereafter \citetalias{pierre16}) 
is particularly well suited to the study of AGN and galaxy clusters.
With a total observing time of 6.9~Ms, it is the largest X-ray survey carried out by XMM-{\it Newton}.  
It covers two fields of 25 square degrees each, one in the southern hemisphere (XXL-South) and the other at an equatorial declination (XXL-North).  
The X-ray observations have been complemented by an extensive  multiwavelength program.\footnote{\url{http://xxlmultiwave.pbworks.com}}

In this paper we discuss two double-lobed radio galaxies (see Fig.~\ref{figCFHTcolGMRTcont}) that we have identified in the  
mosaic of XXL-North obtained with the Giant Metrewave Radio Telescope (GMRT).  
The GMRT-XXL-N 610~MHz survey is presented in this issue (\citealt{smolcic2018}, hereafter \citetalias{smolcic2018}). 
It is of interest to compare the properties of the two sources as they are both at the same redshift ($z\simeq 0.14$) and the exact same data set is available for both. 
Their redshift is near the peak of the redshift distribution of the most 
securely detected clusters in XXL (\citealt{2016A&A...592A...3G}, hereafter \citetalias{2016A&A...592A...3G}). 
As our analysis  shows, the radio galaxies are both part of the same supercluster, but their immediate surroundings are different.

The first radio galaxy appears regular, with symmetric jets leading to two radio lobes. 
It is located at the center of a cluster and hosted by the brightest cluster galaxy (BCG), both of  
 which have been detected and characterized in studies of the sample of the brightest 100 XXL clusters 
 (\citealt{2016A&A...592A...2P}, hereafter \citetalias{2016A&A...592A...2P}; 
\citetalias{2016A&A...592A...3G}; 
\citealt{2016A&A...592A...4L}, hereafter \citetalias{2016A&A...592A...4L}; 
\citealt{2016A&A...592A..12E}, hereafter \citetalias{2016A&A...592A..12E}; 
\citealt{2016MNRAS.462.4141L}, hereafter \citetalias{2016MNRAS.462.4141L}). 
Because of its appearance with twin jets and plumes and its association with an early-type galaxy in a cluster,  
we  refer to this source in the remainder of this paper as {\emph {the Exemplar}}. 
The other radio galaxy stands out for its very large extent (a total length of $7\farcm56$, or 1100~kpc  
in our adopted cosmology\footnote{Throughout this paper and for consistency with the first XXL series of papers, 
we use the WMAP9 cosmology 
($\Omega_{\rm m} = 0.28$, 
$\Omega_\Lambda = 0.72$, 
$H_0 = 70$~km~s$^{-1}$~Mpc$^{-1}$; 
\citealt{2013ApJS..208...19H}). 
At the redshift of the radio galaxies discussed here ($z \simeq 0.138$), this gives a scale of 2.443 kpc/$\arcsec$ and a luminosity distance of 652.5~Mpc.})  
and its peculiar shape reminiscent of two reversed question marks,  the reason why we named it {\emph {the Double Irony.}}\footnote{The percontation point 
(a reversed question mark, \reflectbox?)
was invented by the English printer Henry Denham at the end of the sixteenth century to mark the end of a rhetorical question and was later used to denote irony.} 

\begin{table*}
\caption{Characteristics of the radio images.}
\centering 
\begin{tabular}{lcccc} 
\hline
\hline
Survey      & Frequency & Resolution    & Pixel size    &$\sigma$\\
            & (MHz)     &($\arcsec$)    &($\arcsec$)    & (mJy~beam$^{-1}$)\\
\hline
TGSS ADR1   & 150    & 25               &6.4    &5.0\\
GMRT-XXL-N  & 610    & 6.5              &1.9    &0.060\\
FIRST       & 1400   & $6.8\times 5.4$  &1.8    &0.15\\
NVSS        & 1400   & 45               &15     &0.45 (I), 0.29 (Q, U)\\
\hline
\end{tabular}
\label{tabRadioData}
\end{table*}

This paper is organized as follows. The available data sets are presented in Sect.~\ref{SectData}. 
In Sect.~\ref{SectExemplar} we discuss the Exemplar radio galaxy, its host galaxy, and host cluster; 
the radio images are presented and analyzed.
Section~\ref{SectDbleIrony} is about the Double Irony and follows a similar structure; an overdensity of galaxies is found. 
In Sect.~\ref{SectCompa} the two radio galaxies are compared and discussed. In particular, 
the distributions of surrounding galaxies are examined in relation to the orientation of the radio galaxies to search for environmental effects. 
In Sect.~\ref{SectSuperclu} it is shown that the group of galaxies associated with the Double Irony radio galaxy is part of the same supercluster as the Exemplar's cluster and this large-scale structure is discussed. 
In Sect.~\ref{SectDiscussionRadiogalsSuperclu} we discuss the occurrence of radio galaxies in superclusters and their potential use as tracers of the large-scale structure. 
We conclude in Sect.~\ref{SectSummary}. 
\vfill\eject
\section{Data and data analysis}
\label{SectData}

\subsection{Radio}
\label{SectDataRadio}

As detailed below, we used images at three different frequencies obtained from four different surveys 
carried out with the Very Large Array (VLA) and the GMRT. 
These surveys have different sensitivities and angular resolutions, and are also sensitive to different angular scales.  
The characteristics of the data are summarized in Table~\ref{tabRadioData}. 

The most important data set is the GMRT-XXL-N 610~MHz radio continuum survey described in \citetalias{smolcic2018}. 
    XXL-North contains the XMM-LSS area that was observed earlier with the GMRT (\citealt{tasse06}, \citeyear{tasse07}). 
The rest of the XXL-North field was recently observed at higher sensitivity, and the data were combined. 
The final mosaic has an angular resolution of $6\farcs5$. 
The two radio galaxies discussed here are located outside the XMM-LSS region, in the newly covered region of 12.66 square degrees that has an average sensitivity of 45~$\mu$Jy~beam$^{-1}$. 

The other radio continuum data were taken from the following, publicly available surveys:
\begin{itemize}
    \item the Tata Institute of Fundamental Research (TIFR) GMRT Sky Survey (TGSS) first alternative data release  (TGSS ADR1) 
at 150~MHz and 25$\arcsec$ resolution \citep{TGSS_ADR1}; 

    \item the Faint Images of the Radio Sky at Twenty Centimeters (FIRST) 
at 1.4~GHz, $6\farcs8\times 5\farcs4$ resolution 
 (\citealt{first95}; \citealt{first15}); 
    
    \item the NRAO VLA Sky Survey (NVSS) at 1.4~GHz, 45$\arcsec$ resolution (\citealt{NVSS}). 
NVSS also contains polarization information. We downloaded Stokes $Q$ and $U$ images from the NVSS website\footnote{\url{http://www.cv.nrao.edu/nvss}} and 
produced images of the polarized intensity and of the polarization angle using the task {\tt POLC} in the Astronomical Image Processing System\footnote{\url{http://www.aips.nrao.edu}} 
({\tt AIPS}) that corrects for the Ricean bias in polarized intensity \citep{1974ApJ...194..249W}. 
\end{itemize} 

Total-intensity images do not always do justice to the details of the surface brightness distributions in astronomical sources. Images of the norm of the intensity gradient, $|\nabla I|$, have been used to reveal the paths of jets and substructures in radio galaxies 
(e.g., \citealt{2011MNRAS.417.2789L}). 
We computed images of $|\nabla I|$ at 610~MHz using a Sobel filter (Sobel \& Feldman 1968, unpublished, reported in Wikipedia\footnote{\url{ https://en.wikipedia.org/wiki/Sobel\_operator}, 
page last edited on 22 February 2018, at 14:58}). 

Flux densities and spectral indices were measured within defined regions. The statistical uncertainties on the flux density measurements were calculated as $\sigma \sqrt{N_{\rm beam}}$, where $\sigma$ is the standard deviation of the noise and $N_{\rm beam}$ is the number of beams within the region. 

The survey images were used and no attempt was made to re-image the data to match the $(u,v)$ ranges or the imaging weighting scheme.  
Only the FIRST image of the Double Irony may have extended emission that has been filtered out 
(the largest angular scale of FIRST is $2'$), and this is  discussed in 
Sect.~\ref{SectDbleIronySpectralIndex}.  
The sources are barely resolved in TGSS and in NVSS, 
and some of the features of the Double Irony are only marginally detected in TGSS, leading to large uncertainties in the flux measurements that would not be improved much by re-imaging. 

Throughout the paper we use the following convention for the radio spectral index, $\alpha$: $S_\nu \propto \nu^\alpha$, where $S_\nu$ is the flux density at frequency $\nu$. The uncertainties on the spectral indices measured between two frequencies were calculated using standard error propagation, 
\begin{equation}
    \Delta\alpha = 
\sqrt{
    \left(
    \frac{\Delta S_1}{S_1}
    \right)^2
    + 
    \left(
    \frac{\Delta S_2}{S_2}
    \right)^2
    }
    \Bigg/
        \ln\left(\frac{\nu_2}{\nu_1}\right) \, , 
\label{equncalpha}
\end{equation}
where $\Delta S_i$ is the uncertainty on the flux density $S_{i}$ measured at frequency $\nu_i$ and $\nu_2 > \nu_1$. 

A systematic uncertainty of 10\% on each flux density measurement (due to uncertainties on the absolute flux calibration and from the imaging process) results in the following systematic uncertainties on the spectral indices  
(Eq.~\ref{equncalpha}): 
\begin{equation}
\Delta\alpha_{\rm sys}(\nu_1,\nu_2) = 
\left \{
\begin{tabular}{l}
0.10\\
0.17\\
0.06\\
\end{tabular}
\right.
{\rm if\, } \frac{\nu_1}{\nu_2} = 
\left \{
\begin{tabular}{l}
150 MHz/610 MHz\\
610 MHz/1400 MHz\\
150 MHz/1400 MHz\\
\end{tabular}
.
\right.
\label{eqDeltaAlphaSys}
\end{equation}

\subsection{Multiwavelength}

We used data from the following surveys:   

\begin{itemize}

\item 
The X-ray measurements with XMM-{\it Newton} and the XXL project are described in  
\citetalias{pierre16}.  
The cluster catalogs  are presented in 
\citetalias{2016A&A...592A...2P} and 
\citetalias{adami2018}, 
and the AGN catalogs by 
\citeauthor{2016A&A...592A...5F} (\citeyear{2016A&A...592A...5F}; \citetalias{2016A&A...592A...5F}) 
and by  
\citeauthor{lucio2018} (\citeyear{lucio2018}, hereafter \citetalias{lucio2018}).

\item 
There is a large overlap between XXL-North and the CFHT-LS W1 region covered in several optical bands by the Canada-France-Hawaii Telescope (CFHT). 
All XXL-North clusters but five have $ugriz$ photometry from MegaCam \citep{2012AJ....143...38G}. 

\item 
The Sloan Digital Sky Survey (SDSS) Data Release 14 (SDSS DR14\footnote{\url{http://skyserver.sdss.org/dr14}}) contains optical and near-infrared images of our fields and spectra of some of the sources. 

\item 
The Galaxy and Mass Assembly (GAMA) database\footnote{\url{http://www.gama-survey.org}} contains spectra, spectroscopic redshifts, and information on the detected spectral lines for the host galaxies of our two radio sources and other galaxies in the field \citep{2018MNRAS.474.3875B}. 

\item The Wide-field Infrared Survey Explorer (WISE; \citealt{2010AJ....140.1868W}) 
has detected the host galaxies of our two radio sources in several bands.
\end{itemize}

\section{The Exemplar}
\label{SectExemplar}


\begin{table}[ht]
\caption{Characteristics of the cluster of galaxies XLSSC~112 and of the BCG hosting the Exemplar radio galaxy. The first three blocks give cluster properties, while the others refer to the BCG.}
\label{tabTheOtherOne}
\begin{tabular} {lll}
\hline\hline
\noalign{\smallskip}
Quantity                & Value & Notes \cr
\noalign{\smallskip}
\hline 
\noalign{\smallskip}
Cluster name                    & XLSSC~112                 &(1)\\
RA$_{\rm cluster}$(J2000)       & $32\fdg514$           &(1)\\
Dec$_{\rm cluster}$(J2000)      & $-5\fdg462$           &(1)\\
$z_{\rm cluster}$   & 0.139                         &(1)\\
$N_{\rm galaxies}$      & 14                            &(1)\\
$L_{\rm 500,MT}^{\rm XXL}$   &$(0.61\pm0.08) \cdot 10^{43}$~erg~s$^{-1}$ &(1)\\
\noalign{\smallskip}
\hline 
\noalign{\smallskip}
$F_{60}$            & $(5.89\pm0.62) \cdot 10^{-17}$~Wm$^{-2}$  &(2)\\
$T_{\rm 300~kpc}$   & $1.76^{+0.25}_{-0.15}$~keV                    &(2)\\
$M_{\rm 500,MT}$        & $(9.0 \pm4.1) \cdot 10^{13} M_\odot$      &(2)\\
$r_{\rm 500,MT}$    & 0.653~Mpc                                 &(2)\\
$M_{\rm gas,500}$   & $(0.42\pm0.12)\cdot 10^{13} M_\odot$      &(3)\\
\noalign{\smallskip}
\hline 
\noalign{\smallskip}
$r_{\rm 500,WL}$        & $0.6^{+0.1}_{-0.2}$~Mpc                         &(4)\\
$M_{\rm 500,WL}$        & $0.8^{+0.6}_{-0.5} \cdot 10^{14} M_\odot$   &(4)\\
$M_{\rm 200,WL}$        & $1.2^{+0.9}_{-0.8} \cdot 10^{14} M_\odot$   &(4)\\
\noalign{\smallskip}
\hline 
\noalign{\smallskip}
RA$_{\rm BCG}$ (J2000)  &$32\fdg5093$   &(5)\\
Dec$_{\rm BCG}$ (J2000) &$-5\fdg4678$   &(5)\\
Offset from X-ray center &$24\farcs5$          &(5)\\
$z_{\rm BCG}$               & 0.138                 &(5)\\
$M_{\rm BCG}$               &$5.35^{+0.41}_{-0.29} \cdot 10^{11} M_\odot$           &(5)\\
\noalign{\smallskip}
\hline
\noalign{\smallskip}
$M_{\rm stellar}$       & $2.13^{+0.30}_{-0.56}\cdot10^{11}~M_\odot$   &(6)\\
SFR1 (2--20 Myr ago)     & $0~M_\odot$~yr$^{-1}$        &(6)\\
SFR2 (20--600 Myr ago)   & $5.6~M_\odot$~yr$^{-1}$     &(6)\\
SFR3 (0.6--5.6 Gyr ago)  & $46.2~M_\odot$~yr$^{-1}$    &(6)\\
SFR4 ($> 5.6$~Gyr ago)   & $14.9~M_\odot$~yr$^{-1}$    &(6)\\
LW-age                  & $3.6\times10^9$~yr            &(6)\\
MW-age                  & $5.8\times10^9$~yr            &(6)\\
\noalign{\smallskip}
\hline
\noalign{\smallskip}
$\sigma_V$              &$282.0\pm 8.6$~km~s$^{-1}$     &(7)\\
$z_{\rm SDSS\, DR14}$     &$0.13818\pm 0.00002$           &(7)\\
$z_{\rm GAMA}$          &0.13814                        &(8)\\ 
\noalign{\smallskip}
\hline
\end{tabular}
\tablefoot{
(1) X-ray and optical characteristics from \citetalias{2016A&A...592A...2P}.
$N_{\rm galaxies}$ is the number of galaxies on which the mean redshift, $z_{\rm cluster}$, was based. \\\hspace{\textwidth} 
(2) X-ray--derived quantities from \citetalias{2016A&A...592A...3G}. 
$F_{60}$ is the [0.5--2]~keV flux within a $1'$ radius; 
$T_{\rm 300~kpc}$ is the temperature within 300~kpc; 
the other quantities are the total mass and gas mass within a radius of $r_{500}$.\\\hspace{\textwidth}
(3) Gas mass within $r_{500}$ from \citetalias{2016A&A...592A..12E}.\\\hspace{\textwidth} 
(4) Weak lensing mass and size estimates from \citetalias{2016A&A...592A...4L}.\\\hspace{\textwidth} 
(5) Characteristics of the BCG from \citetalias{2016MNRAS.462.4141L}. 
$M_{\rm BCG}$ refers to the stellar mass of the BCG.\\\hspace{\textwidth} 
(6) This work. 
$M_{\rm stellar}$ is the stellar mass of the BCG estimated from spectral model-fitting.  
SFR refers to the star formation rate in four different periods, derived from model-fitting the GAMA spectrum. 
LW and MW refer to the luminosity-weighted and the mass-weighted age.\\\hspace{\textwidth} 
(7) Velocity dispersion of the galaxy, from the Baryon Oscillation Spectroscopic Survey (BOSS; SDSS DR14, SDSS~ 021002.24 -052804.1) and spectroscopic redshift. \\\hspace{\textwidth}
(8) Spectroscopic redshift from GAMA (GAMA~ J021002.23-052804.1).
}
\end{table}

\begin{figure}[h]
\centering
\includegraphics[width=8.8cm]{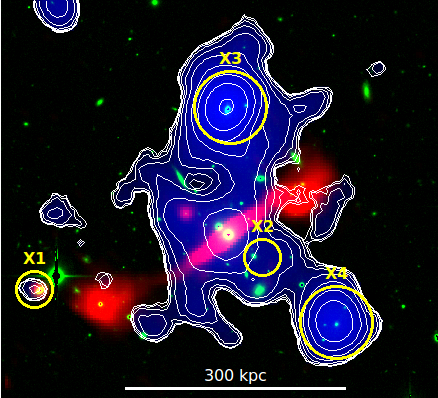}
\caption{Multiband false-color image of the Exemplar: X-ray emission in the [0.5--2]~keV band in blue, optical emission ($i$-band CFHT image) in green, radio emission (GMRT 610 MHz) in red.
   The ten contours showing the X-ray surface brightness increase logarithmically between 
   $3\times 10^{-17}$ and  $6\times 10^{-16}$~erg~s$^{-1}$~cm$^{-2}$. 
   Four X-ray point sources are visible in the image, labeled X1 to X4 above the yellow circles that indicate their positions. 
   The point sources  were subtracted before calculating the flux density of the cluster given in \citetalias{2016A&A...592A...2P} 
   and quoted in Table~\ref{tabTheOtherOne}, but they are shown in this image. 
   X3 corresponds to 3XLSS~J021002.1-052655 and 
   X4 to 3XLSS~J020958.2-052853 in the XXL point source catalog \citepalias{lucio2018}. 
   There are optical counterparts to X3 (a galaxy at photometric redshift of $0.028\pm 0.014$, SDSS J021002.25-052655.5) and to X4 
   (a galaxy at a photometric redshift of $0.351\pm0.118$, SDSS J020958.27-052853.5). 
   The other two  X-ray sources did not match the criteria to be included in the X-ray catalog. 
   The size of the image is about $4'\times4'$. 
   Diffuse X-ray emission is seen, and is associated with the galaxy cluster XLSSC~112. 
}
\label{figExemplarOptXrayradio}
\end{figure}

In Fig.~\ref{figExemplarOptXrayradio} we show a multiband false-color image of the field of the Exemplar: the optical image is shown in green, the X-ray emission in blue and in contours, and the 610~MHz radio emission from the GMRT in red. Several X-ray point sources are seen, in addition to diffuse X-ray emission associated with a cluster.
The two brightest X-ray point-like sources (X3 and X4) are galaxies with optical counterparts and photometric redshifts in SDSS DR14; they are not associated with the cluster (see caption of Fig.~\ref{figExemplarOptXrayradio}). The fainter X-ray source X1 coincides with the radio source S1  
(see discussion in Sect.~\ref{subSectExemplarRadio}). It has an optical galaxy counterpart, but no redshift is available. 

\subsection{Host galaxy in the cluster XLSSC~112 in the supercluster XLSSsC~N03}
\subsubsection{Cluster XLSSC~112}

The Exemplar is located at the center of a galaxy cluster listed among the 100 brightest clusters in XXL (\citetalias{2016A&A...592A...2P}). 
Cataloged as XLSSC~112, the cluster has a mean redshift of $z = 0.139$ calculated from the spectroscopic  redshifts of 14 member galaxies (Table~\ref{tabTheOtherOne}).    
The intracluster gas has a temperature of $\sim1.8$~keV measured within the central radius of 300~kpc \citepalias{2016A&A...592A...3G}. 
The total cluster mass was estimated  from a mass--temperature scaling relation \citepalias{2016A&A...592A...3G} 
and from weak gravitational lensing \citepalias{2016A&A...592A...4L}, 
yielding consistent values of about 10$^{14} M_\odot$ within a radius $r_{500} = 0.6$~Mpc 
(the parameter $r_{500}$ is the radius within which the mean density is 500 times the critical density of the universe at the cluster's redshift).  

\subsubsection{Supercluster XLSSsC~N03}

The cluster was found to be part of a larger structure of seven clusters, all at a redshift of $z \simeq 0.14$,  with a total mass of about $10^{15} M_\odot$ 
(XLSSC-b; \citetalias{2016A&A...592A...2P}). 
Five such superstructures were identified in the sample of the 100 brightest clusters in XXL using a friends-of-friends algorithm in 3D space  (\citetalias{2016A&A...592A...2P}). 
New supercluster candidates were found in the larger sample of 365 clusters cataloged by \citeauthor{adami2018} (\citeyear{adami2018}, hereafter \citetalias{adami2018}). 
The five superclusters identified in \citetalias{2016A&A...592A...2P} were confirmed. However,  
the supercluster hosting the Exemplar, which was described as a double structure in \citetalias{2016A&A...592A...2P}, with an eastern component of four clusters and a western component of three clusters (including XLSSC~112), has been split into two superclusters in the new study: 
the four clusters in the eastern part form XLSSsC~N08, at a mean redshift of $z = 0.141$, and  
the three clusters in the western part joined five other clusters to form a larger structure, the supercluster  
XLSSsC~N03,  
at a mean redshift of $z = 0.139$  
\citepalias{adami2018}. 
XLSSC~112 is therefore affiliated to a supercluster with eight cluster members. 
This is discussed in Sect.~\ref{SectSuperclu} (see also Fig.~\ref{figSuperclu}). 

\begin{figure*}
\centering
   \includegraphics[width=8.8cm]{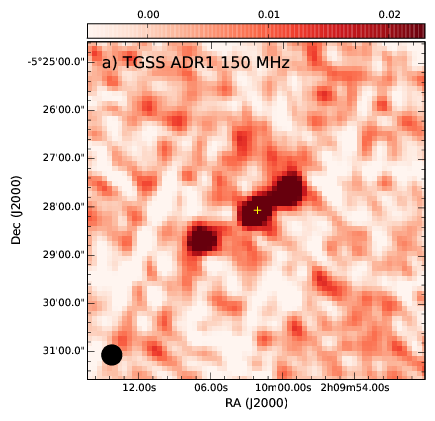}
   \includegraphics[width=8.8cm]{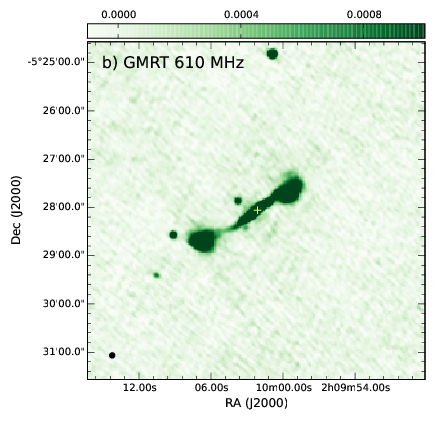}
   \includegraphics[width=8.8cm]{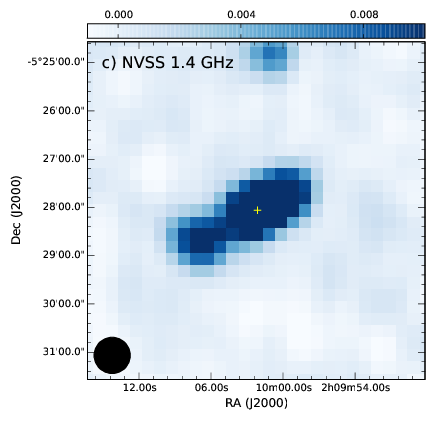}
   \includegraphics[width=8.8cm]{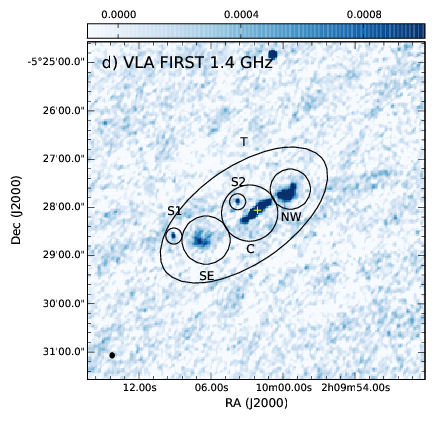}
   \caption{Radio continuum images of the Exemplar: 
   a) from the TGSS ADR1 at 150~MHz; 
   b) from the GMRT at 610~MHz; 
   c) from the NVSS at 1.4~GHz; 
   d) from FIRST at 1.4~GHz. 
  The size of the synthesized beam is indicated in the bottom left corner of each image. 
   The position of the BCG is shown by a yellow cross. 
   The circles and ellipses overlaid on the FIRST 1.4~GHz image 
   show the regions used for the photometry and the spectral index measurements (see Tables~\ref{tabExemplarFlux} and \ref{tabExemplarAlpha}). 
   The NW aperture is a circle with a radius of 25$\arcsec$, centered at  
   ($32\fdg49747$, $-5\fdg46036$). 
  The aperture in the central region (C) is a circle with a radius of 35$\arcsec$, centered at 
   ($32\fdg51157$, $-5\fdg46863$). 
   The SE circular aperture has a radius of 30$\arcsec$, centered at 
   ($32\fdg52672$, $-5\fdg47796$).  
   The ellipse (T) has a semimajor axis of 2$'$, a semiminor axis of 1$'$, and a position angle of 
   $+125^\circ$ 
   measured counterclockwise from the north, centered at 
   ($32\fdg51357$, $-5\fdg46929$). 
   The position of S1 is ($32\fdg53811$,$-5\fdg47619$) and that of S2 is ($32\fdg51574$, $-5\fdg46446$) and the circles have a radius of 10$\arcsec$. 
   }
\label{figSarcasmAllradioimages}  
\end{figure*}

\subsubsection{Brightest cluster galaxies and the nearest neighbor}
\label{subSectExemplarBCG}
The properties of the BCG of the 100 brightest XXL clusters have been studied in \citetalias{2016MNRAS.462.4141L}. 
\citeauthor{guglielmo2018a} (\citeyear{guglielmo2018a}; \citetalias{guglielmo2018a}) 
have assembled a catalog of optically detected galaxies in the X-ray--detected clusters and groups of XXL-North. 
In Fig.~\ref{figGAMAspectra} (top panel) we show the GAMA spectrum of the BCG, typical of an old stellar population. 
We fitted the GAMA spectrum  
using the full spectral fitting code {\tt SINOPSIS} (\citealt{2007A&A...470..137F}) 
and obtained an estimate of its stellar mass of about $2\times10^{11}~M_\odot$
(using a \citealt{2003PASP..115..763C} 
initial mass function). This is lower than  had been estimated from broadband photometry (\citetalias{2016MNRAS.462.4141L}). 
This analysis also provided an estimate of the star formation history of the galaxy.  
The characteristics of the cluster and of its BCG are summarized in Table~\ref{tabTheOtherOne}. 

There is a second, smaller galaxy located 7$\arcsec$ to the NE of the BCG (about 17~kpc on the sky) at a similar redshift 
($z = 0.13623$, galaxy GAMA~J021002.58-052759.8). 
Companions like this are very common for  radio sources of this type (e.g., 3C296). 

\subsection{Radio}
\label{subSectExemplarRadio}

\subsubsection{Radio morphology}
\begin{figure}
\centering
\includegraphics[width=8.8cm]{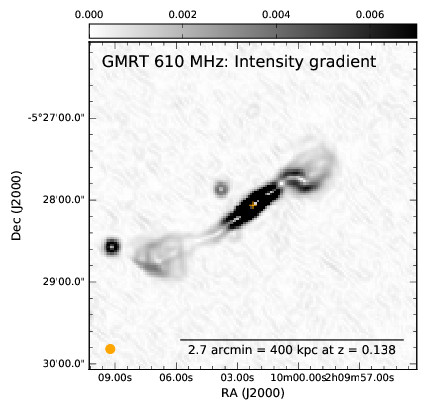}
\caption{Grayscale image of the norm of the intensity gradient of the Exemplar at 610~MHz.  
It reveals the path of the jets and substructures in the brightness distribution of the radio lobes. 
The beam is shown in orange in the bottom left corner. 
}
\label{figExemplarGradI}
\end{figure}

Figure~\ref{figSarcasmAllradioimages} shows the four radio continuum images of the Exemplar used in this work. 
The Exemplar has two jets and lobes that appear clearly in the higher resolution images 
(GMRT 610~MHz and the FIRST 1.4~GHz images, Figs.~\ref{figSarcasmAllradioimages}b and d), 
but are not resolved  
in the lower resolution images (TGSS and NVSS, Figs.~\ref{figSarcasmAllradioimages}a and c). 
The TGSS ADR1 catalog of \cite{TGSS_ADR1} 
lists three individual Gaussian sources.\footnote{
TGSS J021006-052841, 
which corresponds to the SE lobe, is listed as an isolated single-Gaussian source with a flux density of $57.2\pm7.0$~mJy; 
TGSS J021002-052804, 
which corresponds to the core and the inner jets, has $90.5\pm9.9$~mJy; 
TGSS J020959-052738,  
which corresponds to the NW lobe, has $73.3\pm 3.3$~mJy.
The second and third  sources overlap.
} 
The NVSS catalog lists two sources\footnote{
NVSS~J021007-052842 has a total flux density of $15.3\pm1.0$~mJy. 
NVSS~J021001-052758 has a total flux density $58.8\pm 2.2$~mJy, 
} 
\citep{NVSS}. 
The XXL-GMRT-610~MHz catalog\footnote{
The three sources listed as resolved within a radius of 4$'$ from the core of the Exemplar are 
XXL-GMRT J021006.7-052840, 
XXL-GMRT J021004.3-052826, and
XXL-GMRT J021000.9-052753. 
The two compact sources that we named S1 and S2 are 
XXL-GMRT J021009.1-052834 and
XXL-GMRT J021003.7-052751, respectively. 
} 
lists three resolved sources within a radius of 4$'$ from the core of the Exemplar \citepalias{smolcic2018}. 
From end to end the radio source stretches over 160$\arcsec$ (396~kpc) in the GMRT 610~MHz image.

Zooming into the central region, we see that the radio surface brightness peaks about $3\farcs5$ (8.7~kpc) away from the optical center of the BCG on either side of the jets, with a flux density of about 8.8~mJy in the NW and 9.8~mJy to the SE at 610 MHz. 
There are two corresponding sources in the FIRST catalog\footnote{
FIRST~J021001.9-052800 with a peak flux density of 4.15~mJy~beam$^{-1}$ 
(integrated flux density $S_{\rm T} = 5.53$~mJy)  
and FIRST~J021002.4-052806 with a peak flux density of 6.32~mJy~beam$^{-1}$ ($S_{\rm T} = 10.19$~mJy).
From the peak values this gives a spectral index of $-0.90$ for the fainter NW peak and $-0.53$ for the SE peak between 610~MHz and 1.4~GHz.
}. 
We note that these brighter radio spots are located just outside the optical extent of the host galaxy. 
No cataloged FIRST source is found at the central location of the BCG. The radio core is unresolved and is likely embedded in the jet and/or is synchrotron self-absorbed. 

The jets have position angles of 
$+130^\circ$  
and $-50^\circ$  
(measured counterclockwise from the  north).
They are straight out to a distance of 32$\arcsec$ to the NW and 40$\arcsec$ to the SE from the center of the BCG, 
at which point they change direction and turn right (that is, the NW jet turns toward the W and the SE jet turns toward the E). The NW jet 
moves into 
two brighter regions embedded in a diffuse lobe. 
The SE jet continues to grow fainter but remains collimated for another 15$\arcsec$ before 
reaching 
a diffuse lobe. 
This morphology suggests an interaction with the surrounding medium at a distance of about 80~kpc (30$\arcsec$) to the NW 
and about 90~kpc (35$\arcsec$) to the SE. 

Two compact radio sources are seen near the radio galaxy. 
They are named S1 and S2 in Tables~\ref{tabExemplarFlux} and \ref{tabExemplarAlpha} and their flux densities and their spectral index were estimated
between 610~MHz and 1.4~GHz (FIRST image). The spectral indices are used to estimate the flux at 150~MHz. The contributions of those two sources are subtracted from the total flux densities measured within the regions outlined in Fig.~\ref{figSarcasmAllradioimages}. The radio photometry of the Exemplar is presented in the coming section. 

In Fig.~\ref{figExemplarGradI} we show an image of the gradient of the intensity in the GMRT image at 610~MHz, $|\nabla I|$. The path of the jet is clearly delineated in white where the intensity has a local maximum. Strong gradients are seen near the inner jets 
where the jet bends as it enters the NW lobe, and at the extremity of the SE lobe. 
Similar features have been seen in deep images of nearby low-luminosity 
\ion{FR}{I} radio galaxies 
(\citealt{2011MNRAS.417.2789L}). 

\begin{table*}[t]
\centering
\caption{Flux density measurements in the different regions of the Exemplar shown in Fig.~\ref{figSarcasmAllradioimages}.}
\label{tabExemplarFlux}
\begin{tabular}{ccccc}
\hline \hline
Region & $S_{\rm 150~MHz}$ &$S_{\rm 610~MHz}$  &$S_{\rm 1.4~GHz}$ (FIRST) & $S_{\rm 1.4~GHz}$ (NVSS) 
\\
            & (mJy)  & (mJy)   & (mJy)   & (mJy) \\
\hline
S1 & 25.73 $\pm$ 3.11 & 3.27 $\pm$ 0.14 & 0.97 $\pm$ 0.40 & -- \\
S2 & 6.74 $\pm$ 3.30 & 2.48 $\pm$ 0.14 & 1.38 $\pm$ 0.40 & -- \\
A (SE lobe) & 61.14 $\pm$ 9.46 & 32.25 $\pm$ 0.43 & 8.21 $\pm$ 1.19 & 12.98 $\pm$ 0.37\\
B (NW lobe) & 71.98 $\pm$ 7.85 & 38.76 $\pm$ 0.36 & 13.11 $\pm$ 1.00 & 10.91 $\pm$ 0.28\\
C & 97.07 $\pm$ 10.94 & 56.64 $\pm$ 0.50 & 29.89 $\pm$ 1.40 & 30.60 $\pm$ 0.41\\
C $\setminus$ S2    &90.33 $\pm$ 11.42 & 54.16 $\pm$ 0.52 & 28.52 $\pm$ 1.45 & 29.23 $\pm$ 0.44\\
T & 284.79 $\pm$ 26.63 & 138.16 $\pm$ 1.21 & 44.81 $\pm$ 3.38 & 74.95 $\pm$ 0.99\\
T$_{\rm A \cup B \cup C}$ &230.20 $\pm$ 16.45 & 127.66 $\pm$ 0.75 & 51.21 $\pm$ 2.09 & 54.50 $\pm$ 0.63\\
T$_{\rm (A \cup B \cup C) \setminus (S1 \cup S2)}$ &197.72 $\pm$ 28.60 & 121.90 $\pm$ 1.30 & 48.87 $\pm$ 3.63 & 52.16 $\pm$ 1.09\\
T $\setminus$ (S1 $\cup$ S2) &252.32 $\pm$ 27.01 & 132.40 $\pm$ 1.23 & 42.47 $\pm$ 3.43 & 72.61 $\pm$ 1.01\\
\hline
\end{tabular}
\tablefoot{
The coordinates of S1 and S2 are given in the caption of Fig.~\ref{figSarcasmAllradioimages}.  
The flux densities of S1 and S2 at 150~MHz were extrapolated from the measurements at 610~MHz and 1400~MHz (FIRST). 
The symbols $\cup$ and $\setminus$ are taken from set theory and are used to indicate a union and an exclusion, respectively: 
for instance, C $\setminus$ S2 means that the flux density from source S2 was subtracted from that measured in the central (C) region; 
T$_{\rm A \cup B \cup C}$ is the flux density measured by summing  the measurements in the regions A (SE lobe), B (NW lobe), and C (central region). 
The 1.4~GHz flux densities of S1 and S2, which are point-like, were taken from FIRST.
}
\end{table*}

\begin{table}[t]
\caption{Spectral index measurements in the different regions of the Exemplar shown in Fig.~\ref{figSarcasmAllradioimages}.}
\centering
\begin{tabular}{cccc}
\hline \hline
\noalign{\smallskip}
Region & $\alpha_{\rm 150~MHz}^{\rm 610~MHz}$ & $\alpha_{\rm 610~MHz}^{\rm 1400~MHz}$  &$\alpha_{\rm 150~MHz}^{\rm 1400~MHz}$
\\
\noalign{\smallskip}
    &(1)    &(2)    &(3)\\
\hline
$\Delta\alpha_{\rm sys}$       &0.1    &0.17      &0.06\\
A (SE lobe) & $-0.46 \pm 0.11$ & $-1.65 \pm 0.18$ & $-0.69 \pm 0.07$\\
B (NW lobe) & $-0.44 \pm 0.00$  & $-1.30 \pm 0.09$ & $-0.84 \pm 0.05$\\
C           & $-0.38 \pm 0.08$ & $-0.77 \pm 0.06$ & $-0.52 \pm 0.05$\\
C $\setminus$ S2     &$-0.36 \pm 0.09$ & $-0.77 \pm 0.04$ & $-0.51 \pm 0.06$\\
T           & $-0.52 \pm 0.07$         &$-1.36 \pm 0.09$  & $-0.60 \pm 0.04$\\
T $\setminus$ (S1 $\cup$ S2) &$-0.46 \pm 0.08$  & $-1.37 \pm 0.06$ & $-0.56 \pm 0.05$\\
\hline
\end{tabular}
\tablefoot{
The spectral indices were calculated from flux density measurements in 
(1) the TGSS and the GMRT 610~MHz maps; 
(2) the GMRT 610~MHz and the FIRST 1.4~GHz maps; 
(3) the TGSS and the NVSS 1.4~GHz maps. 
The first line gives the systematic uncertainties on $\alpha$ for a 10\% systematic uncertainty on each flux density measurement. The other uncertainties listed in the table are statistical. 
The meanings of the $\setminus$ and $\cup$ symbols were given in the notes of Table~\ref{tabExemplarFlux}.
}
\label{tabExemplarAlpha}
\end{table}

\subsubsection{Luminosity, flux density, spectral index}
\label{SectExemplarSpectralIndex}

Flux densities were measured at the three different frequencies in the regions indicated in Fig.~\ref{figSarcasmAllradioimages}; 
the values are listed in  
Table~\ref{tabExemplarFlux}.  

The spectral luminosity, $L_\nu$, can be calculated following 
\begin{equation}
\left( 
    \frac{L_{\nu}}
    {{\rm W~ Hz}^{-1}} 
\right) = 
5.257\times 10^{22} \,  
\left( 
    \frac{S_{\nu}}
    {{\rm mJy}} 
\right) 
\left( 
    \frac{D_L}
    {{\rm 662.8~Mpc}} 
\right)^2  \, .
\label{eqLnu}
\end{equation}

In the commonly used classification of \cite{fanaroff74}, 
FR~{\sc i} radio galaxies have been observed to have a spectral luminosity lower than $5\times 10^{25}$ W~Hz$^{-1}$ at 178~MHz, 
while the FR~{\sc ii} radio galaxies are more luminous. 
The Exemplar has a total flux density of about 250~mJy at 150~MHz (taking an average of the value obtained by summing up the different regions and by measuring in a larger ellipse). For a spectral index of $-0.7$, this gives $L_{\rm 178~MHz}\simeq 1.5\times 10^{25}$  W~Hz$^{-1}$, making it an FR~{\sc i} radio galaxy. 
Its morphology is also that of an FR~{\sc i}.  

Spectral indices were calculated in the regions overlaid 
on Fig.~\ref{figSarcasmAllradioimages} (Table~\ref{tabExemplarAlpha}). 
The central region has flatter spectral indices than the lobes. 
The spectral indices calculated at high frequencies are systematically steeper 
than those at low frequencies\footnote{
Flux loss in the FIRST 1.4~GHz image would produce a similar effect. 
The largest angular scale to which the FIRST observations are sensitive is 2$'$ and the radio galaxy stretches over $\sim$$2\farcm5$. On the other hand, the lobes themselves are smaller than 1$'$ across. 
For the SE lobe, the flux density measured in the FIRST image within the elliptical region displayed in Fig.~\ref{figSarcasmAllradioimages} is lower than that measured in the NVSS image (about 8.2 vs. 13~mJy, see Table~\ref{tabExemplarFlux}), an indication that some of the extended emission might have been filtered out. For the NW lobe on the other hand, the FIRST flux is higher than the NVSS flux, perhaps because the region 
(within a circle of 50$\arcsec$ diameter) does not include all the flux of the 45$\arcsec$ resolution NVSS image.}
Using the NVSS 1.4~GHz image and the TGSS 150~MHz image gives intermediate values. 

Unfortunately, the quality of the data does not allow a quantitative analysis of the 
evolution of the cosmic-ray electrons, and deeper images with matching angular resolutions are required to measure the curvature of the radio spectrum. 

\subsubsection{Polarization}  

No polarization is seen in the NVSS images of the Exemplar. 
The lobes have a flux density at 1.4~GHz ranging between 8 and 13~mJy, and the flux density of the core/central region  is about 30~mJy (Table~\ref{tabExemplarFlux}). 
The noise level in polarized intensity is about 0.4~mJy in NVSS. So we place a $3\sigma$ upper limit on the fractional polarization of 15\% in the lobes and 4\% in the central region. 

\section{The Double Irony as a giant radio galaxy}
\label{SectDbleIrony}

\begin{table}
\caption{Characteristics of the galaxy hosting the Double Irony radio galaxy.}
\label{tabDbleIrony}
\begin{tabular} {llc}
\hline
\hline
\noalign{\smallskip}
Quantity                & Value & Notes\\
\noalign{\smallskip}
\hline 
\noalign{\smallskip}
RA$_{\rm galaxy}$(J2000)   &$30\fdg97696$   & (1)\\
Dec$_{\rm galaxy}$(2000)   &$-4\fdg23247$   & (1)\\
$z_{\rm spec}$             & 0.13751           & (1)\\
\hline
Central point source:\\
RA$_{\rm X-ray}$(J2000)  & $30\fdg977942$      & (2)\\
Dec$_{\rm X-ray}$(J2000) & $-4\fdg232405$      & (2)\\
$F_{\rm X,[0.5-2]~keV}$  & $2.4\times 10^{-15}$ erg~s$^{-1}$~cm$^{-2}$ & (2)\\
$F_{\rm X,[2-10]~keV}$   & undetected             &(2)\\
$L_{\rm X,[0.5-2]~keV}$  & $1.26\times10^{41}$~erg~s$^{-1}$  &(3)\\
\noalign{\smallskip}
\hline
Limit on cluster:\\
Count rate              & $< 0.032$~ct~s$^{-1}$             &(4)\\
$T_{\rm 300~kpc}$       & $< 1.26$~keV                      &(4)\\
$M_{500}$               & $< 5.13\times 10^{13} M_\odot$    &(4)\\
$L_{500}$               & $< 1.76\times 10^{42}$~erg~s$^{-1}$               &(4)\\
$F_{\rm X,[0.5-2]~keV, < 300~kpc}$ &$< 2.82\times 10^{14}$~erg~s$^{-1}$~cm$^{-2}$ &(4)\\
\noalign{\smallskip}
\hline
$M_{\rm stellar}$        & $(2.97\pm0.25)\times10^{11}~M_\odot$   &(5)\\
SFR1 (2--20 Myr ago)     & $0~M_\odot$~yr$^{-1}$        &(5)\\
SFR2 (20--600 Myr ago)   & $1.48~M_\odot$~yr$^{-1}$     &(5)\\
SFR3 (0.6--5.6 Gyr ago)  & $33.08~M_\odot$~yr$^{-1}$    &(5)\\
SFR4 ($> 5.6$~Gyr ago)   & $42.36~M_\odot$~yr$^{-1}$    &(5)\\
LW-age                   & $5.7\times10^9$~yr           &(5)\\
MW-age                   & $7.9\times10^9$~yr           &(5)\\
\hline
\end{tabular}
\tablefoot{
(1) GAMA~J020354.47-041356.\\\hspace{\textwidth} 
(2) X-ray point-like source 3XLSS~J020354.7-041356 \citepalias{lucio2018}.\\\hspace{\textwidth} 
(3) Derived from (2) using our adopted distance.\\\hspace{\textwidth} 
(4) This work. 1$\sigma$ upper limit on the count rate, converted into limits on flux, mass, and luminosity using scaling relations of \citetalias{2016A&A...592A...4L} and \citetalias{adami2018}.\\\hspace{\textwidth}  
(5) This work. SFR refers to the star formation rate in four different periods, derived from model-fitting the GAMA spectrum. 
LW and MW refer to the luminosity-weighted and the mass-weighted ages.\\\hspace{\textwidth} 
}
\end{table}

The  Double Irony stands out in XXL-North because of its large angular size and peculiar shape. 
As we  show in Sect.~\ref{sectDbleIronyRadio}, it has a total physical length of about 1.1~Mpc, which makes it a member of the rare class of giant radio galaxies (GRGs), i.e.,
Mpc-sized radio galaxies.  
Samples of GRGs have been built based on searches in large radio surveys at GHz frequencies such as  
the Westerbork Northern Sky Survey\footnote{WENSS, \cite{1997A&AS..124..259R}} 
(e.g., \citealt{2001A&A...374..861S}); 
the NVSS\footnote{NVSS, \cite{NVSS}} 
and the FIRST survey\footnote{FIRST, \cite{first95}, \cite{first15}} 
(e.g., 
\citealt{2001A&A...370..409L}, 
\citeyear{2001A&A...378..826L}; 
\citealt{2001A&A...371..445M}; 
\citealt{2016ASSP...42..231S}; 
\citealt{2016ApJS..224...18P}), 
or the Sydney University Molonglo Sky Survey\footnote{SUMSS, \cite{1999AJ....117.1578B}} 
(e.g., \citealt{2005AJ....130..896S}). 

The total number of known GRGs is about 300 \citep{2017MNRAS.469.2886D}, 
but this number is likely to increase rapidly with the advent of radio surveys at lower frequencies where the flux  densities are stronger because most of the emission comes from the more numerous lower energy relativistic electrons that have a longer lifetime. 
Surveys carried out for example with the LOw Frequency ARray  (LOFAR, \citealt{2013A&A...556A...2V}), 
MSSS\footnote{The LOFAR Multifrequency Snapshot Sky Survey, \cite{2015A&A...582A.123H}} 
{\it Herschel}-ATLAS, and LoTSS\footnote{The LOFAR Two-metre Sky Survey,  \cite{2017A&A...598A.104S}} 
have begun to reveal even larger GRGs than those discovered at higher frequencies (e.g., 
\citealt{2016MNRAS.462.1910H}; 
\citealt{2017A&A...601A..25C}). 

GRGs are interesting in their own right, but also as probes of their environment. 
As radio lobes expand beyond the halos of their host galaxies, they may interact with the so-called Warm-Hot Intergalactic Medium (WHIM), a diffuse plasma at a temperature of $10^5$ to $10^7$~K that is associated with galaxies in the large-scale structure  
(e.g., \citealt{2008ApJ...677...63S}; 
\citealt{2009MNRAS.393....2S}). 
They are therefore useful probes of the intergalactic medium and possibly even of its evolution if they can be found at high redshift \citep{2007AcA....57..227M}. 
The large sizes of GRGs do not seem to be due to  stronger nuclear activity (\citealt{1999MNRAS.309..100I}) 
and GRGs do not seem to form a class that is  fundamentally different  from normal-sized radio galaxies 
(\citealt{2006A&A...454...95M};  
\citealt{2012MNRAS.426..851K}). 
The general picture is that GRGs are able to grow to larger sizes because of their lower density environments 
and/or simply because they are older (e.g., \citealt{2008MNRAS.385.1286J}). 

In this section we begin by examining the host galaxy of the Double Irony; then we look at the X-ray information and the galaxy distribution in the optical before presenting the radio maps; finally the possible nature of this unusual radio galaxy is discussed. In the next section the two radio galaxies will be compared and more quantities will be derived.  

\subsection{ Host galaxy}

\begin{figure}[t]
\centering
\includegraphics[width=8.8cm]{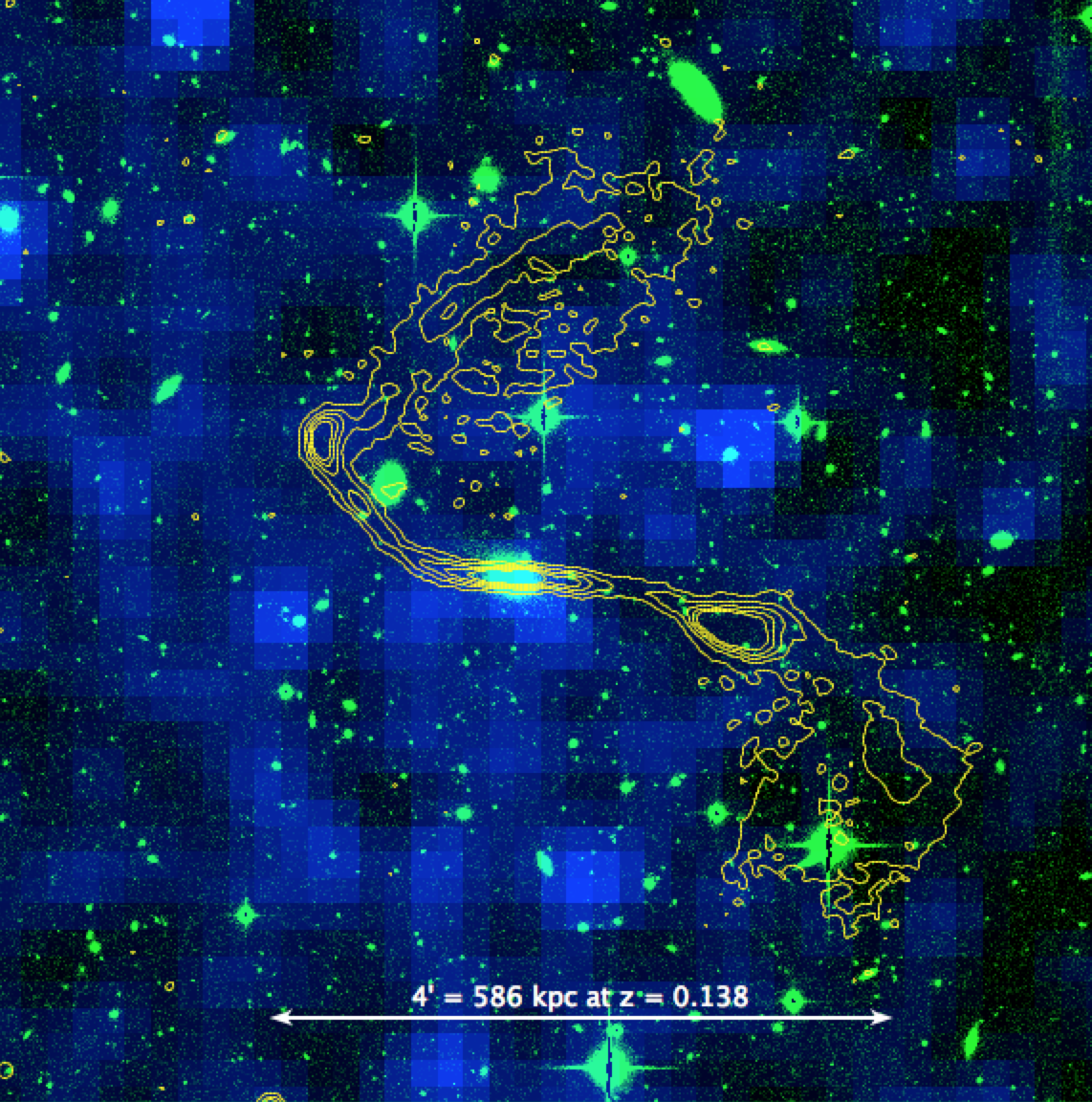}
\caption{Multiband false-color image of the Double Irony: X-ray emission in blue displayed at the noise level, optical emission ($i$-band image from the CFHT) in green, 
 radio emission from the GMRT at 610~MHz as yellow contours. 
}
\label{figDbleIronyOptXrayradio}
\end{figure}

\begin{figure*}
\centering
   \includegraphics[width=8.8cm]{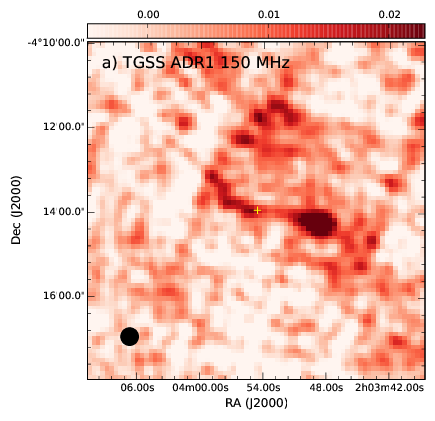}
   \includegraphics[width=8.8cm]{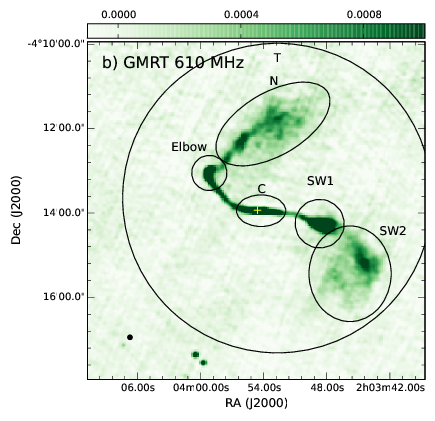}
   \includegraphics[width=8.8cm]{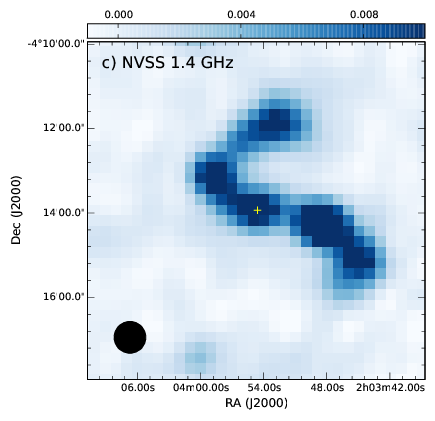}
   \includegraphics[width=8.8cm]{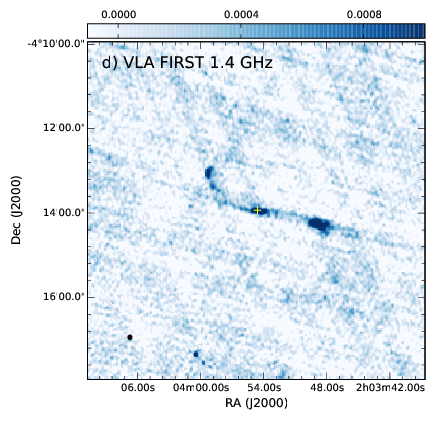}
   \caption{Radio continuum images of the Double Irony: 
   a) from the TGSS ADR1 at 150~MHz; 
   b) from the GMRT at 610~MHz; 
   c) from the NVSS at 1.4~GHz; 
   d) from FIRST at 1.4~GHz. 
   For each image the size of the beam is indicated in the bottom left corner. 
   The cross shows the position of the host galaxy. 
   The northern region (N) is an ellipse of semimajor axis of 90$\arcsec$, semiminor  axis of 45$\arcsec$, position angle of $-60^\circ$ from the north, centered at 
   ($30\fdg97133,-4\fdg19817$).
   The Elbow region is a circular aperture of radius 25$\arcsec$, centered at 
   ($30\fdg99652$,$-4\fdg21751$).
   The central aperture (C) is an ellipse of semimajor axis of 35$\arcsec$, semiminor axis of $22\farcs5$, position angle of $-55^\circ$ from the north, centered at 
   ($30\fdg97601$,$-4\fdg23241$). 
   The SW1 aperture is an ellipse of semimajor axis of $34\farcs935$, 
   a semiminor axis of $34\farcs551$, position angle of $-255^\circ$ from the north, 
   centered at ($30\fdg95282$,$-4\fdg237526$). 
   The SW2 aperture is an ellipse of semimajor axis of $68\farcs025$, semiminor axis of $58\farcs426$, position angle of 0$^\circ$, centered at 
   ($30\fdg94079$,$-4\fdg2573$). 
   The region used for the total flux measured (T) is a circle of radius of 220$\arcsec$, centered at 
   ($30\fdg96956$,$-4\fdg2274$).
   }
\label{figDbleIronyAllradioimages}  
\end{figure*}

We identify a galaxy with a spectroscopic redshift of $z=  0.13751$ (GAMA~J020354.47-041356.8) as the host galaxy of the Double Irony 
based on its position with respect to the radio contours. 
The GAMA spectrum is displayed in 
Fig.~\ref{figGAMAspectra} (bottom panel).  
It is very similar to the spectrum of the host galaxy of the Exemplar, and reflects an old stellar population. 
We derived an estimate of its stellar mass of about $3\times10^{11}~M_\odot$. 
This analysis provides some insight into the star formation history of the galaxy:  
it has no ongoing star formation, but had a star formation rate (SFR) 
of about 1.5~$M_\odot$~yr$^{-1}$ $\sim20$--600~Myr ago, 
and an even higher SFR of $\sim30$--45~$M_\odot$~yr$^{-1}$ in the more distant past 
(see Table~\ref{tabDbleIrony}). 

There is also an X-ray point-like counterpart to the optical/near-IR host  (see Table~\ref{tabDbleIrony}). 
The source is detected only in the [0.5 -- 2]~keV band, and the detection is not of sufficient quality to say anything about the nature of the X-ray spectrum (thermal or non-thermal). The corresponding X-ray luminosity is $1.26\times10^{41}$ erg~s$^{-1}$. 
\cite{2003A&A...399...39R} 
found a relation between the X-ray luminosity and the SFR of nearby star-forming galaxies. 
Using their relation (their Eq.~14) gives a SFR estimate of about 28~$M_\odot$~yr$^{-1}$, which is significantly higher than that inferred from the analysis of the optical spectrum. It is very unlikely that the X-ray emission is related to star formation; 
since the X-ray emission is point-like and the optical major axis of the galaxy is about 20$\arcsec$ 
(more than double the full width at half maximum (FWHM) of the point response function of XMM-{\it Newton}),   
it indicates that the X-ray emission does not come from hot gas surrounding the elliptical galaxy. 
We conclude that it is likely to come from the AGN. The signal is, however, very weak, with only about 12 counts, and a detection likelihood 
of 15.6, just 
above the threshold of 15 below which sources are considered spurious and 
not included in the catalog (\citealt{lucio2018}). 

\subsection{Host cluster/group }
\label{SectDbleIronyHostCluster}
In the following we review the information that can be gathered 
from X-ray and optical observations 
on the possible existence of a cluster or group around the Double Irony. 

-- X-ray. 
No cluster is listed at the location of the Double Irony radio galaxy in the new XXL catalog of X-ray detected clusters \citepalias{adami2018}. 
From the 1$\sigma$ upper limit on the count rate within a radius of 300~kpc, we were able to place an upper limit on the X-ray flux, temperature, and luminosity within a radius $r_{500}$ using scaling relations established for XXL clusters (see Table~\ref{tabDbleIrony}): 
a cluster/group around the Double Irony would have a temperature lower than 1.3~keV and a mass within $r_{500}$ lower than $5.1\times10^{13}~M_\odot$.  
This is significantly lower than the corresponding values measured for the Exemplar's cluster. 

A closer inspection of the X-ray image 
(Fig.~\ref{figDbleIronyOptXrayradio}) does not reveal any clear extended emission. 
The noise varies across the image and the quality of the data does not allow us to say anything about the presence or absence of gas interaction with the radio lobes. 

-- Optical. 

A search in the GAMA database led to the identification of 32 galaxies 
with spectroscopic redshifts in the $0.13 \leq z \leq 0.15$ range and within a radius of $6.3'$ (0.92~Mpc) of the core of the Double Irony. 
This radius was chosen as it  
corresponds to $r_{200}$ (which is approximately the virial radius) of the Exemplar galaxy cluster. 
The actual virial radius of the Double Irony galaxy might be different, but this shows that the radio galaxy is surrounded by a large number of galaxies. 
In Fig.~\ref{figDbleIronyGAMAgalaxies} we show the GMRT 610~MHz image of the Double Irony radio galaxy in red, superimposed on an optical $i$-band image from the CFHT in blue. The yellow circles indicate the positions 
of some of the galaxies in the field cataloged in the GAMA survey;  their spectroscopic redshifts are also indicated. 
We defer the analysis of the spectroscopic redshifts 
to Sect.~\ref{sectDiscusssionClusterMass} 
(cluster mass estimates) 
and Sect.~\ref{sectCompaEnvironment} (study of the anisotropy of the surrounding galaxy distribution).

\subsection{Radio}
\subsubsection{Radio morphology}
\label{sectDbleIronyRadio}
\begin{figure}
\centering
\includegraphics[width=8.8cm]{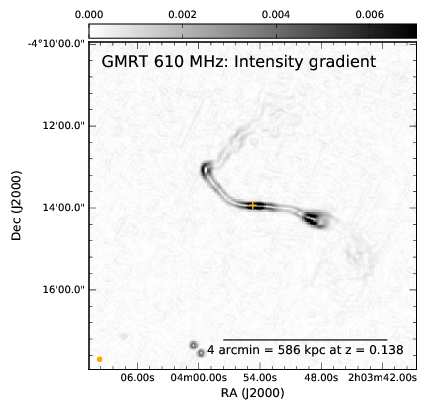}
\caption{Grayscale image of the norm of the intensity gradient of the Double Irony at 610~MHz.  
The path of the jets is clearly visible, in particular the abrupt turn in the region of the Elbow. 
The beam is shown in the bottom left corner. 
}
\label{figDbleIronyGradI}
\end{figure}

In  Fig.~\ref{figDbleIronyAllradioimages} we show the radio images of the Double Irony. 
The structure of the source is best seen in the GMRT 610~MHz image that combines high angular resolution and good 
sensitivity.\footnote{
The Double Irony lies in the northwestern corner of the GMRT 610~MHz mosaic at the intersection of two pointings 
where the noise is slightly higher than average  
(see Fig.~5 of \citetalias{smolcic2018}). 
} 
The yellow cross indicates the location of the center of the host galaxy. 
Several features appear clearly in this image: 
the central east--west jets, 
a bright lobe to the southwest, 
a diffuse plume farther out (SW2), 
a bright region to the northeast (known as the ``Elbow''), 
and a second diffuse plume-like feature to the north that breaks into several lumps. 
These five main features are also seen in the lower resolution (45$\arcsec$) 1.4~GHz image from the NVSS. 
The higher resolution  FIRST image at the same frequency is badly affected by stripes that prevent us from seeing the two plumes in the north and the southwest.  
The TGSS ADR1 image is noisy as well. Only the lobe in the SW1 region is clearly visible. 
While the lumps of emission in the north and in the Elbow are discernible, it is clear that flux measurements in these regions are affected by the artifacts from the imaging. 

Most striking  perhaps are the multiple bends: on the eastern side the nearly horizontal jet turns abruptly, but remains collimated until the bright Elbow; the continued structure to the north of the Elbow is fainter and at about 90 degrees from its original direction. On the SW side, the jet shows a small deviation before entering the SW1 lobe, and the SW2 plume is at another angle. The entire structure covers about 4$'$ and shows a remarkable symmetry by rotation of 180 degrees, which suggests that it is not a superposition of several sources, but forms one system. We examined optical images and could not find any counterparts to the individual features described above.  

The XXL-GMRT-610~MHz catalog\footnote{
Seven sources are listed as resolved in the XXL-GMRT-610~MHz catalog (\citealt{smolcic2018}) 
within 4$'$ of the core of the Double Irony: 
XXL-GMRT J020354.8-041356, 
XXL-GMRT J020353.0-041358,
XXL-GMRT J020357.6-041244,
XXL-GMRT J020359.1-041303, 
XXL-GMRT J020348.5-041416,
XXL-GMRT J020356.1-041216, and 
XXL-GMRT J020344.7-041505. 
} 
lists seven resolved sources within 4$'$ of the core of the Double Irony \citepalias{smolcic2018}. 

Figure~\ref{figDbleIronyGradI} shows the GMRT 610~MHz intensity gradient map.  
The white line (which corresponds to the peaks in the intensity map where the gradient is zero) 
extends far out and traces the path of the jets. 
Areas of strong gradients in the inner region and in the Elbow region 
have a flatter spectral index 
(Table~\ref{tabDbleIronyAlpha}).  
This has been observed in more nearby low-luminosity FR~{\sc i} radio galaxies by \cite{2011MNRAS.417.2789L}. 

Following the jet's path  
out to the 5$\sigma$ contour of the GMRT 610~MHz map, we measure a total length of 
$\sim 7.56'$, which corresponds to a projected physical size of 1100~kpc. 
{\emph{This means that the Double Irony is a GRG.}}

\begin{figure*}
  \includegraphics[width=8.8cm]{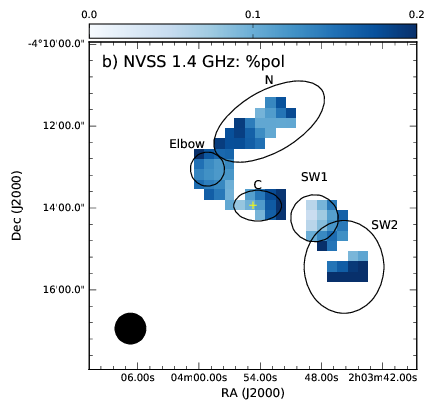}
  \includegraphics[width=8.8cm]{fig8DbleIronyNVSSFPOL.png}
\caption{Polarization in the Double Irony at 1.4~GHz (NVSS). 
{\bf a)}
Polarized intensity image from the NVSS at 1.4~GHz 
(in Jy~beam$^{-1}$)
with vectors indicating the orientation of the electric field of the polarized wave rotated by 90$^\circ$. 
The length of the vectors is proportional to the fractional polarization, and 
$15\arcsec$ corresponds to a polarized fraction of 10\%. 
In the calculation of the fractional polarization, a threshold of 4.5~mJy~beam$^{-1}$ ($10 \sigma$) was used in total intensity 
and of 0.8~mJy~beam$^{-1}$ ($2 \sigma$) in polarized intensity. 
The pixels are not independent: the FWHM of the beam is 45$\arcsec$ and the pixel size is 15$\arcsec$, and vectors are shown for every pixel above the threshold in fractional polarization. 
{\bf b)} Image of the fractional polarization. 
The color scale ranges from 0\% to 20\%. 
The ellipses correspond to the regions shown in Fig.~\ref{figDbleIronyAllradioimages}.   
The highest fractional polarization is found in the southwestern plume. 
The western jet is more polarized than the eastern one, which might be an indication that the eastern jet points away from the observer and is depolarized 
(Laing--Garrington effect), while the western jet is directed toward the observer.
In both figures   
the yellow crosses indicate the location of the center of the host galaxy, and  
the beam is displayed in the bottom left corner. 
} 
\label{figDbleIronyPolar}
\end{figure*}

\subsubsection{Luminosity, flux density, spectral index}
\label{SectDbleIronySpectralIndex}

\begin{table*}[t]
\caption{Flux density measurements in the regions of the Double Irony shown in Fig.~\ref{figDbleIronyAllradioimages}. }
\label{tabDbleIronyFlux}
\centering
\begin{tabular}{ccccc}
\hline \hline
\smallskip
Region & $S_{\rm 150~MHz}$ &$S_{\rm 610~MHz}$  &$S_{\rm 1.4~GHz}$ (FIRST) & $S_{\rm 1.4~GHz}$ (NVSS)\\
            & (mJy)  & (mJy)   & (mJy)   & (mJy) \\
\hline
C & 31.60 $\pm$ 10.06 & 23.75 $\pm$ 0.45    & 10.95 $\pm$ 1.23     & 11.54 $\pm$ 0.84\\
Elbow & 17.09 $\pm$ 9.14 & 20.83 $\pm$ 0.40 & 7.59 $\pm 1.10$   & 8.67 $\pm$ 0.68\\
N & 136.25 $\pm$ 23.15 & 77.76 $\pm$ 1.03   & 7.06 $\pm 2.79^*$ & 29.29 $\pm$ 1.82\\
SW1 & 98.17 $\pm$ 12.54 & 46.29 $\pm$ 0.56  & 14.43 $\pm 1.52^*$ & 23.98 $\pm$ 0.99\\
SW2 & 99.51 $\pm$ 22.84 & 63.97 $\pm$ 1.02  & 22.93 $\pm 2.77^*$ & 27.88 $\pm$ 1.79\\
T & 434.95 $\pm$ 79.70 & 205.76 $\pm$ 3.54  & 21.57 $\pm 9.65^*$ & 131.23 $\pm$ 6.27\\
\hline
\end{tabular}
\tablefoot{
The FIRST values marked by a * are significantly lower than those measured in the NVSS, an indication that extended emission has been filtered out.}
\end{table*}

\begin{table}[t]
\caption{Spectral index measurements in the different regions of the Double Irony shown in Fig.~\ref{figDbleIronyAllradioimages}.}
\label{tabDbleIronyAlpha}
\centering
\begin{tabular}{ccc}
\hline \hline
\noalign{\smallskip}
Region & $\alpha_{\rm 150~MHz}^{\rm 610~MHz}$ 
       &$\alpha_{\rm 150~MHz}^{\rm 1400~MHz}$\\
\noalign{\smallskip}
\hline 
$\Delta \alpha_{\rm sys}$   &0.1    &0.17\\
C   & $-0.20 \pm 0.23$ 
    & $-0.45 \pm 0.15$\\
Elbow & $+0.14 \pm 0.38$ 
    & $-0.30 \pm 0.24$\\
N   & $-0.40 \pm 0.12$ 
    & $-0.69 \pm 0.08$\\
SW1 & $-0.54 \pm 0.09$ 
    & $-0.63 \pm 0.06$\\
SW2 & $-0.31 \pm 0.16$ 
    & $-0.57 \pm 0.11$\\
T   & $-0.53 \pm 0.13$ 
    & $-0.54 \pm 0.08$\\
\hline
\end{tabular}
\tablefoot{At 1.4~GHz the NVSS flux density measurements were used. 
The first line lists the systematic uncertainties on the spectral indices assuming a 10\% systematic uncertainty on the flux densities. The other uncertainties in the table refer to the statistical uncertainties. 
}
\end{table}

From the NVSS flux density measurement (Table~\ref{tabDbleIrony}) we derive a spectral luminosity at 1.4~GHz of $6.9\times10^{24}$~W~Hz$^{-1}$ (Eq.~\ref{eqLnu}), or 
of $2.9\times10^{25}$~W~Hz$^{-1}$ at 178~MHz assuming a spectral index of $-0.7$. 
This is about twice the spectral luminosity of the Exemplar, but still in the range of luminosities of FR~{\sc i} radio galaxies.  
Some GRGs are powerful radio galaxies with FR~{\sc ii} structures 
(e.g., \citealt{1996MNRAS.279..257S}; 
\citealt{1999MNRAS.309..100I}; 
\citealt{2001A&A...371..445M}); 
as the sensitivity of observations increases, more FR~{\sc i} GRGs (or hybrid FR~{\sc i}/FR~{\sc ii}) with 
lobes of lower surface brightness are found 
(e.g., \citealt{2001A&A...370..409L}; 
\citealt{2005AJ....130..896S}). 

In Tables~\ref{tabDbleIronyFlux} and \ref{tabDbleIronyAlpha} we list the values for the flux densities and the spectral indices measured in different regions displayed in Fig.~\ref{figDbleIronyAllradioimages}. 
The spectral indices measured between 150~MHz and 1.4~GHz are systematically steeper than the low-frequency spectral indices measured between 150~MHz and 610~MHz, an indication of spectral aging  of the cosmic-ray electrons. 

 The Elbow and the central region both have a relatively flat spectral index,  
but no systematic pattern is seen. 
We intend to model the evolution of cosmic-ray electrons in the Double Irony and in the Exemplar and to derive spectral age estimates by analyzing more sensitive low-frequency data (pending GMRT observations below 500~MHz). 

\subsubsection{Polarization}

Figure~\ref{figDbleIronyPolar} shows images of the polarization detected by NVSS\footnote{
Four continuum sources associated with the Double Irony are listed in the NVSS catalog \citep{NVSS}; 
all of them are listed as polarized, with polarized flux densities ranging from 1.71 to 5.95~mJy. 
None of these sources appears in the NVSS rotation measure (RM) catalog of \cite{2009ApJ...702.1230T}, 
which had a selection threshold of 5~mJy on the Stokes $I$ intensity and of 8$\sigma$ on the polarized intensity.} at 1.4~GHz. 
The left panel shows the 
polarized 
intensity image overlaid with vectors showing the direction of the electric field of the polarized wave rotated by 90$^\circ$. In the absence of Faraday rotation, this would correspond to the direction of the magnetic field component on the plane of the sky, $B_\perp$, in the source. The length of the vectors is proportional to the 
fractional polarization. 

The vectors follow  the large-scale distribution of the total intensity rather well, except in the very center where they are almost perpendicular to the direction of the inner jets and may trace a poloidal magnetic field.  
We note, however, that one vector per pixel is shown, so the vectors (like the pixels) are correlated, and variations on scales  smaller than the synthesized beam (45$\arcsec$, shown in the bottom left corner) are smoothed out. 

In the right panel of Fig.~\ref{figDbleIronyPolar} we show an image of the fractional polarization. 
The position of the core is indicated by a cross. 
It appears that the fractional polarization peaks on the right side of the core, while the left side is less polarized. This could be a signature of the Laing--Garrington effect 
(\citealt{1988Natur.331..149L}; \citealt{1988Natur.331..147G}): 
stronger external Faraday depolarization \citep{1966MNRAS.133...67B} 
occurs along the line of sight of the counter-jet due to the longer path length through a fluctuating magnetic field component.  
This would mean that 
{the eastern part of jet points away from us, and the more strongly polarized western jet points toward us.}

Dedicated observations of the polarization at several radio frequencies are required to map the magnetic field pattern in this radio galaxy and measure the Faraday rotation. 

\subsection{Nature of the Double Irony}
\label{SectDiscussionNatureDbleIrony}

The nature of the Double Irony radio galaxy remains somewhat of a mystery. 
Given the very large extent on the sky, we suspect that the inclination is not very large, perhaps 30$^\circ$ to 45$^\circ$ relative to the plane of the sky, but 
the structure is probably not entirely coplanar. 
The radio galaxy displays a remarkable symmetry by rotation of 180$^\circ$, except for 
the strong bend near the Elbow in the northeast, 
while the western part is more regular. 

Similar features have been seen in other FR~{\sc i} radio galaxies. For example 3C~31 has two 90-degree elbows, and the overall radio structure could be nicely reproduced in the dynamical model of an unbound gravitational encounter
with a neighboring galaxy (\citealt{1978MNRAS.185..527B}). 
The morphology of the Double Irony is similar to that of NGC~326, a radio galaxy that has been the subject of different interpretations, 
such as jet precession (\citealt{1978Natur.276..588E}), 
jet realignment due to gravitational interaction between two galaxies (\citealt{1982AJ.....87..602W}), 
or buoyancy forces that redirect the lobes into regions of lower pressure in the surrounding medium (\citealt{1995ApJ...449...93W}). 
In the following we discuss some physical effects that may have shaped the Double Irony.  

\subsubsection{A restarted radio galaxy?}

Some other GRGs have a morphology that is indicative of a recurrent nuclear activity. For example \cite{2002ApJ...565..256S} 
found an inner double radio structure within larger edge-brightened lobes. 
The two bright regions of the Double Irony (the SW lobe and the Elbow) and the two diffuse outer plumes beyond the SW lobe and the Elbow may be due to two separate events, where the two brighter inner features would be more recent.  
However, it would be a coincidence for the restarting lobes to have just reached the bends. It seems more likely that it is a quasi-steady structure influenced by the intergalactic medium. 

\subsubsection{Gravitational interaction with the second brightest galaxy} 
\label{sectDiscussion2ndGal}

A second galaxy is clearly visible to the NE of the host of the Double Irony. 
It is the next brightest galaxy within a radius of 1~Mpc and is listed in SDSS (SDSS~J020357.62-041321.5).  
With a spectroscopic redshift (from BOSS) of 0.13433, the line-of-sight velocity difference with 
the host of the Double Irony is 840~km~s$^{-1}$. 
It is at a distance on the sky of 59$\arcsec$, or a projected distance of 146~kpc. 
An analysis of the spectrum in the GAMA database (GAMA~J20357.62-041321.5) yields a stellar mass of about $0.9\times10^{11}~M_\odot$, which is about one-third of the stellar mass of the host of the Double Irony 
(Table~\ref{tabDbleIrony}).  
There is no sign of tidal perturbations in the optical surface brightness  
distribution of this galaxy, which is an argument against a prograde encounter with the host of the Double Irony.   
Nevertheless, given the relative line-of-sight velocity difference and their projected distance, the two galaxies could have been interacting  
gravitationally in the past $\sim100$~Myr, causing a motion of the host of the Double Irony in the general direction of the second galaxy during the lifetime of the radio source. 
The drop-like morphology of the SW1 radio lobe is consistent with {a global motion of the AGN of the Double Irony toward the east, in the general direction of the companion}. 

An observation of the distribution of atomic hydrogen via the H{\sc i} 21~cm line in the companion galaxy may reveal a possible interaction between the two galaxies. 
The galaxy must contain some atomic hydrogen gas since it is disk-like and star forming: from the GAMA spectrum, we estimated a SFR of about 0.8~$M_\odot$~yr$^{-1}$ in the last 2--20 Myr and a higher SFR earlier (8.8~$M_\odot$~yr$^{-1}$ 20--600~Myr ago and 
 about 19~$M_\odot$~yr$^{-1}$ 0.6--5.6~Gyr ago). 
 The detection of H{\sc i} tidal tails would constrain the dynamical history of this galaxy. Unfortunately, its redshift places the H{\sc i} line near 1250~MHz, in a region that is usually affected by radio-frequency interference. 

\subsubsection{Interaction with the gaseous environment}

In the absence of a deeper X-ray image not much can be said about the surrounding medium of the Double Irony. 
The upper limit on the X-ray emission provides an upper limit on the temperature of about 1.3~keV 
and a limit on the mass within $r_{500}$ of about $5\times 10^{13} M_\odot$ (Table~\ref{tabDbleIrony}).  
The ejected radio-emitting plasma may have been able to travel farther out because of the low-density of the surrounding medium.  
The absence of significant X-ray emission may also indicate that the atmosphere of the cluster/group has been so spread out by the radio galaxy's heating and pushing 
that the $n^2 L$ (density squared times pathlength) through the cluster has dropped from what it was when the radio source turned on. 
The change in direction of the outer plumes may also be a sign of backflow, as seen for instance 
in the FR~{\sc i} radio galaxies 3C~296 \citep{2006MNRAS.372..510L}, 
3C~270 \citep{2015MNRAS.450.1732K}, 
or NGC~326 (\citealt{2012ApJ...746..167H}). 

The circumgalactic medium (the gas located outside the optical body of a galaxy but inside the galaxy's virial radius) 
extends in some cases on scales larger than 100~kpc (\citealt{2017ARA&A..55..389T}) 
and the Double Irony may have interacted with the circumgalactic medium of its bright neighbor located beyond the bend and mentioned in the previous section. 

\section{Comparison of the two radio galaxies}
\label{SectCompa}

\begin{figure*}
\centering
\includegraphics[width=17.7cm]{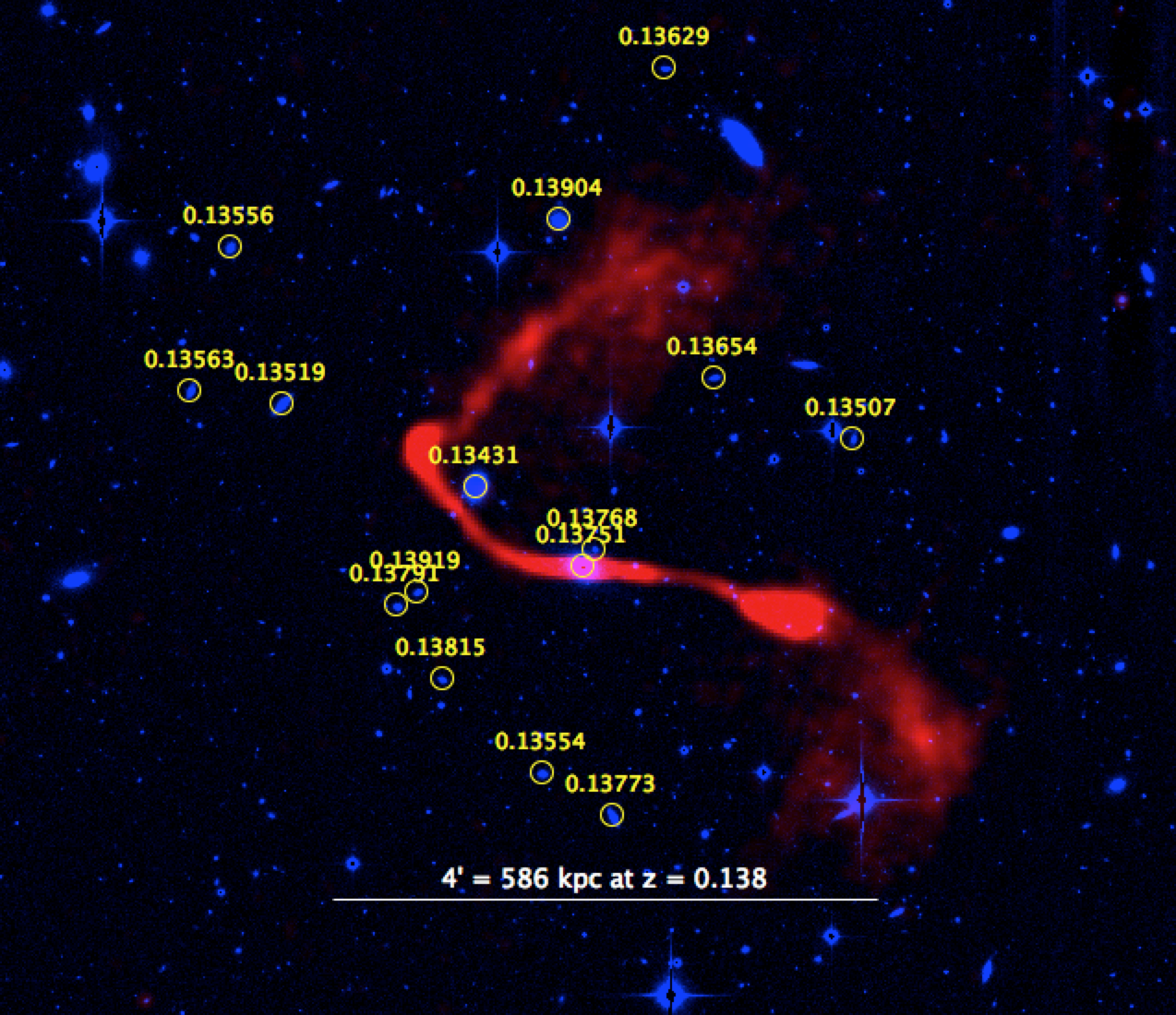}
\caption{Multiband false-color image of the Double Irony: $i$-band image from the CFHT in blue, GMRT 610~MHz image in red. 
The galaxies at a redshift close to that of the host of the Double Irony are shown as yellow circles labeled by their spectroscopic redshift from the GAMA database.}
\label{figDbleIronyGAMAgalaxies}
\end{figure*}

In Sect.~\ref{SectSuperclu} we show that the Double Irony and the Exemplar are part of the same large-scale structure, a supercluster at $z\simeq 0.14$. 
In this context it is of interest to compare the two radio galaxies and their environments. 

In Sect.~\ref{SectDiscussionMBH} we estimate their black hole masses from scaling relations.  
In Sect.~\ref{SectDiscussionLERGs} we discuss  
the properties of the optical spectra of the host galaxies in relation to what is known from large samples of radio galaxies. 
In Sect.~\ref{sectDiscusssionClusterMass}  
we analyze the distribution of galaxies surrounding the radio galaxies and make estimates of the cluster/group masses. 
In Sect.~\ref{SectDiscussionClusterAges} 
we derive an age estimate from the properties of the two brightest galaxies in each cluster/group. 
Finally, in Sect.~\ref{sectCompaEnvironment} we examine the distributions of galaxies that surround each radio galaxy and quantify their degrees of anisotropy in relation to the orientation of the radio jets and lobes. 

\subsection{Estimation of the black hole masses}
\label{SectDiscussionMBH}

There is a well-known correlation  between the mass of the black hole and 
the stellar mass (bulge mass) of the host galaxy (e.g., \citealt{2000ApJ...539L...9F}). 
This correlation is expressed as 
\begin{equation}
\log_{10}\left( 
    \frac{M_{\rm BH}}{M_\odot}   
          \right) = 
a + b \log_{10}(X) \, ,  
\end{equation}
where $X$ is the stellar mass itself or an indicator of it (e.g., the velocity dispersion, the $K$-band absolute magnitude). 

We have stellar mass estimates for both galaxies  
(Tables~\ref{tabTheOtherOne} and \ref{tabDbleIrony}).  
Using the parameters given by \cite{2013ApJ...764..184M} 
in their fits to the stellar mass--black hole mass relation 
($a = 8.56\pm 0.10$; $b = 1.34\pm0.15$), we obtain 
\begin{equation}
\begin{array}{lcl}
M_{\rm BH}^{\rm Exemplar}     &\simeq &1.0\times10^9 M_\odot\\
\medskip
M_{\rm BH}^{\rm Double~Irony} &\simeq &1.6\times10^9 M_\odot \, .\\
\end{array}
\end{equation}

For the Exemplar we can also use the value of the velocity dispersion from SDSS DR~14 (Table~\ref{tabTheOtherOne}). 
The fits of \cite{2013ApJ...764..184M} 
for elliptical galaxies  
when $X$ is the velocity dispersion in units of 200~km~s$^{-1}$ 
have the following parameters: 
$a = 8.39 \pm 0.06$  and $b = 5.20 \pm 0.36$.  
This gives a black hole mass of about 
$1.5\times10^9$~$M_\odot$. 

We see that {the black holes of the Double Irony and Exemplar both have a mass on the order of $10^9$~$M_\odot$}. 

\subsection{Brightest cluster galaxies}
\label{SectDiscussionLERGs}

The host galaxy of the Double Irony is somewhat brighter than that of the Exemplar in all bands
except the WISE W1 and W2 bands 
(the difference is largest in the $u$ band, Tables~\ref{tabOptNIRphot} and \ref{tabWISEphot}). 

The two host galaxies have very similar optical spectra (Fig.~\ref{figGAMAspectra}) characteristic of old stellar populations.  
Detection of H$\alpha$, H$\beta$, and [O{\sc ii}] is reported in the GAMA database for both galaxies. 
However, there is no detection of high-excitation emission lines such as [O{\sc iii}] that is seen in the host galaxies of  high-excitation radio galaxies (HERGs). 
It has been shown that the most luminous radio galaxies (FR~{\sc ii}) are predominantly HERGs, 
while the hosts of the less powerful FR~{\sc i} radio galaxies do not exhibit such spectral lines. 
Low-excitation radio galaxies (LERGs) are in general hosted by galaxies that have a lower star formation rate, are located in clusters, and are more massive 
than the host galaxies of HERGs. They are believed to have different modes of accreting gas: the LERGs accrete mostly hot gas, whereas the HERGs accrete cold gas and radiate more efficiently 
(e.g., 
\citealt{hardcastle07}; 
\citealt{2008A&A...490..893T}; 
\citealt{smolcic09}; 
\citealt{best12}; 
\citealt{best14}). 
{The Exemplar and the Double Irony can both be classified as LERGs.}

Both radio galaxies can be classified as FR~{\sc i} from their morphologies. 
By examining the mid-infrared photometry and colors of the host galaxies of radio galaxies,
\cite{2014MNRAS.438..796S} 
was able to show that FR~{\sc i} and FR~{\sc ii} fall in different parts of the (W1-W2)-(W2-W3) color-color diagram (([3.4~$\mu$m] - [4.6~$\mu$m])-([4.6~$\mu$m]-[12~$\mu$m]), their Fig.~13). 
The WISE colors of the hosts of our radio galaxies are consistent with those of FR~{\sc i} radio galaxies (Table~\ref{tabWISEphot}).

\subsection{Mass of the clusters}
\label{sectDiscusssionClusterMass}

The spectroscopic redshifts obtained from the GAMA database allow us to obtain a new mass estimate of the galaxy cluster that hosts the Exemplar, 
and to search for a cluster/group of galaxies around the Double Irony. 
The velocities $v_i$ in the rest frame of a cluster at redshift $z_{\rm cluster}$ are simply
\begin{equation}
    v_i = c \frac{z_i - z_{\rm cluster}}{1 + z_{\rm cluster}}\, , 
\end{equation}
where $z_i$ are the redshifts of the galaxies.

The velocity dispersion within the virial radius can be used to obtain an estimate of the virial mass. However, this mass estimate relies on a single number and 
does not take into account the full distribution of galaxies on the sky and in redshift space.  
A more sophisticated method is the caustic method developed by \cite{1997ApJ...481..633D} and \cite{1999MNRAS.309..610D}. 
The idea is to estimate the group's escape velocity from the characteristic trumpet shape of the distribution of galaxies in the plane defined by the group/cluster-centric distance and the line-of-sight velocities with respect to the median recession velocity of the group. 
We used the modified caustic mass estimation algorithm written by \cite{2012MNRAS.426.2832A} 
to calculate the total masses of galaxy groups in the GAMA group catalog (G$^3$Cv1; \citealt{2011MNRAS.416.2640R}). 
 
\subsubsection{Cluster/group membership}

The first step is to identify the member galaxies of our two clusters/groups. We simply selected all the galaxies located within an estimated radius (which corresponds to $r_{200}$ for the Exemplar's cluster) and within a certain spectroscopic redshift range: 
\[
\left \{ 
\begin{tabular}{l}
$R < r_{200} = r_{500}/0.7 = 6.3' = 934$~kpc\\
$0.13 < z < 0.15\, .$\\
\end{tabular}
\right.
\]

To estimate the radius $r_{200}$, we used the relation of \cite{2009A&A...496..343E}: 
$r_{200} = r_{500}/0.7$, which gives $r_{200} = 934$~kpc~$= 6.3'$ for the Exemplar's cluster, using the value of $r_{500}$ given in 
Table~\ref{tabTheOtherOne}. 
In the absence of other information on the Double Irony, we took the same value of $r_{200}$ 
as for the Exemplar.  
The number of galaxies selected in this manner is given in Table~\ref{tabClusterMasses}.
It should be noted that mass estimates depend on how accurately the cluster members have been selected. 
However, as we  see below, the caustic mass method provides a way to assess cluster membership that is complete to a very large degree for massive clusters 
(\citealt{2013ApJ...768..116S}). 

\subsubsection{Caustic mass method}

The caustic mass method requires the cluster's redshift and velocity dispersion as inputs in addition to the catalog of positions and redshifts of the individual member galaxies. 
Rather than using a simple median value and standard deviation, 
we used   the  ``gapper estimator'' introduced by \cite{1990AJ....100...32B}. 
This estimator has been shown to be unbiased, even for groups with few member galaxies and robust to weak variations in group memberships. 
The estimated ``scale'' in velocity space (an estimator of the ``velocity dispersion'') is calculated as 
\begin{equation}
\sigma_{\rm gap} = \frac{\sqrt{\pi}}{N(N-1)} \sum_{i = 1}^{N-1} w_i g_i \, ,
\label{eqBeersGap}
\end{equation}
where $g_i = v_{i+1} - v_{i}$ is the velocity difference between each velocity pair computed after having sorted the galaxies' recession velocities in the catalog in increasing order; the weights are $w_i = i(N-i)$, where $N$ is the number of galaxy redshifts and $i$ varies between 1 and $N-1$. 

Following \cite{2011MNRAS.416.2640R}, 
we increased the velocity dispersion by a factor $\sqrt{N/(N-1)}$ to take into account the fact that the central galaxy is moving with the center of mass of the halo (associated cluster); we also corrected for the measurement uncertainties on the recession velocities of the individual galaxies  (50~km~s$^{-1}$ for GAMA, \citealt{2011MNRAS.413..971D}) 
by subtracting the contribution for the $N$ galaxies in the cluster 
($\sigma_{\rm err} = 50 \sqrt{N}$~km~s$^{-1}$):  
\begin{equation}
    \sigma = \left(
    \frac{N}{N-1} \sigma_{\rm gap}^2 - \sigma_{\rm err}^2 
    \right)^{1/2} \, .
\label{eqBeersGapcorr}
\end{equation}For the Exemplar we used the position of the BCG as the central position. 
For the Double Irony we used the position of the host galaxy, which is also the brightest galaxy within a radius of about 1~Mpc. 

\begin{figure}
    \centering
    \includegraphics[width=8.8cm]{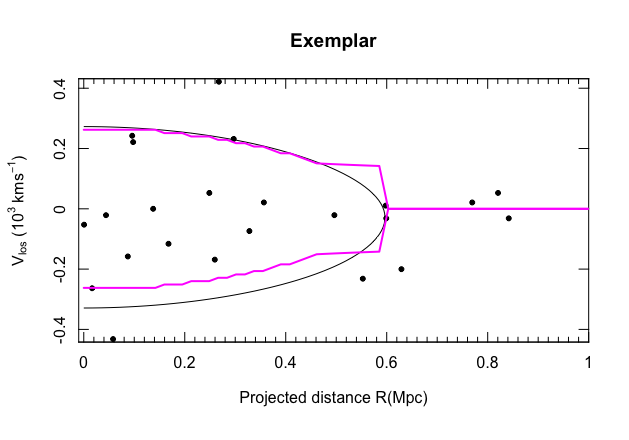}
    \includegraphics[width=8.8cm]{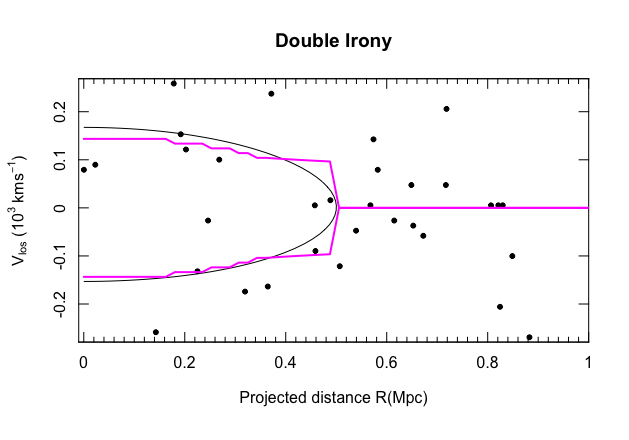}
    \caption{Result of the caustic mass analysis for the Exemplar (top panel) and the Double Irony (bottom panel). 
    The dots shows the positions of the galaxies in phase space ($R$ is the distance on the sky to the brightest galaxy, and $v_{\rm los}$ is the line-of-sight velocity relative to the mean line-of-sight velocity of the galaxies in the region. The magenta lines show the caustic lines. Galaxies outside the caustic lines have a $v_{\rm los}$ greater than the escape velocity of the cluster/group. 
    }
    \label{figCaustic}
\end{figure}

In Fig.~\ref{figCaustic}  
we show the results of the caustic mass analysis. The dots show the distribution of the galaxies in 
phase space, and the magenta lines show the caustic lines fitted to the black parabola. 
The dots outside the caustics are, by definition, beyond the turn-around radius of the cluster and have a velocity that is greater than 
the escape velocity. 
The results are summarized in Table~\ref{tabClusterMasses}. 

While there are more galaxies in the area around the Double Irony, the estimated mass of the Double Irony cluster 
is about three times smaller than that of the Exemplar. 
The caustic mass analysis shows that many galaxies fall outside the caustic line (see Fig.~\ref{figCaustic}), which indicates that they are not cluster members.  

\subsubsection{Scaling relations}
Now let us estimate the 
mass within $r_{200}$ using a simple scaling relation with the velocity dispersion within the same radius, $\sigma_{200}$. 
\cite{2004cgpc.symp....1E} 
showed that the following relation is an excellent fit to a wide range of simulated clusters over a wide range of redshifts,  
\begin{equation}
M_{200} = \frac{10^{15} h^{-1} M_\odot}{H(z)/H_0} 
    \left( 
    \frac{\sigma_{200}}{1080~{\rm km~s}^{-1}} 
    \right)^3 \, , 
\label{eqM200}
\end{equation}
where $H(z)$ is the value of the Hubble parameter at the redshift $z$ of the cluster. 
This relation is very close to the mass--velocity dispersion relation for a singular isothermal sphere truncated at the virial radius. 
For such a model in hydrostatic equilibrium, there is a simple relation between mass and temperature (e.g., \citealt{2005RvMP...77..207V}) 
\begin{equation}
k_B T_{200} = ({\rm 8.2~keV}) 
\left(
\frac{M_{200}}
{10^{15} h^{-1} M_\odot} 
\right)^{2/3} 
\left( 
\frac{H(z)}{H_0}
\right)^{2/3} \,  ,
\label{eqIsoSphereT200M200}
\end{equation} 
where $k_B$ is the Boltzmann constant and $T_{200}$ is the temperature within $r_{200}$. 

$M_{200}$ and $M_{500}$ are related by the  factor 
\begin{equation}
    M_{500} = M_{200}/1.35
    \label{eqM500}
\end{equation}
for an appropriate value of the concentration parameter. 

The mass estimates are given in Table~\ref{tabClusterMasses}.  
The caustic mass estimates are somewhat larger than the estimates from a single scaling relation with the velocity dispersion, and from the X-ray derived masses (an upper limit in the case of the Double Irony). 
Several galaxies in the field of the Double Irony
fall outside the caustic line  
(e.g., the brightest neighbor on the sky 
has a relatively high velocity relative to the mean velocity of the cluster/group). 
Given the relatively small number of galaxies, the caustic method may not be reliable to assess cluster membership for individual galaxies. 
It is also difficult to attach a realistic uncertainty to the caustic mass estimate when the number of galaxies is small. 
For the GAMA group catalog (G3Cv1), \cite{2012MNRAS.426.2832A} 
showed that ``on average, the caustic mass estimates agree with dynamical mass estimates within a factor of 2 in about 90\% of the groups and compares equally well to velocity dispersion based mass estimates for both high and low multiplicity groups over the full range of masses probed by the G3Cv1.''
{According to all indicators, the cluster/group of the Double Irony is less massive than that of the Exemplar.} 

The  temperature estimates derived from Eq.~\ref{eqIsoSphereT200M200} are 2.1~keV for the Exemplar and 0.7~keV for the Double Irony. 
These values are consistent with the X-ray estimates ($\approx 1.8$~keV for the Exemplar, Table~\ref{tabTheOtherOne}, and 
$\lessapprox 1.3$~keV for the Double Irony, Table~\ref{tabDbleIrony}). 

From the estimates of $M_{500}$ in Table~\ref{tabClusterMasses}, we can estimate that the corresponding radius, $r_{500}$, for the Double Irony's group is about 0.6 times that of the Exemplar's cluster, or about 380~kpc, which corresponds to $2\farcm6$. A circle of that radius roughly encompasses the entire radio structure. 

\begin{table*}
\centering
\begin{tabular}{cccl}
\hline
\hline
                            & Exemplar  & Double Irony  & Note\\
\hline
$N (R < 6\farcm3)$               &  46               & 55             &(1)\\
$N (< 6\farcm3, 0.13 < z < 0.15)$&  23               & 32             &(2)\\
$z_{\rm cluster}$             & 0.1387               & 0.1368              &(3)\\
$\sigma_v [{\rm km~s}^{-1}]$  & 475                  & 275                 &(4)\\
$M_{\rm Caustic} [M_\odot]$   &$3.2\times10^{14}$    & $9.1\times10^{13}$  &(5)\\
$M_{\rm 200} [M_\odot]$       &$1.7\times10^{14}$    & $3.3\times10^{13}$  &(6)\\
$M_{\rm 500} [M_\odot]$       &$1.3\times10^{14}$    & $2.5\times10^{13}$  &(7)\\
$M_{\rm 500}^{MT} [M_\odot]$  &$(0.9\pm0.4)\times10^{14}$    & $<5.1\times10^{13}$  &(8)\\
\hline
\end{tabular}
\caption[caption]{Mass estimates of the clusters.\\\hspace{\textwidth} 
    (1) Total number of galaxies within a radius 6.3$'$ of the cluster center.\\\hspace{\textwidth} 
    (2) Number of galaxies within a radius 6.3$'$ of the cluster center and with $0.13 < z < 0.15$.\\\hspace{\textwidth} 
    (3) Redshift estimate of the cluster.\\\hspace{\textwidth} 
    (4) Velocity dispersion estimated via the gapper method (Eq.~\ref{eqBeersGapcorr}).\\\hspace{\textwidth} 
    (5) Mass derived via the caustic method.\\\hspace{\textwidth} 
    (6) Mass within $r_{200}$ derived via the mass--$\sigma_{200}$ scaling relation (Eq.~\ref{eqM200}).\\\hspace{\textwidth}
    (7) Mass within $r_{500}$ scaled from $M_{200}$ estimated via the  scaling relation (Eq.~\ref{eqM500}).\\\hspace{\textwidth}
    (8) Mass within $r_{500}$ derived from  X-rays (Table~\ref{tabTheOtherOne} for the Exemplar, and upper limit for the Double Irony, Table~\ref{tabDbleIrony}).
    The caustic mass estimates are higher than those derived via a simple scaling relation between the mass and the velocity dispersion, and also higher than those derived from  X-rays. This apparent discrepancy is not surprising given the relatively small number of galaxies within the structures. 
}
\label{tabClusterMasses}
\end{table*}

\subsection{Cluster ages and dynamical states}
\label{SectDiscussionClusterAges}

The luminosity difference between the BCG and the second brightest galaxy in a cluster is used as an indicator of the evolutionary state of a cluster (e.g., \citetalias{2016MNRAS.462.4141L}). In the hierarchical model of galaxy formation, BCGs located at the center of a cluster will grow by accretion and mergers faster than other galaxies. The difference in mass (and luminosity) is expected to increase with time, and galaxy groups/clusters with a large magnitude difference $\Delta m_{12}$ between their 
two brightest galaxies are expected to have assembled early. 

\cite{2014MNRAS.442.1578R} 
used galaxies drawn from semi-analytic models of \cite{2011MNRAS.413..101G} 
based on the Millennium simulations, and investigated how some measurable parameters are related to the age of the groups. 
They defined ``young'' galaxy groups as those that have assembled up to 30\% of their mass by redshift $z = 1$, 
and ``old'' groups as the ones that had assembled more than 50\%  of their mass at  
redshift  $z = 1$. 
Plotting the distributions of the magnitude gap $\Delta m_{12}$ versus the absolute magnitude in the $r$ band ($M_r$) of the BCG, 
they were able to quantify the fraction of old and young groups in different parts of the diagram.

For the Exemplar's cluster (XLSSC~112), the difference in the $r$-band magnitude is 0.97, 
while for the BCG of the Double Irony group it is much lower, 0.35. 
The second brightest galaxy in the Double Irony environment is a late-type galaxy located near the Elbow radio structure 
(see Fig.~\ref{figDbleIronyOptXrayradio} and Sect.~\ref{sectDiscussion2ndGal}). 
The Exemplar has  $M_r = -22.73$ and the Double Irony has $M_r = -22.85$. 
Placing them in the diagram of \citeauthor{2014MNRAS.442.1578R}, 
we see that the Double Irony group has a higher probability of being a young galaxy group 
(box~(3) with 63\% probability of being young and 4\% probability of being old). 
The Exemplar group, XLSSC~112, falls in box~(7) with a 30\% probability of being young
and a 22\% probability of being old. 
This means that the probability that the Exemplar's cluster is older that the Double Irony's group/cluster is 51\%,  
while the probability that the Double Irony is older than the Exemplar is 13\%; 
there is a 36\% probability that they are in the same age group (young, old, or in between). 
While this is not conclusive in itself, it is consistent with the idea that the Double Irony's group/cluster is not as evolved as XLSSC~112.

\subsection{Comparing the surrounding galaxy distributions}
\label{sectCompaEnvironment}
\begin{figure*}
  \includegraphics[height=6.5cm]{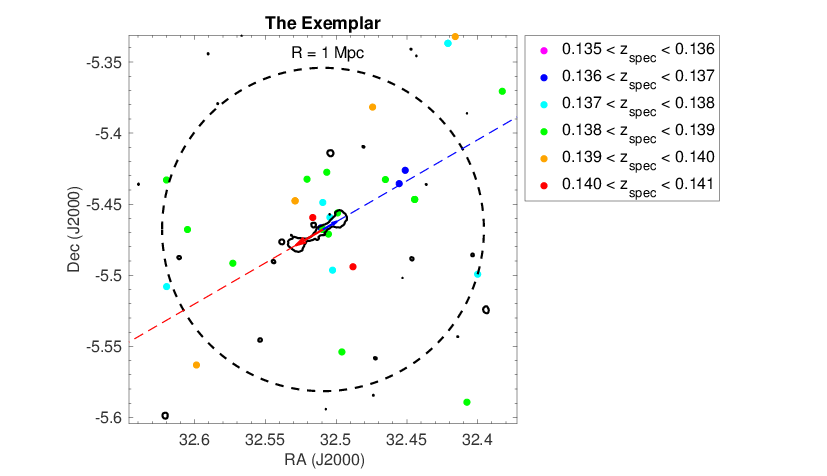}
  \includegraphics[height=6.5cm]{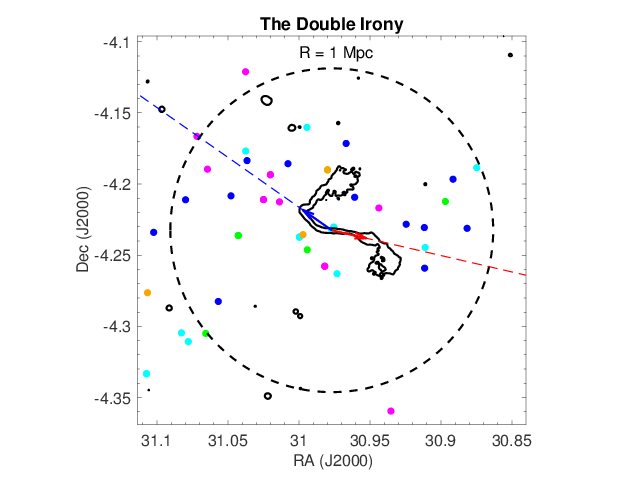}
\caption{Distribution of galaxies (colored dots) in the fields of the Exemplar (left panel) and of the Double Irony (right panel). The dots are color-coded according to the galaxies' spectroscopic redshifts in the GAMA database as indicated by the labels. 
The black contours correspond to the surface brightness at 610~MHz and increase by a factor of two from 0.3 to 4.8~mJy~beam$^{-1}$. 
The arrows point toward the brightest areas of the radio lobes; the lines were extended to show the directions of the lobes more clearly. 
In the Exemplar, the SE lobe is longest ($80\arcsec$ at a position angle of 120$^\circ$ measured from the N toward the E, shown in red); the shorter lobe ($40\arcsec$, blue arrow) points in the opposite direction. 
In the Double Irony, the arrows point toward the SW lobe (PA = $-103^\circ$, in red) 
and toward the NE Elbow (PA = $+55^\circ$ degrees, in blue) and have the same length ($90\arcsec$). 
Some asymmetries can be seen: more galaxies lie ahead of the shorter (blue) lobe than on the side of the longer lobe in the Exemplar;
in the Double Irony, there is an overdensity of galaxies in the NE, beyond  the Elbow and the northern  plume. 
}
\label{figRadiogalsEnv}
\end{figure*}

\begin{table*}
\centering
\begin{tabular}{lccccrcccc}
\hline\hline
Name    
& $R_{\rm max}$
& $N$ & $n$ 
& $\bar{N}_{\rm env}$ 
& $N/\bar{N}_{\rm env}$  
& $A_2$ &$A_3$  &$A_4$  &$A_5$\cr
& (Mpc)
& & (Mpc$^{-3}$)
& \cr
%
\hline
Exemplar &   0.5 & 15  & 0.794 &  1.12 &  13.3      & ${\bf -0.337 \pm 0.014}$ & ${\bf -0.086 \pm 0.012}$ & $+0.048 \pm 0.012$       & ${\bf -0.250 \pm 0.015}$\cr 
Exemplar &   1.0 & 24  & 0.318 &  3.62 &   6.6      & ${\bf -0.183 \pm 0.006}$ & ${\bf -0.188 \pm 0.006}$ & $-0.017 \pm 0.006$       & $+0.041 \pm 0.007$ \cr 
Double Irony &   0.5 & 19  & 1.006 &  0.75 &  25.3  & ${\bf +0.119 \pm 0.009}$ & ${\bf -0.178 \pm 0.009}$ & ${\bf +0.161 \pm 0.008}$ & $-0.007 \pm 0.010$ \cr 
Double Irony &   1.0 & 38  & 0.503 &  3.38 &  11.3  & ${\bf +0.187 \pm 0.003}$ & ${\bf -0.173 \pm 0.003}$ & ${\bf +0.097 \pm 0.003}$ & ${\bf +0.261 \pm 0.003}$ \cr 
\hline
\end{tabular}
\caption{Fourier components of the galaxy distribution around the two radio galaxies.
$N$ is the number of galaxies  
in a cylindrical volume of radius $R_{\rm max}$ and depth of $2\Delta z$, 
where $\Delta z = 0.003$, centered at the position and redshift of the host galaxy; 
$n$ is the number density of galaxies in the same volume; 
$\bar{N}_{\rm env}$ is the average number of galaxies in the environment, calculated in similar volumes placed at eight locations around the radio galaxy (see text).  
The most significantly detected anisotropy parameters are in bold. 
}
\label{tabFourierComps}
\end{table*}

\begin{figure*}
\includegraphics[width=17.7cm]{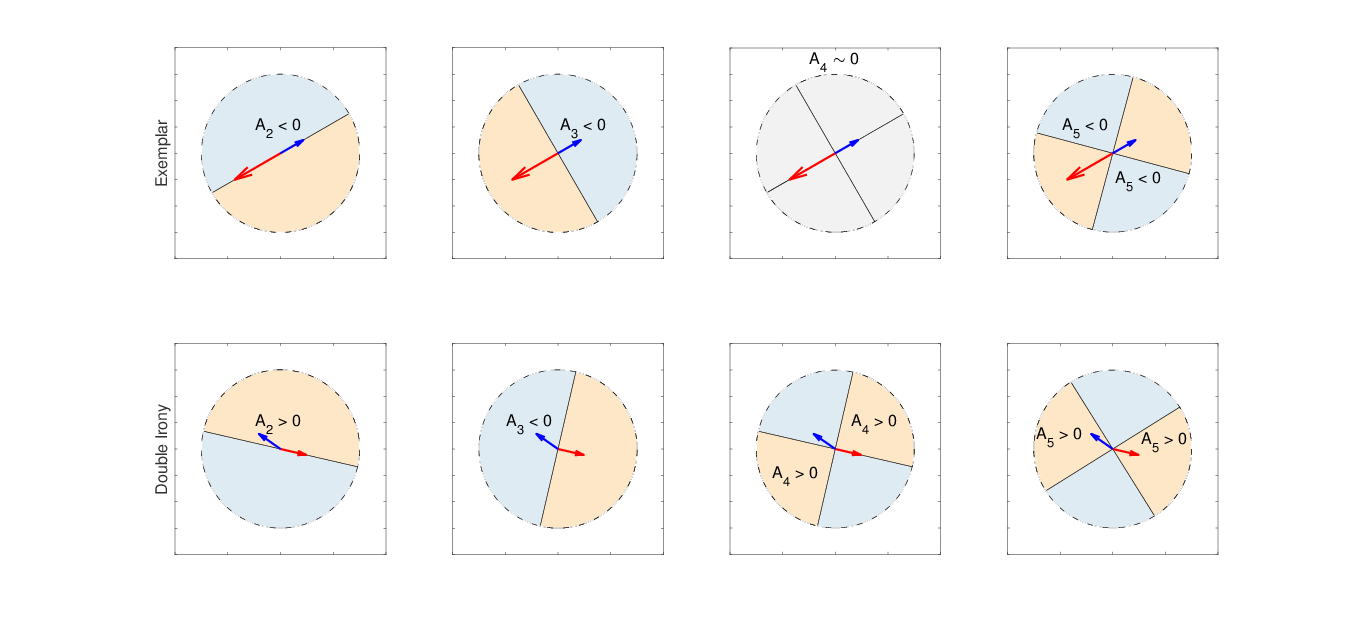}
\caption{Illustration of the Fourier components used to quantify the anisotopy in the distribution of galaxies surrounding the Exemplar radio galaxy (top panel) and the Double Irony (bottom panel). The $A_2$ and $A_3$ parameters are sensitive to a dipolar distribution and the $A_4$ and $A_5$ to a quadrupolar distribution. The arrows indicate the orientation of the radio jets. For the Exemplar the red arrow shows the direction of the longer radio lobe. 
In the Double Irony the blue and red arrows have the same length and point toward the brightest regions in the lobes. 
In the calculation of the $A_k$ parameters, the red arrow (SW lobe) is taken as the origin of the angles $\theta_i$. 
The blue regions correspond to negative values of the $A_k$ parameters and the orange regions to positive values. 
In each panel the measured sign of each $A_k$ parameter is indicated in the corresponding region. 
}
\label{figSketchAnisotropies}
\end{figure*}

A number of studies have been carried out to search for a relation between the large-scale distribution of galaxies surrounding GRGs and the orientation of the radio lobes and plumes 
(e.g., \citealt{1986A&A...170...20S}; 
\citealt{2008ApJ...677...63S}; 
\citealt{2009MNRAS.393....2S}; 
\citealt{2013MNRAS.434.2877T}; 
\citealt{2015MNRAS.449..955M}). 
There is increased evidence that overdensities of galaxies are found on the side of the shorter radio lobes, 
and that jets are oriented in directions perpendicular to the overall galaxy distribution surrounding the host galaxy; also, non-colinear radio jets and lobes seem to have directions perpendicular to the surrounding galaxy distribution, 
suggesting that they have been deflected \citep{2015MNRAS.449..955M}. 

XXL-North has been covered by the GAMA spectroscopic survey (G02 region in the GAMA Data Release 3; \citealt{2018MNRAS.474.3875B}) 
that is 95.5\%\ redshift complete to a magnitude $r < 19.8$~mag.  
In this section, we begin with a qualitative description of the distribution of galaxies located within a radius of 1~Mpc of the centers of the two radio galaxies and with a spectroscopic redshift from the GAMA database close to that of the radio galaxy (see Fig.~\ref{figRadiogalsEnv}). 
Then we carry out a more quantitative analysis in terms of Fourier components of the distribution of the galaxies surrounding the radio sources. 

In Fig.~\ref{figRadiogalsEnv} 
we show the distribution of the galaxies with a spectroscopic redshift $z_{\rm RG} -\Delta z < z < z_{\rm RG} +\Delta z$, where $z_{\rm RG} = 0.138$ is the redshift of the radio galaxy and $\Delta z = 0.003$. The galaxies are color-coded in redshift bins of 0.001, increasing from blue to red. 
The overlaid contour corresponds to the 5$\sigma$ level (0.3~mJy~beam$^{-1}$) in the 610~MHz images of the radio galaxies. The arrows point in the directions of the brightest regions in the lobes. 
The lines were extended to separate the regions above and below the radio axes.

In the Exemplar, the jets and lobes point in opposite directions, but the SW lobe is farther away from the core than the NE lobe. 
The red arrow points to the brightest region in the longer lobe and the blue arrow to the shorter one. We note the following: 

-- more galaxies are clearly situated above the Exemplar's radio axis; 

--  an overdensity of galaxies are in the direction of the shorter lobe relative to the opposite direction;  

-- in the central region, the galaxy distribution is roughly perpendicular to the radio axis. 

The Double Irony radio galaxy is not a linear structure. 
The red arrow points to the brightest region in the SW lobe and the blue arrow to the Elbow; 
these two regions are at about the same distance from the center of the host galaxy ($90\arcsec$), but have different position angles. We see: 

-- an overdensity of galaxies above  the radio axis (at positive angles, measured  counterclockwise). 

--  in the direction of the blue arrow,  an overdensity of galaxies beyond the Elbow and to the east of the northern plume; 

--  an overdensity to the northwest of the SW plume. 

To quantify the distribution of the galaxies around our two radio galaxies and those apparent features, 
let us use the Fourier component method 
proposed by \cite{2013MNRAS.434.2877T} 
and applied to a sample of GRGs by \cite{2015MNRAS.449..955M}. 

The method consists of estimating   five components defined as
\begin{equation}
A_1 = \sum\limits_{i=1}^{N} f_k (\theta_i) \, , \\
\end{equation}
and 
\begin{equation}
A_k = \frac{1}{N} \sum\limits_{i=1}^{N} f_k (\theta_i) \,\,\,\,\,  {\rm for}\,\, k = 2, ..., 5, 
\end{equation}
where the summation is over the number of galaxies, $N$, 
within a radius $R$ (taken to 0.5 or 1~Mpc) 
and a redshift $\Delta z = \pm 0.003$ of 
the position and redshift of the radio galaxy; the values of 
 $\theta_i$ are the angles measured counterclockwise on the sky 
between the direction of galaxy $i$ and that of the longer radio lobe; the $f_k$ functions are  
$f_1 = 1$, 
$f_2 = \sin(\theta_i)$, 
$f_3 = \cos(\theta_i)$, 
$f_4 = \sin(2 \theta_i)$,
$f_5 = \cos(2 \theta_i)$.
 
We use a slightly different notation from that of \cite{2015MNRAS.449..955M} 
($A_k$ rather than $a_k$) to make it clear that   
our $A_k$ are normalized by dividing by the number of galaxies $N$ in the corresponding volume surrounding the radio galaxy,  
while \cite{2015MNRAS.449..955M} divided by the number of galaxies 
located farther out in the environment of the radio sources. 
The absolute values of their $a_k$ are therefore a mixture of information related to the actual local anisotropy of the galaxy distribution and the overdensity with respect to the more distant environment. 
The difference is therefore simply a normalization factor (which is different for every radio galaxy)  
and we find the use of $A_k$ more intuitive because they are actual mean quantities.   

The $A_k$ components are illustrated in Fig.~\ref{figSketchAnisotropies} and can be understood as follows:  
\begin{enumerate} 
\item $A_1 = N$ is simply the number of galaxies located within the pre-defined radius from the core of the radio galaxy. 
\item $A_2$ and $A_3$ are sensitive to a dipolar distribution of galaxies. The 
 $A_3$ parameter (average of the cosines) is sensitive to asymmetries in the density distribution of galaxies along the radio axis (e.g., a positive $A_3$ indicates an overdensity of galaxies along the longer radio lobe compared to the 
opposite direction; a negative $A_3$ indicates an overdensity along the shorter radio lobe). 
The $A_2$ parameter (average of the sines) expresses asymmetries between the two sides of the radio axis (i.e., in the direction perpendicular to that of the radio axis, above or below the radio axis);
\item 
$A_4$ and $A_5$ are sensitive to a quadrupolar distribution. A positive $A_5$ (average of $\cos(2\theta_i)$) corresponds to 
overdensities of galaxies along the radio axis compared to the direction perpendicular to the radio axis; 
a negative $A_5$ will indicate an overdensity above and below the radio axis compared to along the radio axis. 
$A_4$ (average of $\sin(2\theta_i)$ also indicates a quadrupolar anisotropy, but at a $45^\circ$ angle compared to the previous distribution (see Fig.~\ref{figSketchAnisotropies}). 
\end{enumerate}
In the case of a non-uniform distribution of surrounding galaxies,  
all $A_k$ components will differ significantly from zero. 

The quoted uncertainties on the $A_k$ Fourier coefficients ($k\geq 2)$ are standard errors and were calculated using jackknives: the $A_k$ coefficients were recalculated by removing one galaxy at a time; then the standard error on the $A_k$ values was calculated.  

To estimate the overdensity of galaxies around the radio galaxy, we defined a $3\times3$ grid centered on the radio galaxy host at (RA$_{\rm RG}$,DEC$_{\rm RG}$) with eight points 
placed at RA = RA$_{\rm RG}\pm \Delta\theta$ and DEC = DEC$_{\rm RG}\pm\Delta\theta$, with $\theta = 0.5^\circ$. Then we calculated the mean number of galaxies in eight 24~Mpc$^3 $ volumes 
(radius of 1~Mpc and $z = z_{RG}\pm 0.003$), 
$\bar{N}_{\rm env}$. The overdensity is $N/N_{\rm env}$, where $N$ is the number of galaxies within the same volume centered on the radio galaxy (in the middle of the grid).

The distribution of galaxies with spectroscopic redshifts from the GAMA database close to that of the two radio galaxies 
is displayed in Fig.~\ref{figRadiogalsEnv}. 
The Fourier analysis was applied to this data set and the resulting $A_k$ components are listed in Table~\ref{tabFourierComps}. 
The main results can be summarized as follows: 

\begin{figure*}
\includegraphics[width=9.0cm]{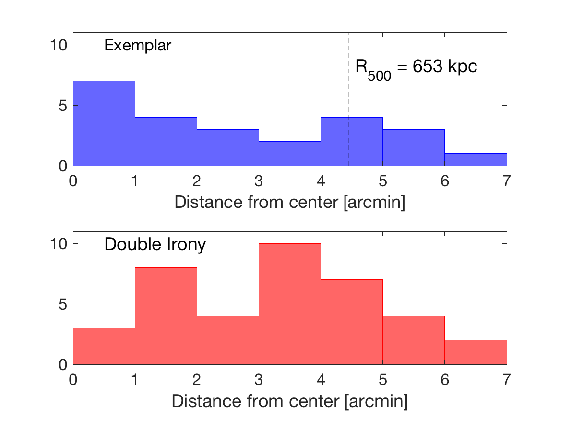}
\includegraphics[width=8.6cm]{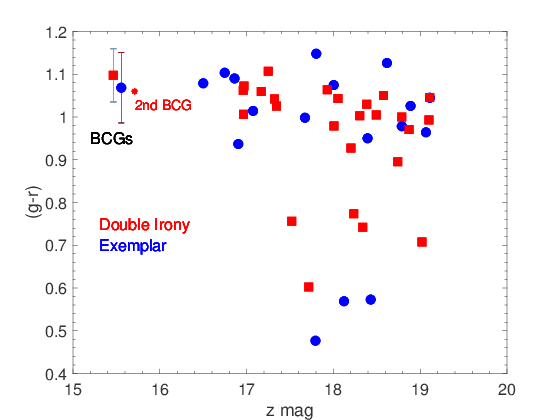}
\caption{{\it Left panel:} Histograms of the radial distribution of galaxies with spectroscopic redshifts from GAMA within $7'$ (about 1~Mpc) of the host of each radio galaxy and with a spectroscopic redshift within $\pm$0.003 from the redshift of the radio galaxies ($z = 0.138$). The galaxy distribution in the field of the Exemplar is more concentrated than that in the field of the Double Irony.
{\it Right panel:} 
SDSS color-magnitude diagram of the galaxies located within a radius of 1~Mpc of the host of each radio galaxy and with a spectroscopic redshift within $\pm$0.003 from the redshift of the radio galaxies ($z = 0.138$). 
Also shown (red star) is the second brightest galaxy located about $1'$ to the NE of  Double Irony's host and discussed in Sect.~\ref{sectDiscussion2ndGal}. 
This galaxy has a redshift that falls just below the selected redshift range and is not included in the Fourier component analysis.
Except for the presence of the  second bright galaxy near the Double Irony, the distributions are very similar (red squares for  Double Irony and blue dots for  Exemplar). 
}
\label{figColorDiagSDSS}
\end{figure*}

-- $N$ ($= A_1$) and $N/\bar{N}_{env}$: There are significantly more galaxies within a radius of 1~Mpc of the host of the Double Irony than around the Exemplar. The overdensity compared to the local environment is almost twice as large for the Double Irony as for the Exemplar. Given such a large number of galaxies in the Double Irony, the non-detection of X-ray emitting gas may seem surprising. Could it be that the galaxies around the Double Irony are less massive than in and around the Exemplar's cluster? To answer this question we retrieved the $ugriz$ magnitudes (model magnitudes from SDSS) of the galaxies with spectroscopic redshifts from GAMA. Figure~\ref{figColorDiagSDSS} (right panel) is a color-magnitude diagram for both regions. 
Error bars are shown only for the two BCGs so as not to clutter the figure (the error bars on the $z$-magnitudes are smaller than the widths of the points). The distributions of galaxies in the radio galaxies' environment are similar: most galaxies are red, with a $(g-r)$ color around 1.05. On the $x$-axis we show the $z$-band magnitude which can be used as a proxy for the mass. The two BCGs stand out in the upper left corner of the plot as the brightest (and likely most massive galaxies), with a $z$-band magnitude difference of at least one compared to the other galaxies, except for 
the second  brightest galaxy (marked as a red star) discussed in Sect.~\ref{sectDiscussion2ndGal} and located about 1$'$ to the NE of the Double Irony host; we note that this galaxy was not included in the Fourier component analysis because its redshift falls just below the selected redshift interval of $0.138\pm0.003$. 
{The distribution of galaxies is more concentrated around the Exemplar than around the Double Irony}
(Fig.~\ref{figRadiogalsEnv}; 
left panel of Fig.~\ref{figColorDiagSDSS}). 

-- Dipole ($A_2$ and $A_3$): $A_2 >0$ for the Exemplar and $<0$ for the Double Irony. In both cases this means that there are more galaxies above the radio axes (see Fig.~\ref{figSketchAnisotropies}; we note that for the Examplar the origin of the angles is taken as the longer lobe (red arrow), directed toward the SE, so the positive $A_2$ are under the radio axis; this is the opposite for the Double Irony where the red arrow points toward the SW). 
There is also an anisotropy in the other direction ($A_3$). 

-- Quadrupole ($A_4$ and $A_5$): Most interesting are the negative $A_5$ values for the Exemplar that reflect overdensities in the direction perpendicular to the radio axis, as noted through visual examination. This anisotropy is considerably stronger within a radius of 0.5~Mpc.  For the Double Irony, the effect is the opposite: the anisotropy corresponds to overdensities in the direction of the radio axis that corresponds to the orange arrow). Interestingly, the jets seems to be deflected in both directions and redirected into regions of lower galaxy densities (see right panel of Fig.~\ref{figRadiogalsEnv}). 

It seems that the Double Irony radio galaxy has been able to expand and become a GRG in an environment that may not be significantly less dense in galaxies than that of the Exemplar, and has been strongly affected by this environment: the jets deviated from their paths and the lobes expanded away from regions of higher galaxy density into  sparser regions. 

\begin{figure*}[t]
\includegraphics[width=8.8cm]{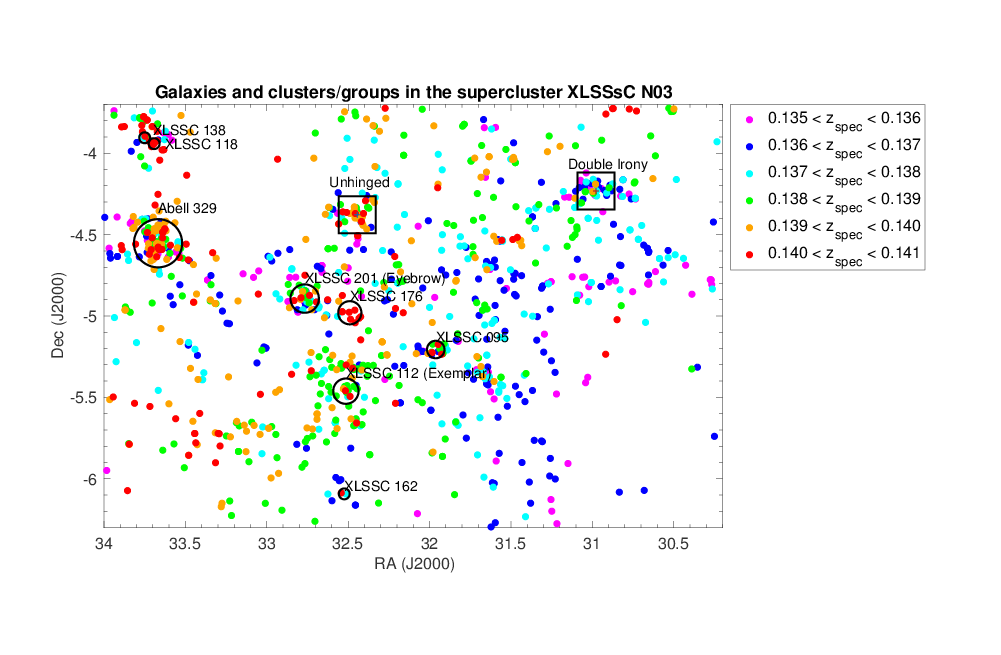}
\includegraphics[width=8.8cm]{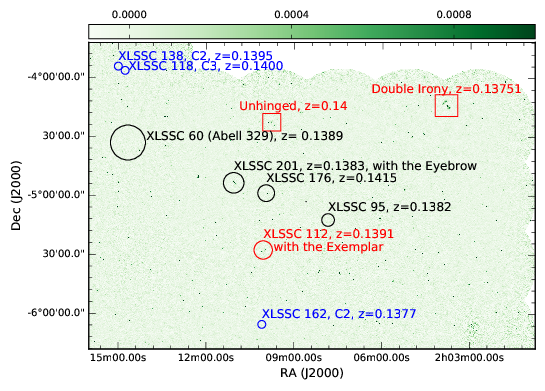}
\caption{
{\bf Top panel:}
Distribution of galaxies in the area of the supercluster XLSSsC~N03  
identified in \citetalias{adami2018}. 
The galaxies are color-coded according to their spectroscopic redshifts in the GAMA database. 
The black circles indicate the locations of the eight clusters/groups of galaxies that are known members of the supercluster. 
The two black squares indicate the location of the newly identified group associated with the Double Irony (in the upper right corner) 
and that of an overdensity of galaxies where a peculiar radio source (the ``Unhinged'') is found.  
{\bf Bottom panel:} 
GMRT 610~MHz image of the field. 
As in the upper panel, the size of the black circles indicates the extent (radius $r_{500}$) of the securely detected XXL clusters (C1 class). The three clusters circled in blue are C2 or C3 clusters whose 
sizes were not determined (in the C2 sample of XXL clusters, 50\% of the sources are likely to be real detections; the C3 sample contains the most uncertain sources). 
The Double Irony radio galaxy can be seen in the upper right corner inside a square of 1~Mpc on the side. 
The location of the Unhinged radio source is also indicated by a red square. 
} 
\label{figSuperclu}
\end{figure*}

\begin{figure*}
\includegraphics[height=8.2cm]{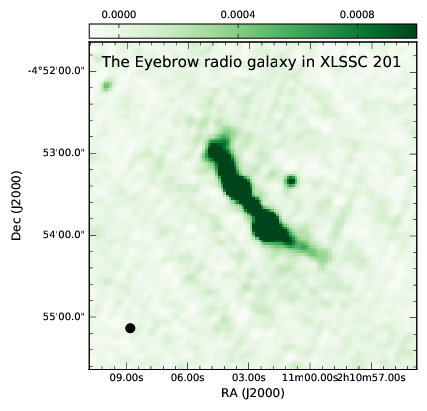}
\includegraphics[height=8.2cm]{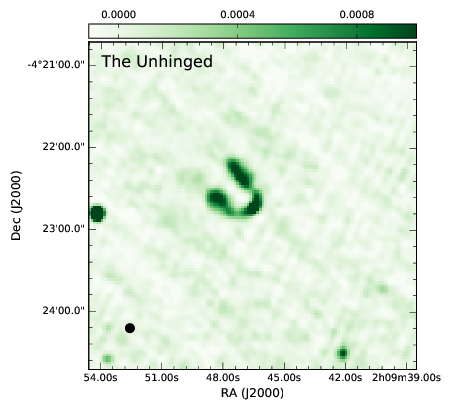}
\caption{
GMRT 610~MHz images. 
{\it Left panel:} Radio galaxy located in the cluster XLSSC~201, a known member of the supercluster. 
There are three radio sources within a radius of $1'$ of the center of the figure in the GMRT-XXL-N 610~MHz catalog of \cite{smolcic2018}: 
XXL-GMRT J021102.8-045338 and XXL-GMRT J021102.9-045335 (same ``groupid'' 22), and XXL-GMRT J021100.9-045320. 
{\it Right panel:} Peculiar radio source (possibly a composite) located in the region of a newly discovered overdensity of galaxies in the same supercluster. 
The grayscale ranges from $-10^{-4}$ to $10^{-3}$~Jy~beam$^{-1}$ and the beam is displayed in the bottom left corner. 
The GMRT-XXL-N 610~MHz contains four sources within a radius of 1$'$ at the location of the Unhinged: 
XXL-GMRT J020946.5-042242, XXL-GMRT J020947.0-042223, XXL-GMRT J020948.1-042240, and XXL-GMRT J020947.5-042213 (``groupid'' 157).
}
\label{figMoreRadiogals}
\end{figure*}

\section{Large-scale structure at {\it z} = 0.14}
\label{SectSuperclu}

In Fig.~\ref{figSuperclu} we show 
the distribution of the galaxies with a spectroscopic redshift from the GAMA database between 0.135 and 0.141 (top panel) and 
the GMRT 610~MHz image (bottom panel) of the region of XXL-North that encompasses the supercluster XLSSsC~N03 identified in \citetalias{adami2018}. 
The supercluster contains eight cluster members,  five of which are securely detected clusters in X-rays (C1 class). 
The cluster hosting the Exemplar, XLSSC~112, is a secure X-ray detection and its position is circled in red. 
There is an interesting alignment on the sky of four clusters, with a decrease in size and mass from the more massive Abell cluster (XLSSC~60) to the other clusters on the right. This large structure might lead to the formation of a massive cluster if infall and merger occur. 

The Double Irony radio galaxy is seen in the top right corner of Fig.~\ref{figSuperclu} (left panel). 
The redshift of its surrounding cluster/group is very similar to the redshifts of the members of the supercluster ($z \simeq 0.14$). 
We performed a new friends-of-friends analysis with exactly the same parameters as those used 
in the study presented in   
\citetalias{adami2018}, which used 326 clusters with $0.03 < z < 1.0$, but adding the Double Irony's cluster to the list of clusters. 
Very interestingly, {\emph{the Double Irony's cluster was identified as a new member (the ninth) of the supercluster XLSSsC~N0}}. 

We note an overdensity of galaxies at RA$\simeq 32\fdg445$, DEC$\simeq -4\fdg38$ (upper panel) where there is no reported XXL cluster. 
When this structure is added to the list of identified clusters/groups, 
{the friends-of-friends algorithm 
identifies it as a new member (the tenth) of the same supercluster.} 
{Even at this location, a radio source is found in the GMRT 610~MHz mosaic;}  
we call it ``the Unhinged'' because of its bizarre appearance 
(see Fig.~\ref{figMoreRadiogals}). 

\section{Radio galaxies in superclusters}
\label{SectDiscussionRadiogalsSuperclu}

We note that XLSSC~201, another cluster member of the supercluster, also hosts an extended radio source, shown in Fig.~\ref{figMoreRadiogals} 
and denoted the ``Eyebrow.'' 
It is remarkable that four out of ten clusters/groups in this structure contain well-developed radio galaxies. 
The Eyebrow and the Unhinged may be the superposition of radio sources at different redshifts and deserve a separate detailed study. 

More work is clearly needed to quantify the incidence of radio galaxies in clusters residing inside superclusters. 
A number of superclusters have been identified in XXL 
(21 in XXL-North and 14 in XXL-South; 
\citetalias{adami2018}).  
\cite{guglielmo2018b} \citepalias{guglielmo2018b}  
characterized the stellar populations of the galaxies in the richest XXL supercluster (at $z\simeq 0.3$) and found evidence for an active role of the environment on the star formation rates and stellar masses.   
\cite{2016A&A...592A..11K} \citepalias{2016A&A...592A..11K} 
detected a number of X-ray AGN in three XXL superclusters, but a larger study is needed to reach firm conclusions on the frequency of AGN in superclusters. 
The first supercluster detected in XXL is at $z = 0.43$  
(\citealt{2016A&A...592A...6P}; \citetalias{2016A&A...592A...6P}). 
Its galaxy and AGN populations have been studied in the optical and in the radio, but no large radio galaxies were found within the supercluster's overdensities 
(\citealt{baran16}; \citetalias{baran16}). 
The GMRT-XXL-N 610~MHz survey 
(\citetalias{smolcic2018}) 
can be used to search for radio galaxies in superclusters in XXL-North;  
XXL-South has been covered  
with the Australia Telescope Compact Array 
(\citealt{butler2018a}; \citetalias{butler2018a}, 
\citealt{butler2018b}; \citetalias{butler2018b}).  

In the hierarchical model of structure formation,  
structures grow by accretion and mergers, small structures forming first.   
Superclusters of X-ray-detected clusters/groups are large structures that may still be collapsing. 
Both the AGN of a radio galaxy and its host galaxy (usually a massive elliptical) have accreted significant amounts of material, and 
the jets and lobes trace the relatively recent (a few $10^7$ years) ejection of plasma. 
Because of their sizes, radio galaxies are easily seen and might be a powerful way to find 
new clusters/groups that have escaped detection in X-ray surveys
(e.g., \citealt{2016MNRAS.460.2376B}, from the Radio Galaxy Zoo). 

\section{Summary and conclusion}
\label{SectSummary}

We have presented an analysis of two radio galaxies at $z\sim 0.14$ identified in 
the GMRT-XXL-N 610~MHz survey of XXL-North. We have made use of the extensive multiwavelength coverage of XXL to gain a better understanding of those sources. We were able to identify their host galaxies, and in the case of the Exemplar the host cluster that had been detected in the X-ray and cataloged as an XXL cluster. 
The second and more spectacular source, the Double Irony, is a giant radio galaxy with a linear size of about 1100~kpc 
on the sky.  
We were able to show that it is hosted by the BCG of 
a lower mass cluster with member galaxies detected in the optical and listed in the GAMA database. 
Using a friends-of-friends algorithm, we were able to show that both clusters are part of the same supercluster.
We identified another overdensity of galaxies in the supercluster that is associated with a radio galaxy of peculiar shape. 
This brings the total number of members of this supercluster to ten. Four of these clusters/group host a radio galaxy. 
Anisotropies were found in the distribution of the surrounding galaxies, possibly indicating that the jets and lobes/plumes  
reside in lower density  regions.  
This shows the potential of the XXL survey to lead to the discovery of new structures, and the use of radio galaxies as tracers of large-scale structure. 

Future work on the two clusters presented here will require more sensitive, higher angular resolution radio and X-ray data. 
In particular, the JVLA (including polarization) at GHz frequencies and the GMRT below 500~MHz 
will allow a detailed modeling of the aging of the cosmic-ray electrons in the radio jets and lobes and will constrain timescales. 
In the X-ray, a deeper XMM-{\it Newton} observation of the Double Irony is necessary to detect the hot gas that is likely to be associated with  
the optically detected group; higher resolution images with {\it Chandra} may reveal the structure of the gaseous atmosphere around the radio galaxies and 
possibly shocks. 

The XXL fields contain other remarkable radio galaxies that will be analyzed following an approach similar to the one presented here. 
Special attention will be given to their potential relation to 
clusters and superclusters. 

\begin{acknowledgements}

XXL is an international project based on an XMM Very Large Programme
surveying two 25~deg$^{2}$ extragalactic fields at a depth of $\sim5\times10^{-15}$~erg~cm$^{-2}$~s$^{-1}$
in the {[}0.5--2{]}~keV band for point-like sources. The XXL website
is {\url{http://irfu.cea.fr/xxl}}. Multiband information and spectroscopic
follow-up of the X-ray sources are obtained through a number of survey
programs, summarized at {\url{http://xxlmultiwave.pbworks.com/}}.

We thank the staff of the GMRT who  made these observations possible. The GMRT is run by the National Centre for Radio Astrophysics of the Tata Institute of Fundamental Research.

This work is based on observations obtained with XMM-Newton, an ESA science mission with instruments and contributions directly funded by ESA Member States and NASA. 

It is also based on observations obtained with MegaPrime/MegaCam, a joint project of CFHT and CEA/IRFU, at the Canada-France-Hawaii Telescope (CFHT) which is operated by the National Research Council (NRC) of Canada, the Institut National des Science de l'Univers of the Centre National de la Recherche Scientifique (CNRS) of France, and the University of Hawaii. This work is based in part on data products produced at Terapix available at the Canadian Astronomy Data Centre as part of the Canada-France-Hawaii Telescope Legacy Survey, a collaborative project of NRC and CNRS.

This research has made use of the NASA/IPAC Extragalactic Database (NED) which is operated by the Jet Propulsion Laboratory, California Institute of Technology, under contract with the National Aeronautics and Space Administration.

This research has made use of the VizieR catalog access tool, CDS, Strasbourg, France. The original description of the VizieR service was published by \cite{2000A&AS..143...23O}. 

We have also made use of the table analysis software {\sc topcat} \citep{topcat}
and of the caustic mass estimation algorithm written by \cite{2012MNRAS.426.2832A}. 

This research made use of Astropy, a community-developed core Python package for Astronomy \citep{astropy}.

This research also made use of the Matplotlib plotting library \citep{matplotlib}.

GAMA is a joint European-Australasian project based around a spectroscopic campaign using the Anglo-Australian Telescope. The GAMA input catalog is based on data taken from the Sloan Digital Sky Survey and the UKIRT Infrared Deep Sky Survey. Complementary imaging of the GAMA regions is being obtained by a number of independent survey programs including GALEX MIS, VST KiDS, VISTA VIKING, WISE, Herschel-ATLAS, GMRT, and ASKAP providing UV to radio coverage. GAMA is funded by the STFC (UK), the ARC (Australia), the AAO, and the participating institutions. The GAMA website is {\url{http://www.gama-survey.org/}}.



This research has made use of the NASA/ IPAC Infrared Science Archive, which is operated by the Jet Propulsion Laboratory, California Institute of Technology, under contract with the National Aeronautics and Space Administration. 

VS acknowledges support from the European Union's Seventh Frame-work program under grant agreement 337595 (ERC Starting Grant, ``CoSMass''). 
MERC acknowledges support from the German Aerospace Agency (DLR) with funds from the Ministry of Economy and Technology (BMWi) through grant 50 OR 1514. 
The Saclay group acknowledges long-term support from the Centre National d'Etudes Spatiales (CNES). 
SF acknowledges financial support from the Swiss National Science Foundation. 
CH thanks Michael Olberg for his help with the $R$ software and John~H. Black for useful comments and moral support. 
We thank the referee for pointing out relevant references and for other constructive comments. 
\end{acknowledgements}
\bibliographystyle{aa}
\bibliography{Horellou_final_20180723.bib}

\newcommand{\noop}[1]{}
\begin{thebibliography}{101}
\expandafter\ifx\csname natexlab\endcsname\relax\def\natexlab#1{#1}\fi

\bibitem[{{Adami} \& {et al.}(2018)}]{adami2018}
{Adami}, C. \& {et al.} 2018, \aap, 999, XXL Survey, (XXL Paper XX)

\bibitem[{{Alpaslan} {et~al.}(2012){Alpaslan}, {Robotham}, {Driver}, {Norberg},
  {Peacock}, {Baldry}, {Bland-Hawthorn}, {Brough}, {Hopkins}, {Kelvin},
  {Liske}, {Loveday}, {Merson}, {Nichol}, \& {Pimbblet}}]{2012MNRAS.426.2832A}
{Alpaslan}, M., {Robotham}, A.~S.~G., {Driver}, S., {et~al.} 2012, \mnras, 426,
  2832

\bibitem[{{Astropy Collaboration} {et~al.}(2013){Astropy Collaboration},
  {Robitaille}, {Tollerud}, {Greenfield}, {Droettboom}, {Bray}, {Aldcroft},
  {Davis}, {Ginsburg}, {Price-Whelan}, {Kerzendorf}, {Conley}, {Crighton},
  {Barbary}, {Muna}, {Ferguson}, {Grollier}, {Parikh}, {Nair}, {Unther},
  {Deil}, {Woillez}, {Conseil}, {Kramer}, {Turner}, {Singer}, {Fox}, {Weaver},
  {Zabalza}, {Edwards}, {Azalee Bostroem}, {Burke}, {Casey}, {Crawford},
  {Dencheva}, {Ely}, {Jenness}, {Labrie}, {Lim}, {Pierfederici}, {Pontzen},
  {Ptak}, {Refsdal}, {Servillat}, \& {Streicher}}]{astropy}
{Astropy Collaboration}, {Robitaille}, T.~P., {Tollerud}, E.~J., {et~al.} 2013,
  \aap, 558, A33

\bibitem[{{Baldry} {et~al.}(2018){Baldry}, {Liske}, {Brown}, {Robotham},
  {Driver}, {Dunne}, {Alpaslan}, {Brough}, {Cluver}, {Eardley}, {Farrow},
  {Heymans}, {Hildebrandt}, {Hopkins}, {Kelvin}, {Loveday}, {Moffett},
  {Norberg}, {Owers}, {Taylor}, {Wright}, {Bamford}, {Bland-Hawthorn},
  {Bourne}, {Bremer}, {Colless}, {Conselice}, {Croom}, {Davies}, {Foster},
  {Grootes}, {Holwerda}, {Jones}, {Kafle}, {Kuijken}, {Lara-Lopez},
  {L{\'o}pez-S{\'a}nchez}, {Meyer}, {Phillipps}, {Sutherland}, {van Kampen}, \&
  {Wilkins}}]{2018MNRAS.474.3875B}
{Baldry}, I.~K., {Liske}, J., {Brown}, M.~J.~I., {et~al.} 2018, \mnras, 474,
  3875

\bibitem[{{Banfield} {et~al.}(2016){Banfield}, {Andernach}, {Kapi{\'n}ska},
  {Rudnick}, {Hardcastle}, {Cotter}, {Vaughan}, {Jones}, {Heywood}, {Wing},
  {Wong}, {Matorny}, {Terentev}, {L{\'o}pez-S{\'a}nchez}, {Norris}, {Seymour},
  {Shabala}, \& {Willett}}]{2016MNRAS.460.2376B}
{Banfield}, J.~K., {Andernach}, H., {Kapi{\'n}ska}, A.~D., {et~al.} 2016,
  \mnras, 460, 2376

\bibitem[{{Baran} {et~al.}(2016){Baran}, {Smol{\v c}i{\'c}}, {Milakovi{\'c}},
  {Novak}, {Delhaize}, {Gastaldello}, {Ramos-Ceja}, {Pacaud}, {Bourke},
  {Carilli}, {Ettori}, {Hallinan}, {Horellou}, {Koulouridis}, {Chiappetti},
  {Miettinen}, {Melnyk}, {Mooley}, {Pierre}, {Pompei}, \&
  {Schinnerer}}]{baran16}
{Baran}, N., {Smol{\v c}i{\'c}}, V., {Milakovi{\'c}}, D., {et~al.} 2016, \aap,
  592, A8, (XXL Paper IX)

\bibitem[{{Becker} {et~al.}(1995){Becker}, {White}, \& {Helfand}}]{first95}
{Becker}, R.~H., {White}, R.~L., \& {Helfand}, D.~J. 1995, \apj, 450, 559

\bibitem[{{Beers} {et~al.}(1990){Beers}, {Flynn}, \&
  {Gebhardt}}]{1990AJ....100...32B}
{Beers}, T.~C., {Flynn}, K., \& {Gebhardt}, K. 1990, \aj, 100, 32

\bibitem[{{Best} \& {Heckman}(2012)}]{best12}
{Best}, P.~N. \& {Heckman}, T.~M. 2012, \mnras, 421, 1569

\bibitem[{{Best} {et~al.}(2014){Best}, {Ker}, {Simpson}, {Rigby}, \&
  {Sabater}}]{best14}
{Best}, P.~N., {Ker}, L.~M., {Simpson}, C., {Rigby}, E.~E., \& {Sabater}, J.
  2014, \mnras, 445, 955

\bibitem[{{Blandford} \& {Icke}(1978)}]{1978MNRAS.185..527B}
{Blandford}, R.~D. \& {Icke}, V. 1978, \mnras, 185, 527

\bibitem[{{Bock} {et~al.}(1999){Bock}, {Large}, \&
  {Sadler}}]{1999AJ....117.1578B}
{Bock}, D.~C.-J., {Large}, M.~I., \& {Sadler}, E.~M. 1999, \aj, 117, 1578

\bibitem[{{Burn}(1966)}]{1966MNRAS.133...67B}
{Burn}, B.~J. 1966, \mnras, 133, 67

\bibitem[{{Butler} {et~al.}(2017){Butler}, {Huynh}, {Delhaize}, {Smol{\v
  c}i{\'c}}, {Kapi{\'n}ska}, {Milakovi{\'c}}, {Novak}, {Baran}, {O'Brien},
  {Chiappetti}, {Desai}, {Fotopoulou}, {Horellou}, {Lidman}, \&
  {Pierre}}]{butler2018a}
{Butler}, A., {Huynh}, M., {Delhaize}, J., {et~al.} 2017, ArXiv e-prints, (XXL
  Paper XVIII)

\bibitem[{{Butler} {et~al.}(2018){Butler}, {Huynh}, {Delvecchio}, {Kapinska},
  {Ciliegi}, {Jurlin}, {Delhaize}, {Smolcic}, {Desai}, {Fotopoulou}, {Lidman},
  {Pierre}, \& {Plionis}}]{butler2018b}
{Butler}, A., {Huynh}, M., {Delvecchio}, I., {et~al.} 2018, ArXiv e-prints,
  (XXL Paper XXXI)

\bibitem[{{Chabrier}(2003)}]{2003PASP..115..763C}
{Chabrier}, G. 2003, \pasp, 115, 763

\bibitem[{{Chiappetti} \& {et al.}(2018)}]{lucio2018}
{Chiappetti}, L. \& {et al.} 2018, \aap, 999, XXL Survey, (XXL Paper XXVII)

\bibitem[{{Clarke} {et~al.}(2017){Clarke}, {Heald}, {Jarrett}, {Bray},
  {Hardcastle}, {Cantwell}, {Scaife}, {Brienza}, {Bonafede}, {Breton},
  {Broderick}, {Carbone}, {Croston}, {Farnes}, {Harwood}, {Heesen},
  {Horneffer}, {van der Horst}, {Iacobelli}, {Jurusik}, {Kokotanekov},
  {McKean}, {Morabito}, {Mulcahy}, {Nikiel-Wroczy{\~n}ski}, {Orr{\'u}},
  {Paladino}, {Pandey-Pommier}, {Pietka}, {Pizzo}, {Pratley}, {Riseley},
  {Rottgering}, {Rowlinson}, {Sabater}, {Sendlinger}, {Shulevski}, {Sridhar},
  {Stewart}, {Tasse}, {van Velzen}, {van Weeren}, \&
  {Wise}}]{2017A&A...601A..25C}
{Clarke}, A.~O., {Heald}, G., {Jarrett}, T., {et~al.} 2017, \aap, 601, A25

\bibitem[{{Condon} {et~al.}(1998){Condon}, {Cotton}, {Greisen}, {Yin},
  {Perley}, {Taylor}, \& {Broderick}}]{NVSS}
{Condon}, J.~J., {Cotton}, W.~D., {Greisen}, E.~W., {et~al.} 1998, \aj, 115,
  1693

\bibitem[{{Croton} {et~al.}(2006){Croton}, {Springel}, {White}, {De Lucia},
  {Frenk}, {Gao}, {Jenkins}, {Kauffmann}, {Navarro}, \& {Yoshida}}]{croton06}
{Croton}, D.~J., {Springel}, V., {White}, S.~D.~M., {et~al.} 2006, \mnras, 365,
  11

\bibitem[{{Dabhade} {et~al.}(2017){Dabhade}, {Gaikwad}, {Bagchi},
  {Pandey-Pommier}, {Sankhyayan}, \& {Raychaudhury}}]{2017MNRAS.469.2886D}
{Dabhade}, P., {Gaikwad}, M., {Bagchi}, J., {et~al.} 2017, \mnras, 469, 2886

\bibitem[{{Diaferio}(1999)}]{1999MNRAS.309..610D}
{Diaferio}, A. 1999, \mnras, 309, 610

\bibitem[{{Diaferio} \& {Geller}(1997)}]{1997ApJ...481..633D}
{Diaferio}, A. \& {Geller}, M.~J. 1997, \apj, 481, 633

\bibitem[{{Driver} {et~al.}(2011){Driver}, {Hill}, {Kelvin}, {Robotham},
  {Liske}, {Norberg}, {Baldry}, {Bamford}, {Hopkins}, {Loveday}, {Peacock},
  {Andrae}, {Bland-Hawthorn}, {Brough}, {Brown}, {Cameron}, {Ching}, {Colless},
  {Conselice}, {Croom}, {Cross}, {de Propris}, {Dye}, {Drinkwater}, {Ellis},
  {Graham}, {Grootes}, {Gunawardhana}, {Jones}, {van Kampen}, {Maraston},
  {Nichol}, {Parkinson}, {Phillipps}, {Pimbblet}, {Popescu}, {Prescott},
  {Roseboom}, {Sadler}, {Sansom}, {Sharp}, {Smith}, {Taylor}, {Thomas},
  {Tuffs}, {Wijesinghe}, {Dunne}, {Frenk}, {Jarvis}, {Madore}, {Meyer},
  {Seibert}, {Staveley-Smith}, {Sutherland}, \& {Warren}}]{2011MNRAS.413..971D}
{Driver}, S.~P., {Hill}, D.~T., {Kelvin}, L.~S., {et~al.} 2011, \mnras, 413,
  971

\bibitem[{{Duah Asabere} {et~al.}(2016){Duah Asabere}, {Horellou}, {Jarrett},
  \& {Winkler}}]{2016A&A...592A..20D}
{Duah Asabere}, B., {Horellou}, C., {Jarrett}, T.~H., \& {Winkler}, H. 2016,
  \aap, 592, A20

\bibitem[{{Eckert} {et~al.}(2016){Eckert}, {Ettori}, {Coupon}, {Gastaldello},
  {Pierre}, {Melin}, {Le Brun}, {McCarthy}, {Adami}, {Chiappetti}, {Faccioli},
  {Giles}, {Lavoie}, {Lef{\`e}vre}, {Lieu}, {Mantz}, {Maughan}, {McGee},
  {Pacaud}, {Paltani}, {Sadibekova}, {Smith}, \&
  {Ziparo}}]{2016A&A...592A..12E}
{Eckert}, D., {Ettori}, S., {Coupon}, J., {et~al.} 2016, \aap, 592, A12, (XXL
  Paper XIII)

\bibitem[{{Ekers} {et~al.}(1978){Ekers}, {Fanti}, {Lari}, \&
  {Parma}}]{1978Natur.276..588E}
{Ekers}, R.~D., {Fanti}, R., {Lari}, C., \& {Parma}, P. 1978, \nat, 276, 588

\bibitem[{{Ettori} \& {Balestra}(2009)}]{2009A&A...496..343E}
{Ettori}, S. \& {Balestra}, I. 2009, \aap, 496, 343

\bibitem[{{Evrard}(2004)}]{2004cgpc.symp....1E}
{Evrard}, A.~E. 2004, Clusters of Galaxies: Probes of Cosmological Structure
  and Galaxy Evolution, 1

\bibitem[{{Fanaroff} \& {Riley}(1974)}]{fanaroff74}
{Fanaroff}, B.~L. \& {Riley}, J.~M. 1974, \mnras, 167, 31P

\bibitem[{{Ferrarese} \& {Merritt}(2000)}]{2000ApJ...539L...9F}
{Ferrarese}, L. \& {Merritt}, D. 2000, \apjl, 539, L9

\bibitem[{{Fotopoulou} {et~al.}(2016){Fotopoulou}, {Pacaud}, {Paltani},
  {Ranalli}, {Ramos-Ceja}, {Faccioli}, {Plionis}, {Adami}, {Bongiorno},
  {Brusa}, {Chiappetti}, {Desai}, {Elyiv}, {Lidman}, {Melnyk}, {Pierre},
  {Piconcelli}, {Vignali}, {Alis}, {Ardila}, {Arnouts}, {Baldry}, {Bremer},
  {Eckert}, {Guennou}, {Horellou}, {Iovino}, {Koulouridis}, {Liske},
  {Maurogordato}, {Menanteau}, {Mohr}, {Owers}, {Poggianti}, {Pompei},
  {Sadibekova}, {Stanford}, {Tuffs}, \& {Willis}}]{2016A&A...592A...5F}
{Fotopoulou}, S., {Pacaud}, F., {Paltani}, S., {et~al.} 2016, \aap, 592, A5,
  (XXL Paper VI)

\bibitem[{{Fritz} {et~al.}(2007){Fritz}, {Poggianti}, {Bettoni}, {Cava},
  {Couch}, {D'Onofrio}, {Dressler}, {Fasano}, {Kj{\ae}rgaard}, {Moles}, \&
  {Varela}}]{2007A&A...470..137F}
{Fritz}, J., {Poggianti}, B.~M., {Bettoni}, D., {et~al.} 2007, \aap, 470, 137

\bibitem[{{Garrington} {et~al.}(1988){Garrington}, {Leahy}, {Conway}, \&
  {Laing}}]{1988Natur.331..147G}
{Garrington}, S.~T., {Leahy}, J.~P., {Conway}, R.~G., \& {Laing}, R.~A. 1988,
  \nat, 331, 147

\bibitem[{{Giles} {et~al.}(2016){Giles}, {Maughan}, {Pacaud}, {Lieu}, {Clerc},
  {Pierre}, {Adami}, {Chiappetti}, {D{\'e}mocl{\'e}s}, {Ettori}, {Le
  F{\'e}vre}, {Ponman}, {Sadibekova}, {Smith}, {Willis}, \&
  {Ziparo}}]{2016A&A...592A...3G}
{Giles}, P.~A., {Maughan}, B.~J., {Pacaud}, F., {et~al.} 2016, \aap, 592, A3,
  (XXL Paper III)

\bibitem[{{Guglielmo} {et~al.}(2017){Guglielmo}, {Poggianti}, {Vulcani},
  {Adami}, {Gastaldello}, {Ettori}, {Fotoupoulou}, {Koulouridis}, {Ramos Ceja},
  {Giles}, {McGee}, {Altieri}, {Baldry}, {Birkinshaw}, {Bolzonella},
  {Bongiorno}, {Brown}, {Chiappetti}, {Driver}, {Elyiv}, {Evrard}, {Garilli},
  {Grootes}, {Guennou}, {Hopkins}, {Horellou}, {Iovino}, {Lidman}, {Liske},
  {Maurogordato}, {Owers}, {Pacaud}, {Paltani}, {Pierre}, {Plionis}, {Ponman},
  {Robotham}, {Sadibekova}, {Scodeggio}, {Sereno}, {Smolvcic}, {Tuffs},
  {Valtchanov}, {Vignali}, \& {Willis}}]{guglielmo2018a}
{Guglielmo}, V., {Poggianti}, B.~M., {Vulcani}, B., {et~al.} 2017, ArXiv
  e-prints, (XXL Paper XXII)

\bibitem[{{Guglielmo} {et~al.}(2018){Guglielmo}, {Poggianti}, {Vulcani},
  {Moretti}, {Fritz}, {Gastaldello}, {Adami}, {Caretta}, {Willis},
  {Koulouridis}, {Ramos Ceja}, {Giles}, {Baldry}, {Birkinshaw}, {Bongiorno},
  {Brown}, {Chiappetti}, {Driver}, {Elyiv}, {Evrard}, {Grootes}, {Guennou},
  {Hopkins}, {Horellou}, {Iovino}, {Maurogordato}, {Owers}, {Pacaud},
  {Paltani}, {Pierre}, {Plionis}, {Ponman}, {Robotham}, {Sadibekova}, {Smol{\v
  c}i{\'c}}, {Tuffs}, \& {Vignali}}]{guglielmo2018b}
{Guglielmo}, V., {Poggianti}, B.~M., {Vulcani}, B., {et~al.} 2018, ArXiv
  e-prints, (XXL Paper XXX)

\bibitem[{{Guo} {et~al.}(2011){Guo}, {White}, {Boylan-Kolchin}, {De Lucia},
  {Kauffmann}, {Lemson}, {Li}, {Springel}, \& {Weinmann}}]{2011MNRAS.413..101G}
{Guo}, Q., {White}, S., {Boylan-Kolchin}, M., {et~al.} 2011, \mnras, 413, 101

\bibitem[{{Gwyn}(2012)}]{2012AJ....143...38G}
{Gwyn}, S.~D.~J. 2012, \aj, 143, 38

\bibitem[{{Hardcastle} {et~al.}(2007){Hardcastle}, {Evans}, \&
  {Croston}}]{hardcastle07}
{Hardcastle}, M., {Evans}, D., \& {Croston}, J. 2007, \mnras, 376, 1849

\bibitem[{{Hardcastle} {et~al.}(2016){Hardcastle}, {G{\"u}rkan}, {van Weeren},
  {Williams}, {Best}, {de Gasperin}, {Rafferty}, {Read}, {Sabater}, {Shimwell},
  {Smith}, {Tasse}, {Bourne}, {Brienza}, {Br{\"u}ggen}, {Brunetti},
  {Chy{\.z}y}, {Conway}, {Dunne}, {Eales}, {Maddox}, {Jarvis}, {Mahony},
  {Morganti}, {Prandoni}, {R{\"o}ttgering}, {Valiante}, \&
  {White}}]{2016MNRAS.462.1910H}
{Hardcastle}, M.~J., {G{\"u}rkan}, G., {van Weeren}, R.~J., {et~al.} 2016,
  \mnras, 462, 1910

\bibitem[{{Heald} {et~al.}(2015){Heald}, {Pizzo}, {Orr{\'u}}, {Breton},
  {Carbone}, {Ferrari}, {Hardcastle}, {Jurusik}, {Macario}, {Mulcahy},
  {Rafferty}, {Asgekar}, {Brentjens}, {Fallows}, {Frieswijk}, {Toribio},
  {Adebahr}, {Arts}, {Bell}, {Bonafede}, {Bray}, {Broderick}, {Cantwell},
  {Carroll}, {Cendes}, {Clarke}, {Croston}, {Daiboo}, {de Gasperin}, {Gregson},
  {Harwood}, {Hassall}, {Heesen}, {Horneffer}, {van der Horst}, {Iacobelli},
  {Jeli{\'c}}, {Jones}, {Kant}, {Kokotanekov}, {Martin}, {McKean}, {Morabito},
  {Nikiel-Wroczy{\'n}ski}, {Offringa}, {Pandey}, {Pandey-Pommier}, {Pietka},
  {Pratley}, {Riseley}, {Rowlinson}, {Sabater}, {Scaife}, {Scheers},
  {Sendlinger}, {Shulevski}, {Sipior}, {Sobey}, {Stewart}, {Stroe}, {Swinbank},
  {Tasse}, {Tr{\"u}stedt}, {Varenius}, {van Velzen}, {Vilchez}, {van Weeren},
  {Wijnholds}, {Williams}, {de Bruyn}, {Nijboer}, {Wise}, {Alexov}, {Anderson},
  {Avruch}, {Beck}, {Bell}, {van Bemmel}, {Bentum}, {Bernardi}, {Best},
  {Breitling}, {Brouw}, {Br{\"u}ggen}, {Butcher}, {Ciardi}, {Conway}, {de
  Geus}, {de Jong}, {de Vos}, {Deller}, {Dettmar}, {Duscha}, {Eisl{\"o}ffel},
  {Engels}, {Falcke}, {Fender}, {Garrett}, {Grie{\ss}meier}, {Gunst},
  {Hamaker}, {Hessels}, {Hoeft}, {H{\"o}randel}, {Holties}, {Intema},
  {Jackson}, {J{\"u}tte}, {Karastergiou}, {Klijn}, {Kondratiev}, {Koopmans},
  {Kuniyoshi}, {Kuper}, {Law}, {van Leeuwen}, {Loose}, {Maat}, {Markoff},
  {McFadden}, {McKay-Bukowski}, {Mevius}, {Miller-Jones}, {Morganti}, {Munk},
  {Nelles}, {Noordam}, {Norden}, {Paas}, {Polatidis}, {Reich}, {Renting},
  {R{\"o}ttgering}, {Schoenmakers}, {Schwarz}, {Sluman}, {Smirnov}, {Stappers},
  {Steinmetz}, {Tagger}, {Tang}, {ter Veen}, {Thoudam}, {Vermeulen}, {Vocks},
  {Vogt}, {Wijers}, {Wucknitz}, {Yatawatta}, \& {Zarka}}]{2015A&A...582A.123H}
{Heald}, G.~H., {Pizzo}, R.~F., {Orr{\'u}}, E., {et~al.} 2015, \aap, 582, A123

\bibitem[{{Helfand} {et~al.}(2015){Helfand}, {White}, \& {Becker}}]{first15}
{Helfand}, D.~J., {White}, R.~L., \& {Becker}, R.~H. 2015, \apj, 801, 26

\bibitem[{{Hinshaw} {et~al.}(2013){Hinshaw}, {Larson}, {Komatsu}, {Spergel},
  {Bennett}, {Dunkley}, {Nolta}, {Halpern}, {Hill}, {Odegard}, {Page}, {Smith},
  {Weiland}, {Gold}, {Jarosik}, {Kogut}, {Limon}, {Meyer}, {Tucker}, {Wollack},
  \& {Wright}}]{2013ApJS..208...19H}
{Hinshaw}, G., {Larson}, D., {Komatsu}, E., {et~al.} 2013, \apjs, 208, 19

\bibitem[{{Hodges-Kluck} \& {Reynolds}(2012)}]{2012ApJ...746..167H}
{Hodges-Kluck}, E.~J. \& {Reynolds}, C.~S. 2012, \apj, 746, 167

\bibitem[{Hunter(2007)}]{matplotlib}
Hunter, J.~D. 2007, Computing In Science \& Engineering, 9, 90

\bibitem[{{Intema} {et~al.}(2017){Intema}, {Jagannathan}, {Mooley}, \&
  {Frail}}]{TGSS_ADR1}
{Intema}, H.~T., {Jagannathan}, P., {Mooley}, K.~P., \& {Frail}, D.~A. 2017,
  \aap, 598, A78

\bibitem[{{Ishwara-Chandra} \& {Saikia}(1999)}]{1999MNRAS.309..100I}
{Ishwara-Chandra}, C.~H. \& {Saikia}, D.~J. 1999, \mnras, 309, 100

\bibitem[{{Jamrozy} {et~al.}(2008){Jamrozy}, {Konar}, {Machalski}, \&
  {Saikia}}]{2008MNRAS.385.1286J}
{Jamrozy}, M., {Konar}, C., {Machalski}, J., \& {Saikia}, D.~J. 2008, \mnras,
  385, 1286

\bibitem[{{Kolokythas} {et~al.}(2015){Kolokythas}, {O'Sullivan}, {Giacintucci},
  {Raychaudhury}, {Ishwara-Chandra}, {Worrall}, \&
  {Birkinshaw}}]{2015MNRAS.450.1732K}
{Kolokythas}, K., {O'Sullivan}, E., {Giacintucci}, S., {et~al.} 2015, \mnras,
  450, 1732

\bibitem[{{Koulouridis} {et~al.}(2016){Koulouridis}, {Poggianti}, {Altieri},
  {Valtchanov}, {Jaff{\'e}}, {Adami}, {Elyiv}, {Melnyk}, {Fotopoulou},
  {Gastaldello}, {Horellou}, {Pierre}, {Pacaud}, {Plionis}, {Sadibekova}, \&
  {Surdej}}]{2016A&A...592A..11K}
{Koulouridis}, E., {Poggianti}, B., {Altieri}, B., {et~al.} 2016, \aap, 592,
  A11, (XXL Paper XII)

\bibitem[{{Ku{\'z}micz} \& {Jamrozy}(2012)}]{2012MNRAS.426..851K}
{Ku{\'z}micz}, A. \& {Jamrozy}, M. 2012, \mnras, 426, 851

\bibitem[{{Laing}(1988)}]{1988Natur.331..149L}
{Laing}, R.~A. 1988, \nat, 331, 149

\bibitem[{{Laing} {et~al.}(2006){Laing}, {Canvin}, {Bridle}, \&
  {Hardcastle}}]{2006MNRAS.372..510L}
{Laing}, R.~A., {Canvin}, J.~R., {Bridle}, A.~H., \& {Hardcastle}, M.~J. 2006,
  \mnras, 372, 510

\bibitem[{{Laing} {et~al.}(2011){Laing}, {Guidetti}, {Bridle}, {Parma}, \&
  {Bondi}}]{2011MNRAS.417.2789L}
{Laing}, R.~A., {Guidetti}, D., {Bridle}, A.~H., {Parma}, P., \& {Bondi}, M.
  2011, \mnras, 417, 2789

\bibitem[{{Lara} {et~al.}(2001{\natexlab{a}}){Lara}, {Cotton}, {Feretti},
  {Giovannini}, {Marcaide}, {M{\'a}rquez}, \& {Venturi}}]{2001A&A...370..409L}
{Lara}, L., {Cotton}, W.~D., {Feretti}, L., {et~al.} 2001{\natexlab{a}}, \aap,
  370, 409

\bibitem[{{Lara} {et~al.}(2001{\natexlab{b}}){Lara}, {M{\'a}rquez}, {Cotton},
  {Feretti}, {Giovannini}, {Marcaide}, \& {Venturi}}]{2001A&A...378..826L}
{Lara}, L., {M{\'a}rquez}, I., {Cotton}, W.~D., {et~al.} 2001{\natexlab{b}},
  \aap, 378, 826

\bibitem[{{Lavoie} {et~al.}(2016){Lavoie}, {Willis}, {D{\'e}mocl{\`e}s},
  {Eckert}, {Gastaldello}, {Smith}, {Lidman}, {Adami}, {Pacaud}, {Pierre},
  {Clerc}, {Giles}, {Lieu}, {Chiappetti}, {Altieri}, {Ardila}, {Baldry},
  {Bongiorno}, {Desai}, {Elyiv}, {Faccioli}, {Gardner}, {Garilli}, {Groote},
  {Guennou}, {Guzzo}, {Hopkins}, {Liske}, {McGee}, {Melnyk}, {Owers},
  {Poggianti}, {Ponman}, {Scodeggio}, {Spitler}, \&
  {Tuffs}}]{2016MNRAS.462.4141L}
{Lavoie}, S., {Willis}, J.~P., {D{\'e}mocl{\`e}s}, J., {et~al.} 2016, \mnras,
  462, 4141, (XXL Paper XV)

\bibitem[{{Lieu} {et~al.}(2016){Lieu}, {Smith}, {Giles}, {Ziparo}, {Maughan},
  {D{\'e}mocl{\`e}s}, {Pacaud}, {Pierre}, {Adami}, {Bah{\'e}}, {Clerc},
  {Chiappetti}, {Eckert}, {Ettori}, {Lavoie}, {Le Fevre}, {McCarthy},
  {Kilbinger}, {Ponman}, {Sadibekova}, \& {Willis}}]{2016A&A...592A...4L}
{Lieu}, M., {Smith}, G.~P., {Giles}, P.~A., {et~al.} 2016, \aap, 592, A4, (XXL
  Paper IV)

\bibitem[{{Machalski} \& {Jamrozy}(2006)}]{2006A&A...454...95M}
{Machalski}, J. \& {Jamrozy}, M. 2006, \aap, 454, 95

\bibitem[{{Machalski} {et~al.}(2001){Machalski}, {Jamrozy}, \&
  {Zola}}]{2001A&A...371..445M}
{Machalski}, J., {Jamrozy}, M., \& {Zola}, S. 2001, \aap, 371, 445

\bibitem[{{Machalski} {et~al.}(2007){Machalski}, {Koziel-Wierzbowska}, \&
  {Jamrozy}}]{2007AcA....57..227M}
{Machalski}, J., {Koziel-Wierzbowska}, D., \& {Jamrozy}, M. 2007, \actaa, 57,
  227

\bibitem[{{Malarecki} {et~al.}(2015){Malarecki}, {Jones}, {Saripalli},
  {Staveley-Smith}, \& {Subrahmanyan}}]{2015MNRAS.449..955M}
{Malarecki}, J.~M., {Jones}, D.~H., {Saripalli}, L., {Staveley-Smith}, L., \&
  {Subrahmanyan}, R. 2015, \mnras, 449, 955

\bibitem[{{McConnell} \& {Ma}(2013)}]{2013ApJ...764..184M}
{McConnell}, N.~J. \& {Ma}, C.-P. 2013, \apj, 764, 184

\bibitem[{{McNamara} {et~al.}(2005){McNamara}, {Nulsen}, {Wise}, {Rafferty},
  {Carilli}, {Sarazin}, \& {Blanton}}]{2005Natur.433...45M}
{McNamara}, B.~R., {Nulsen}, P.~E.~J., {Wise}, M.~W., {et~al.} 2005, \nat, 433,
  45

\bibitem[{{Ochsenbein} {et~al.}(2000){Ochsenbein}, {Bauer}, \&
  {Marcout}}]{2000A&AS..143...23O}
{Ochsenbein}, F., {Bauer}, P., \& {Marcout}, J. 2000, \aaps, 143, 23

\bibitem[{{Pacaud} {et~al.}(2016){Pacaud}, {Clerc}, {Giles}, {Adami},
  {Sadibekova}, {Pierre}, {Maughan}, {Lieu}, {Le F{\`e}vre}, {Alis}, {Altieri},
  {Ardila}, {Baldry}, {Benoist}, {Birkinshaw}, {Chiappetti},
  {D{\'e}mocl{\`e}s}, {Eckert}, {Evrard}, {Faccioli}, {Gastaldello}, {Guennou},
  {Horellou}, {Iovino}, {Koulouridis}, {Le Brun}, {Lidman}, {Liske},
  {Maurogordato}, {Menanteau}, {Owers}, {Poggianti}, {Pomar{\`e}de}, {Pompei},
  {Ponman}, {Rapetti}, {Reiprich}, {Smith}, {Tuffs}, {Valageas}, {Valtchanov},
  {Willis}, \& {Ziparo}}]{2016A&A...592A...2P}
{Pacaud}, F., {Clerc}, N., {Giles}, P.~A., {et~al.} 2016, \aap, 592, A2, (XXL
  Paper II)

\bibitem[{{Pierre} {et~al.}(2016){Pierre}, {Pacaud}, {Adami}, {Alis},
  {Altieri}, {Baran}, {Benoist}, {Birkinshaw}, {Bongiorno}, {Bremer}, {Brusa},
  {Butler}, {Ciliegi}, {Chiappetti}, {Clerc}, {Corasaniti}, {Coupon}, {De
  Breuck}, {Democles}, {Desai}, {Delhaize}, {Devriendt}, {Dubois}, {Eckert},
  {Elyiv}, {Ettori}, {Evrard}, {Faccioli}, {Farahi}, {Ferrari}, {Finet},
  {Fotopoulou}, {Fourmanoit}, {Gandhi}, {Gastaldello}, {Gastaud},
  {Georgantopoulos}, {Giles}, {Guennou}, {Guglielmo}, {Horellou}, {Husband},
  {Huynh}, {Iovino}, {Kilbinger}, {Koulouridis}, {Lavoie}, {Le Brun}, {Le
  Fevre}, {Lidman}, {Lieu}, {Lin}, {Mantz}, {Maughan}, {Maurogordato},
  {McCarthy}, {McGee}, {Melin}, {Melnyk}, {Menanteau}, {Novak}, {Paltani},
  {Plionis}, {Poggianti}, {Pomarede}, {Pompei}, {Ponman}, {Ramos-Ceja},
  {Ranalli}, {Rapetti}, {Raychaudury}, {Reiprich}, {Rottgering}, {Rozo},
  {Rykoff}, {Sadibekova}, {Santos}, {Sauvageot}, {Schimd}, {Sereno}, {Smith},
  {Smol{\v c}i{\'c}}, {Snowden}, {Spergel}, {Stanford}, {Surdej}, {Valageas},
  {Valotti}, {Valtchanov}, {Vignali}, {Willis}, \& {Ziparo}}]{pierre16}
{Pierre}, M., {Pacaud}, F., {Adami}, C., {et~al.} 2016, \aap, 592, A1, (XXL
  Paper I)

\bibitem[{{Pompei} {et~al.}(2016){Pompei}, {Adami}, {Eckert}, {Gastaldello},
  {Lavoie}, {Poggianti}, {Altieri}, {Alis}, {Baran}, {Benoist}, {Jaff{\'e}},
  {Koulouridis}, {Maurogordato}, {Pacaud}, {Pierre}, {Sadibekova}, {Smol{\v
  c}i{\'c}}, \& {Valtchanov}}]{2016A&A...592A...6P}
{Pompei}, E., {Adami}, C., {Eckert}, D., {et~al.} 2016, \aap, 592, A6

\bibitem[{{Proctor}(2016)}]{2016ApJS..224...18P}
{Proctor}, D.~D. 2016, \apjs, 224, 18

\bibitem[{{Ranalli} {et~al.}(2003){Ranalli}, {Comastri}, \&
  {Setti}}]{2003A&A...399...39R}
{Ranalli}, P., {Comastri}, A., \& {Setti}, G. 2003, \aap, 399, 39

\bibitem[{{Raouf} {et~al.}(2014){Raouf}, {Khosroshahi}, {Ponman}, {Dariush},
  {Molaeinezhad}, \& {Tavasoli}}]{2014MNRAS.442.1578R}
{Raouf}, M., {Khosroshahi}, H.~G., {Ponman}, T.~J., {et~al.} 2014, \mnras, 442,
  1578

\bibitem[{{Rengelink} {et~al.}(1997){Rengelink}, {Tang}, {de Bruyn}, {Miley},
  {Bremer}, {Roettgering}, \& {Bremer}}]{1997A&AS..124..259R}
{Rengelink}, R.~B., {Tang}, Y., {de Bruyn}, A.~G., {et~al.} 1997, \aaps, 124,
  259

\bibitem[{{Robotham} {et~al.}(2011){Robotham}, {Norberg}, {Driver}, {Baldry},
  {Bamford}, {Hopkins}, {Liske}, {Loveday}, {Merson}, {Peacock}, {Brough},
  {Cameron}, {Conselice}, {Croom}, {Frenk}, {Gunawardhana}, {Hill}, {Jones},
  {Kelvin}, {Kuijken}, {Nichol}, {Parkinson}, {Pimbblet}, {Phillipps},
  {Popescu}, {Prescott}, {Sharp}, {Sutherland}, {Taylor}, {Thomas}, {Tuffs},
  {van Kampen}, \& {Wijesinghe}}]{2011MNRAS.416.2640R}
{Robotham}, A.~S.~G., {Norberg}, P., {Driver}, S.~P., {et~al.} 2011, \mnras,
  416, 2640

\bibitem[{{Sadler} {et~al.}(2014){Sadler}, {Ekers}, {Mahony}, {Mauch}, \&
  {Murphy}}]{2014MNRAS.438..796S}
{Sadler}, E.~M., {Ekers}, R.~D., {Mahony}, E.~K., {Mauch}, T., \& {Murphy}, T.
  2014, \mnras, 438, 796

\bibitem[{{Safouris} {et~al.}(2009){Safouris}, {Subrahmanyan}, {Bicknell}, \&
  {Saripalli}}]{2009MNRAS.393....2S}
{Safouris}, V., {Subrahmanyan}, R., {Bicknell}, G.~V., \& {Saripalli}, L. 2009,
  \mnras, 393, 2

\bibitem[{{Santiago-Bautista} {et~al.}(2016){Santiago-Bautista},
  {Rodr{\'{\i}}guez-Rico}, {Andernach}, {Coziol}, {Torres-Papaqui},
  {Jim{\'e}nez Andrade}, {Plauchu-Frayn}, \& {Momjian}}]{2016ASSP...42..231S}
{Santiago-Bautista}, I.~d.~C., {Rodr{\'{\i}}guez-Rico}, C.~A., {Andernach}, H.,
  {et~al.} 2016, The Universe of Digital Sky Surveys, 42, 231

\bibitem[{{Saripalli} {et~al.}(1986){Saripalli}, {Gopal-Krishna}, {Reich}, \&
  {Kuehr}}]{1986A&A...170...20S}
{Saripalli}, L., {Gopal-Krishna}, {Reich}, W., \& {Kuehr}, H. 1986, \aap, 170,
  20

\bibitem[{{Saripalli} {et~al.}(2005){Saripalli}, {Hunstead}, {Subrahmanyan}, \&
  {Boyce}}]{2005AJ....130..896S}
{Saripalli}, L., {Hunstead}, R.~W., {Subrahmanyan}, R., \& {Boyce}, E. 2005,
  \aj, 130, 896

\bibitem[{{Saripalli} {et~al.}(2002){Saripalli}, {Subrahmanyan}, \& {Udaya
  Shankar}}]{2002ApJ...565..256S}
{Saripalli}, L., {Subrahmanyan}, R., \& {Udaya Shankar}, N. 2002, \apj, 565,
  256

\bibitem[{{Schoenmakers} {et~al.}(2001){Schoenmakers}, {de Bruyn},
  {R{\"o}ttgering}, \& {van der Laan}}]{2001A&A...374..861S}
{Schoenmakers}, A.~P., {de Bruyn}, A.~G., {R{\"o}ttgering}, H.~J.~A., \& {van
  der Laan}, H. 2001, \aap, 374, 861

\bibitem[{{Serra} \& {Diaferio}(2013)}]{2013ApJ...768..116S}
{Serra}, A.~L. \& {Diaferio}, A. 2013, \apj, 768, 116

\bibitem[{{Shimwell} {et~al.}(2017){Shimwell}, {R{\"o}ttgering}, {Best},
  {Williams}, {Dijkema}, {de Gasperin}, {Hardcastle}, {Heald}, {Hoang},
  {Horneffer}, {Intema}, {Mahony}, {Mandal}, {Mechev}, {Morabito}, {Oonk},
  {Rafferty}, {Retana-Montenegro}, {Sabater}, {Tasse}, {van Weeren},
  {Br{\"u}ggen}, {Brunetti}, {Chy{\.z}y}, {Conway}, {Haverkorn}, {Jackson},
  {Jarvis}, {McKean}, {Miley}, {Morganti}, {White}, {Wise}, {van Bemmel},
  {Beck}, {Brienza}, {Bonafede}, {Calistro Rivera}, {Cassano}, {Clarke},
  {Cseh}, {Deller}, {Drabent}, {van Driel}, {Engels}, {Falcke}, {Ferrari},
  {Fr{\"o}hlich}, {Garrett}, {Harwood}, {Heesen}, {Hoeft}, {Horellou},
  {Israel}, {Kapi{\'n}ska}, {Kunert-Bajraszewska}, {McKay}, {Mohan},
  {Orr{\'u}}, {Pizzo}, {Prandoni}, {Schwarz}, {Shulevski}, {Sipior}, {Smith},
  {Sridhar}, {Steinmetz}, {Stroe}, {Varenius}, {van der Werf}, {Zensus}, \&
  {Zwart}}]{2017A&A...598A.104S}
{Shimwell}, T.~W., {R{\"o}ttgering}, H.~J.~A., {Best}, P.~N., {et~al.} 2017,
  \aap, 598, A104

\bibitem[{{Smolcic} {et~al.}(2018){Smolcic}, {Intema}, {Slaus}, {Raychaudhury},
  {Novak}, {Horellou}, {Chiappetti}, {Delhaize}, {Birkinshaw}, {Bondi},
  {Bremer}, {Ciliegi}, {Ferrari}, {Kolokythas}, {Lidman}, {McGee}, {Norris},
  {Pierre}, {Rottgering}, {Tasse}, \& {Williams}}]{smolcic2018}
{Smolcic}, V., {Intema}, H., {Slaus}, B., {et~al.} 2018, ArXiv e-prints, (XXL
  Paper XXIX)

\bibitem[{{Smol{\v c}i{\'c}}(2009)}]{smolcic09}
{Smol{\v c}i{\'c}}, V. 2009, \apjl, 699, L43

\bibitem[{{Subrahmanyan} {et~al.}(1996){Subrahmanyan}, {Saripalli}, \&
  {Hunstead}}]{1996MNRAS.279..257S}
{Subrahmanyan}, R., {Saripalli}, L., \& {Hunstead}, R.~W. 1996, \mnras, 279,
  257

\bibitem[{{Subrahmanyan} {et~al.}(2008){Subrahmanyan}, {Saripalli}, {Safouris},
  \& {Hunstead}}]{2008ApJ...677...63S}
{Subrahmanyan}, R., {Saripalli}, L., {Safouris}, V., \& {Hunstead}, R.~W. 2008,
  \apj, 677, 63

\bibitem[{{Tasse} {et~al.}(2008){Tasse}, {Best}, {R{\"o}ttgering}, \& {Le
  Borgne}}]{2008A&A...490..893T}
{Tasse}, C., {Best}, P.~N., {R{\"o}ttgering}, H., \& {Le Borgne}, D. 2008,
  \aap, 490, 893

\bibitem[{{Tasse} {et~al.}(2006){Tasse}, {Cohen}, {R{\"o}ttgering}, {Kassim},
  {Pierre}, {Perley}, {Best}, {Birkinshaw}, {Bremer}, \& {Liang}}]{tasse06}
{Tasse}, C., {Cohen}, A.~S., {R{\"o}ttgering}, H.~J.~A., {et~al.} 2006, \aap,
  456, 791

\bibitem[{{Tasse} {et~al.}(2007){Tasse}, {R{\"o}ttgering}, {Best}, {Cohen},
  {Pierre}, \& {Wilman}}]{tasse07}
{Tasse}, C., {R{\"o}ttgering}, H.~J.~A., {Best}, P.~N., {et~al.} 2007, \aap,
  471, 1105

\bibitem[{{Taylor} {et~al.}(2009){Taylor}, {Stil}, \&
  {Sunstrum}}]{2009ApJ...702.1230T}
{Taylor}, A.~R., {Stil}, J.~M., \& {Sunstrum}, C. 2009, \apj, 702, 1230

\bibitem[{{Taylor}(2005)}]{topcat}
{Taylor}, M.~B. 2005, in Astronomical Society of the Pacific Conference Series,
  Vol. 347, Astronomical Data Analysis Software and Systems XIV, ed.
  P.~{Shopbell}, M.~{Britton}, \& R.~{Ebert}, 29

\bibitem[{{Thorat} {et~al.}(2013){Thorat}, {Saripalli}, \&
  {Subrahmanyan}}]{2013MNRAS.434.2877T}
{Thorat}, K., {Saripalli}, L., \& {Subrahmanyan}, R. 2013, \mnras, 434, 2877

\bibitem[{{Tumlinson} {et~al.}(2017){Tumlinson}, {Peeples}, \&
  {Werk}}]{2017ARA&A..55..389T}
{Tumlinson}, J., {Peeples}, M.~S., \& {Werk}, J.~K. 2017, \araa, 55, 389

\bibitem[{{van Haarlem} {et~al.}(2013){van Haarlem}, {Wise}, {Gunst}, {Heald},
  {McKean}, {Hessels}, {de Bruyn}, {Nijboer}, {Swinbank}, {Fallows},
  {Brentjens}, {Nelles}, {Beck}, {Falcke}, {Fender}, {H{\"o}randel},
  {Koopmans}, {Mann}, {Miley}, {R{\"o}ttgering}, {Stappers}, {Wijers},
  {Zaroubi}, {van den Akker}, {Alexov}, {Anderson}, {Anderson}, {van Ardenne},
  {Arts}, {Asgekar}, {Avruch}, {Batejat}, {B{\"a}hren}, {Bell}, {Bell}, {van
  Bemmel}, {Bennema}, {Bentum}, {Bernardi}, {Best}, {B{\^i}rzan}, {Bonafede},
  {Boonstra}, {Braun}, {Bregman}, {Breitling}, {van de Brink}, {Broderick},
  {Broekema}, {Brouw}, {Br{\"u}ggen}, {Butcher}, {van Cappellen}, {Ciardi},
  {Coenen}, {Conway}, {Coolen}, {Corstanje}, {Damstra}, {Davies}, {Deller},
  {Dettmar}, {van Diepen}, {Dijkstra}, {Donker}, {Doorduin}, {Dromer}, {Drost},
  {van Duin}, {Eisl{\"o}ffel}, {van Enst}, {Ferrari}, {Frieswijk}, {Gankema},
  {Garrett}, {de Gasperin}, {Gerbers}, {de Geus}, {Grie{\ss}meier}, {Grit},
  {Gruppen}, {Hamaker}, {Hassall}, {Hoeft}, {Holties}, {Horneffer}, {van der
  Horst}, {van Houwelingen}, {Huijgen}, {Iacobelli}, {Intema}, {Jackson},
  {Jelic}, {de Jong}, {Juette}, {Kant}, {Karastergiou}, {Koers}, {Kollen},
  {Kondratiev}, {Kooistra}, {Koopman}, {Koster}, {Kuniyoshi}, {Kramer},
  {Kuper}, {Lambropoulos}, {Law}, {van Leeuwen}, {Lemaitre}, {Loose}, {Maat},
  {Macario}, {Markoff}, {Masters}, {McFadden}, {McKay-Bukowski}, {Meijering},
  {Meulman}, {Mevius}, {Middelberg}, {Millenaar}, {Miller-Jones}, {Mohan},
  {Mol}, {Morawietz}, {Morganti}, {Mulcahy}, {Mulder}, {Munk}, {Nieuwenhuis},
  {van Nieuwpoort}, {Noordam}, {Norden}, {Noutsos}, {Offringa}, {Olofsson},
  {Omar}, {Orr{\'u}}, {Overeem}, {Paas}, {Pandey-Pommier}, {Pandey}, {Pizzo},
  {Polatidis}, {Rafferty}, {Rawlings}, {Reich}, {de Reijer}, {Reitsma},
  {Renting}, {Riemers}, {Rol}, {Romein}, {Roosjen}, {Ruiter}, {Scaife}, {van
  der Schaaf}, {Scheers}, {Schellart}, {Schoenmakers}, {Schoonderbeek},
  {Serylak}, {Shulevski}, {Sluman}, {Smirnov}, {Sobey}, {Spreeuw}, {Steinmetz},
  {Sterks}, {Stiepel}, {Stuurwold}, {Tagger}, {Tang}, {Tasse}, {Thomas},
  {Thoudam}, {Toribio}, {van der Tol}, {Usov}, {van Veelen}, {van der Veen},
  {ter Veen}, {Verbiest}, {Vermeulen}, {Vermaas}, {Vocks}, {Vogt}, {de Vos},
  {van der Wal}, {van Weeren}, {Weggemans}, {Weltevrede}, {White}, {Wijnholds},
  {Wilhelmsson}, {Wucknitz}, {Yatawatta}, {Zarka}, {Zensus}, \& {van
  Zwieten}}]{2013A&A...556A...2V}
{van Haarlem}, M.~P., {Wise}, M.~W., {Gunst}, A.~W., {et~al.} 2013, \aap, 556,
  A2

\bibitem[{{Voit}(2005)}]{2005RvMP...77..207V}
{Voit}, G.~M. 2005, Reviews of Modern Physics, 77, 207

\bibitem[{{Wardle} \& {Kronberg}(1974)}]{1974ApJ...194..249W}
{Wardle}, J.~F.~C. \& {Kronberg}, P.~P. 1974, \apj, 194, 249

\bibitem[{{Wirth} {et~al.}(1982){Wirth}, {Smarr}, \&
  {Gallagher}}]{1982AJ.....87..602W}
{Wirth}, A., {Smarr}, L., \& {Gallagher}, J.~S. 1982, \aj, 87, 602

\bibitem[{{Worrall} \& {Birkinshaw}(2006)}]{2006LNP...693...39W}
{Worrall}, D.~M. \& {Birkinshaw}, M. 2006, in Lecture Notes in Physics, Berlin
  Springer Verlag, Vol. 693, Physics of Active Galactic Nuclei at all Scales,
  ed. D.~{Alloin}, 39

\bibitem[{{Worrall} {et~al.}(1995){Worrall}, {Birkinshaw}, \&
  {Cameron}}]{1995ApJ...449...93W}
{Worrall}, D.~M., {Birkinshaw}, M., \& {Cameron}, R.~A. 1995, \apj, 449, 93

\bibitem[{{Wright} {et~al.}(2010){Wright}, {Eisenhardt}, {Mainzer}, {Ressler},
  {Cutri}, {Jarrett}, {Kirkpatrick}, {Padgett}, {McMillan}, {Skrutskie},
  {Stanford}, {Cohen}, {Walker}, {Mather}, {Leisawitz}, {Gautier}, {McLean},
  {Benford}, {Lonsdale}, {Blain}, {Mendez}, {Irace}, {Duval}, {Liu}, {Royer},
  {Heinrichsen}, {Howard}, {Shannon}, {Kendall}, {Walsh}, {Larsen}, {Cardon},
  {Schick}, {Schwalm}, {Abid}, {Fabinsky}, {Naes}, \&
  {Tsai}}]{2010AJ....140.1868W}
{Wright}, E.~L., {Eisenhardt}, P.~R.~M., {Mainzer}, A.~K., {et~al.} 2010, \aj,
  140, 1868

\end{thebibliography}
\begin{appendix}
\renewcommand\thefigure{\thesection.\arabic{figure}}    
\section{Mid-infrared photometry and optical spectra of the host galaxies}
Table~\ref{tabWISEphot} contains the WISE photometry and Table~\ref{tabOptNIRphot} the optical and near-IR photometry. 
The optical spectra of the host galaxies are shown in Fig.~\ref{figGAMAspectra}. 


\begin{table*}[htbp]
\caption{WISE photometry and colors of the host galaxies, taken from the cross-identification in SDSS~DR14.}
\label{tabWISEphot}
\centering
\begin{tabular} {lcccccc}
\hline\hline
Source          &W1     &W2             &W3             
                        &W1-W2  &W2-W3\cr
\hline 
Exemplar                &12.93  &12.79  &12.04  &0.14   &0.75\cr
Double Irony    &13.10  &12.95  &11.56  &0.15   &1.39\cr
Difference      &$-0.17$  &$-0.16$  & 0.48  &$-0.01$ &$-0.64$\cr
\hline
\end{tabular}
\tablefoot{
The galaxies are not detected in the W4 (22~$\mu$m) band. 
The detections are weak in the W3 (12~$\mu$m) band (signal-to-noise ratio, S/N, of 3.4 for the Exemplar, and S/N = 2.3 for the Double Irony) but the galaxies are clearly detected in the W1 (3.4~$\mu$m) and W2 (4.6~$\mu$m) bands.
The WISE colors of the Exemplar and  the Double Irony are consistent with those of FR~{\sc i} radio galaxies. 
}
\end{table*}

\begin{table*}
\caption[]{Optical and near-infrared photometry of the host galaxies.}
\label{tabOptNIRphot}
\centering
\begin{tabular} {ccccccccc}
\hline\hline
\noalign{\smallskip}
Source          &$u$    &$g$    &$r$    &$i$    &$z$    &$J$    &$H$    &$K_s$\cr
\noalign{\smallskip}
\hline 
\noalign{\smallskip}
Exemplar       & $19.59\pm0.08$ &$17.43\pm0.01$   &$16.37\pm0.00$   &$15.90\pm0.00$   &$15.56\pm0.01$   &15.5       &14.742 &14.176\cr
Double Irony    &$19.26\pm0.06$ &$17.35\pm0.91$ &16.26          &15.82          &$15.47\pm0.01$  &15.223 &14.634 &13.987\cr
Mag. difference & 0.33          &0.08           & 0.11          &0.08           &0.09           &0.28   &0.11   &0.189\cr
\hline
\end{tabular}
\tablefoot{
The $ugriz$ photometry is from the SDSS~DR14 and the NIR photometry from 2MASS. 
The last line shows the magnitude difference between the Exemplar and the Double Irony. 
}
\end{table*}

\setcounter{figure}{0}    
\begin{figure*}
\centering
   \includegraphics[width=17.7cm]{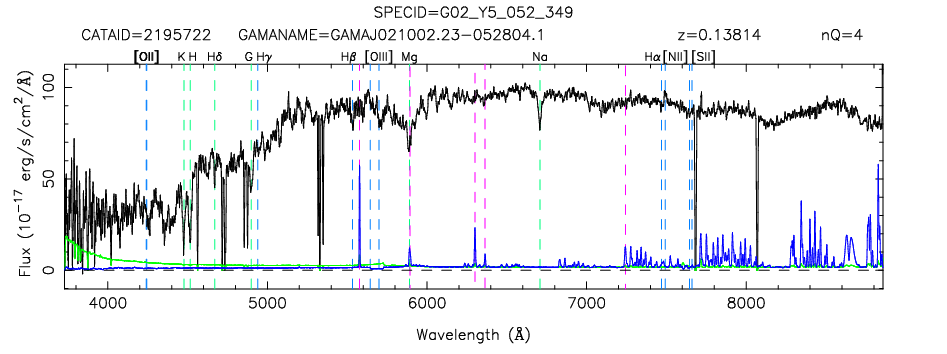}
   \includegraphics[width=17.7cm]{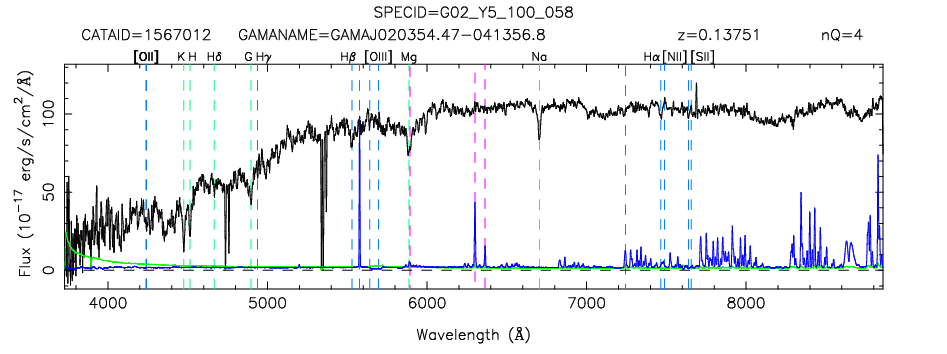}
   \caption{GAMA optical spectra of the host galaxies, displayed in the observer's frame.
   The green curve shows the noise level and the blue line is the sky spectrum.   
   {\it Top panel:} The Exemplar. 
   {\it Bottom panel:} The Double Irony. 
The two galaxies have remarkably similar optical spectra.}
\label{figGAMAspectra}
\end{figure*}

\end{appendix}
\end{document}